\documentclass[twocolumn]{aastex631}
\graphicspath{{FIGURES}}

\DeclareRobustCommand{\ion}[2]{%
\relax\ifmmode
\ifx\testbx\f@series
{\mathbf{#1\,\mathsc{#2}}}\else
{\mathrm{#1\,\mathsc{#2}}}\fi
\else\textup{#1\,{\mdseries\textsc{#2}}}%
\fi}

\DeclareTextFontCommand{\bred}{\color{red}\bfseries}

\newcommand{\JHU}{Department of Physics and Astronomy, The Johns Hopkins University, Baltimore, MD 21218.}
\newcommand{\STScI}{Space Telescope Science Institute, Baltimore, MD 21218.}

\newcommand{\Harvard}{Harvard-Smithsonian Center for Astrophysics, 60 Garden Street, Cambridge, MA 02138, USA}
\newcommand{\IfA}{Institute for Astronomy, University of Hawaii, 2680 Woodlawn Drive, Honolulu, HI 96822, USA}
\newcommand{\IfADavid}{Institute for Astronomy, University of Hawaii, 640 N. A'ohoku Pl., Hilo, HI 96720, USA}

\newcommand{\UCSC}{Department of Astronomy and Astrophysics, University of California, Santa Cruz, CA 95064, USA}
\newcommand{\QUB}{Astrophysics Research Centre, School of Mathematics and Physics, Queen's University Belfast, Belfast BT7 1NN, UK}

\newcommand{\Northwestern}{Center for Interdisciplinary Exploration and Research in Astrophysics (CIERA) and Department of Physics and Astronomy, Northwestern University, Evanston, IL 60208, USA}
\newcommand{\DARK}{DARK, Niels Bohr Institute, University of Copenhagen, Jagtvej 128, 2200 Copenhagen, Denmark}
\newcommand{\Illinois}{Department of Astronomy, University of Illinois at Urbana-Champaign, 1002 W. Green St., IL 61801, USA}

\newcommand{\Melbourne}{School of Physics, The University of Melbourne, VIC 3010, Australia}

\newcommand{\NSFAI}{The NSF AI Institute for Artificial Intelligence and Fundamental Interactions}
\newcommand{\MIT}{Department of Physics, Massachusetts Institute of Technology, Cambridge, MA 02139, USA}
\newcommand{\INAFNAP}{INAF-Osservatorio Astronomico di Capodimonte, salita Moiariello 16, I-80121, Naples, Italy}
\newcommand{\Thai}{National Astronomical Research Institute of Thailand: Chiang Mai, TH}
\newcommand{\Thacher}{The Thacher School, 5025 Thacher Rd., Ojai, CA 93023, USA}
\newcommand{\Queens}{School of Mathematics and Physics, The University of Queensland, QLD 4072, Australia}

\begin{document}

\title{SN 2021foa: The ``Flip-Flop'' Type IIn / Ibn supernova}

\author[0000-0002-6886-269X]{D.~Farias}
\affiliation{\DARK}

\author[0000-0002-8526-3963]{C.~Gall}
\affiliation{\DARK}

\author[0000-0001-6022-0484]{G.~Narayan}
\affiliation{\Illinois}

\author[0000-0002-3825-0553]{S.~Rest}
\affiliation{\JHU}

\author[0000-0002-5814-4061]{V.~A.~Villar}
\affiliation{\Harvard}
\affiliation{\NSFAI}

\author[0000-0002-4269-7999]{C.~R.~Angus}
\affiliation{\DARK}
\affiliation{\QUB}

\author[0000-0002-4449-9152]{K.~Auchettl}
\affiliation{\UCSC}
\affiliation{\Melbourne}

\author[0000-0002-5680-4660]{K.~W.~Davis}
\affiliation{\UCSC}

\author[0000-0002-2445-5275]{R.~J.~Foley}
\affiliation{\UCSC}

\author[0000-0003-4906-8447]{A.~Gagliano}
\affiliation{\Harvard}
\affiliation{\MIT}
\affiliation{\NSFAI}

\author[0000-0002-4571-2306]{J.~Hjorth}
\affiliation{\DARK}

\author[0000-0001-9695-8472]{L.~Izzo}
\affiliation{\DARK}
\affiliation{\INAFNAP}

\author[0000-0002-5740-7747]{C.~D.~Kilpatrick}
\affiliation{\Northwestern}

\author[0009-0000-5561-9116]{H~.M.~L.~Perkins}
\affiliation{\Illinois}

\author[0000-0003-2558-3102]{E.~Ramirez-Ruiz}
\affiliation{\UCSC}

\author[0000-0003-4175-4960]{C.~L.~Ransome}
\affiliation{\Harvard}

\author[0000-0002-9820-679X]{A.~Sarangi}
\affiliation{\DARK}

\author[0000-0003-0381-1039]{R.~Yarza}
\affiliation{\UCSC}

\author[0000-0003-4263-2228]{D.~A.~Coulter}
\affiliation{\STScI}

\author[0000-0002-6230-0151]{D.~O.~Jones}
\affiliation{\IfADavid}

\author[0000-0003-2720-8904]{N.~Khetan}
\affiliation{\Queens}

\author[0000-0002-3825-0553]{A.~Rest}
\affiliation{\STScI}
\affiliation{\JHU}

\author[0000-0003-2445-3891]{M.~R.~Siebert}
\affiliation{\STScI}

\author[0000-0002-9486-818X]{J.~J.~Swift}
\affiliation{\Thacher}

\author[0000-0002-5748-4558]{K.~Taggart}
\affiliation{\UCSC}

\author[0000-0002-1481-4676]{S.~Tinyanont}
\affiliation{\UCSC}
\affiliation{\Thai}

\author[0009-0009-2891-9151]{P.~Wrubel}
\affiliation{\Thacher}

\author[0000-0001-5486-2747]{T.~J.~L.~de~Boer}
\affiliation{\IfA}

\author[0000-0001-8756-1262]{K.~E.~Clever}
\affiliation{\UCSC}

\author[0000-0002-5950-1702]{A.~Dhara}
\affiliation{\UCSC}

\author[0000-0003-1015-5367]{H.~Gao}
\affiliation{\IfA}

\author[0000-0002-7272-5129]{C.-C.~Lin}
\affiliation{\IfA}


\correspondingauthor{D.~Farias}
\email{diego.farias@nbi.ku.dk}

\begin{abstract}
We present a comprehensive analysis of the photometric and spectroscopic evolution of SN~2021foa,
 unique among the class of transitional supernovae for repeatedly changing its spectroscopic appearance from hydrogen-to-helium-to-hydrogen-dominated (IIn-to-Ibn-to-IIn) within 50 days past peak brightness.
The spectra exhibit multiple narrow ($\approx$ 300--600~km~s$^{-1}$)
absorption lines of hydrogen, helium, calcium and iron together with broad helium emission lines with a full-width-at-half-maximum (FWHM) of $\sim 6000$~km~s$^{-1}$. 
For a steady, wind-mass loss regime, light curve modeling 
results in an ejecta mass of $\sim 8$ M$_{\odot}$ and CSM mass below 1 M$_{\odot}$, and an ejecta velocity consistent with the FWHM of the broad helium lines. 
We obtain a mass-loss rate of $\approx 2$ M$_{\odot} {\rm yr}^{-1}$. This mass-loss rate is three orders of magnitude larger than derived for normal Type II SNe. 
We estimate that the bulk of the CSM of SN~2021foa must have been expelled within half a year, about 12 years ago. 
Our analysis 
suggests that SN~2021foa had a helium rich ejecta which swept up a dense shell of hydrogen rich CSM shortly after explosion.
At about 60 days past peak brightness, the photosphere recedes through the dense ejecta-CSM region, occulting much of the red-shifted emission of the hydrogen and helium lines, which results in observed blue-shift ($\sim -3000$~km~s$^{-1}$). 
Strong mass loss activity prior to explosion, such as those seen in SN~2009ip-like objects and SN~2021foa as precursor emission, are the likely origin of a complex, multiple-shell CSM close to the progenitor star. 

\end{abstract}


\vspace{1cm}
\section{Introduction} \label{sec:intro}

Massive stars ($\gtrsim$ 8 M$_{\odot}$) undergo different mass-loss phases, losing material from their outer layers shortly before core-collapse~\citep[CC,][]{Smartt_progenitor,smith_review}. Analysis of this expelled material, termed circumstellar material (CSM), can provide important information about the progenitor system
and thus the late-stage of massive stellar evolution~\citep[see e.g.,][]{Morozova_progenitormass}. 
The CSM surrounds the progenitor and thus, the supernova (SN) radiation and ejecta inevitably interact with the CSM. The emanating signatures arising from the interaction appear at a variety of phases during the evolution of the SN, depending primarily upon the mass distribution of the CSM~\citep[][]{Dessart_interactionsignature}. 
This interaction produces SN spectra that can be dominated by narrow ($\sim 100 - 500$~km~s$^{-1}$) or intermediately broad ($\sim 1000$~km~s$^{-1}$) emission lines and P-Cygni profiles~\citep[see][and references therein]{Fraser_2020rev}. 

Depending upon the progenitor system, as well as the composition and radial distribution  of the CSM, different classes of core-collapse supernovae (CCSNe) have been identified.

CCSNe with a hydrogen (H) rich CSM and little helium (He) emission in their spectra are commonly classified as Type IIn \citep{Schlegel_IIn}. 
Classical examples of such events are e.g., SN~1998S, SN~2005ip and SN~2010jl~\citep{Fox_2005ip,Mauerhan_1998S,Fransson_2010jl,Gall_2010jl}. 
However, if the CSM is He-rich with little-to-no H emission in the SN spectra, then the CCSNe are classified as Type Ibn~\citep{Pastorello_2007_2006jc,Foley_2006jc}.
In recent years, another class of interacting SNe has emerged, the Type Icn~\citep[][]{GalYam_2022,Pellegrino_Icn,Davis2022ann}. 
These SNe exhibit narrow oxygen (O) and carbon (C) emission lines in their spectra. 

In the local universe Type IIn and Ibn SNe comprise about 5\% and 1\% of the volumetric rate of CCSNe, respectively~\citep{Maeda_Moriya2022,Cecilie}.
Among the interacting CCSNe, Type Icn are the rarest, with only five members known thus far~\citep[see][]{Davis2022ann}. 
However, the classification of several interacting SNe is ambiguous, as some CCSNe appear to change their type, e.g., from Type IIn to Ibn or vice versa. Prominent examples are SN~2005la~\citep{Pastorello_2008_2005la}, SN~2011hw~\citep{Smith_2012_2011hw,Pastorello_2015_2011hw}, iPTF15akq~\citep{Hosseinzadeh_2017} and SN~2020bqj~\citep{Kool_2021_2020bqj}.
These objects form the unique group of transitional IIn/Ibn SNe. 

Determining the exact nature of the progenitor of interacting SNe is challenging due to the complexities of the interaction between the ejecta and the CSM. Thus, SNe IIn have diverse light curves (LCs), spanning a broad range of peak magnitudes~\citep{Nyholm2020} and LC shapes. 
This has led to suggest a wide range of plausible progenitors systems for Type IIn SNe, such as
red supergiants (RSG) in binary systems~\citep[SN~1998S-like objects;][]{smith_review},  Luminous Blue Variables (LBV) for e.g., SN~2005gl~\citep[][]{GalYam_2005gl}, SN~2009ip~\citep{Foley2011_2009ip,Smith_2009ip_2010mc}, SN~2010jl~\citep{Smith_2010jl} and SN~2015bh~\citep{Boian_2015bh}, while a $\sim20$ M$_{\odot}$ star is preferred for SN~2016jbu~\citep{Kilpatrick_2018,Brennan_2022b}. 
The progenitor of SN~2015bh is also proposed to be a yellow supergiant~\citep[$\sim 50$ M$_{\odot}$;][]{Thone_2017}.

Contrary, LCs of Type Ibn SNe show a high degree of homogeneity~\citep{Hosseinzadeh_2017}. Thus, the most plausible progenitor is an anevolved Wolf-Rayet star~\citep{Pastorello_Ibn_2008}, which is consistent with the majority of Ibn SNe being found in star-forming galaxies~\citep[although see PS1-12sk;][]{Hosseinzadeh_2019}. Furthermore, the mass loss rates derived from LC modeling of e.g., OGLE-2014-SN-131 and SN~2020bqj~\citep[e.g.,][]{Karamehmetoglu_2017,Kool_2021_2020bqj} favour such a progenitor.
Alternatively, Type Ibn SNe may be the explosion of a low mass helium star in a binary system~\citep{Dessart_2022}. 
Unlike Type IIn SNe~\citep[see SN~2005gl, ][]{GalYam_2005gl}, there are no detection of any progenitor system of Type Ibn SNe in archival data. 
However, late-time photometry at the location of SN~2006jc has shown a potential companion associated with the exploding star~\citep{Maund_2016,Sun_2020}.

A handful of CCSNe exhibited pre-explosion activities or outbursts up to two decades prior to their terminal explosion.
Precursor emission in Type IIn SNe is common~\citep[e.g.,][]{Mauerhan_2009ip_finalexp,Ofek_2010mc,Thone_2017,Elias_2016,Pastorello_2018_09iplike,Hiramatsu_2024}, such as the case of SN~2009ip~\citep{Pastorello_2013_09ip, Margutti_2014_2009ip}.
This transient was first 
classified as a SN impostor~\citep[e.g.,][]{Foley2011_2009ip}. 
However, after two more outbursts in 2011 and 2012, its ``final'' re-brightening in 2012 reached $M_R \approx -18$ mag, which was proposed as the terminal explosion of a SN~IIn~\citep{Mauerhan_2009ip_finalexp}.
On the contrary, precursor emission has only been observed for three Type Ibn supernovae such as SNe~2006jc~\citep{Foley_2006jc}, 2019uo~\citep{Strotjohann_2021} and 2023fyq~\citep{Brennan_23fyq,Dong_23fyq}.

Interacting SNe with observed signatures of pre-explosion outbursts and a photometric and spectroscopic evolution similar to that of SN~2009ip, are termed 2009ip-like objects~\citep{Pastorello_2018_09iplike, Brennan_2022a}. 
Prominent examples of this class include SN~2015bh~\citep{Elias_2016,Thone_2017}, SN~2016jbu~\citep{Kilpatrick_2018,Brennan_2022a} and SN~2019zrk~\citep{Fransson_2019zrk}.
Typically, photometric and spectroscopic data obtained around the epochs of the outbursts suggest that these are LBV-like eruptions prior to the presumed, terminal explosion of the progenitor~\citep{Pastorello_2013_09ip, Thone_2017, Brennan_2022b}. 
However, whether or not the latter are indeed, stellar explosions remains unclear~\citep{Smith_2022}.

Here, we present unpublished multi-band photometry and time-series spectroscopy of the fifth transitional Type IIn/Ibn SN~2021foa. The data were collected by the Young Supernova Experiment~\citep{Jones_2021,Aleo_DR1YSE,Coulter_YSEPZ}.
The SN (RA=13:17:12.29, DEC=$-17$:15:24.19) was discovered by ASAS-SN~\citep[][]{ASASN} on March 15, 2021 ($g\sim 15.9$) in the galaxy~IC~086~(ASASSN-21dg). It was initially classified as a Type IIn due to the strong, narrow Balmer lines in the optical spectrum~\citep{2021TNSCR1133....1A}. An analysis of its light curve and spectra until $+79$ days \citep{Reguitti_2022} suggests that 
SN~2021foa is photometrically similar to 
SN~2009ip-like objects such as SN~2005gl~\citep{GalYam_2005gl}, SN~2009ip~\citep{Pastorello_2013_09ip} and notably, SN~2016jbu~\citep{Kilpatrick_2018,Brennan_2022a}, while spectroscopically resembles the transitional IIn/Ibn SNe.
In this work, we present a comprehensive analysis of the spectroscopic and photometric evolution out to +427 days. We detail a physical picture of this unusual SN, which, amongst its peers in the 2009ip-like class, exhibits some unique characteristics.   

We determine the time of maximum light in $r-$band using  a second-degree polynomial fit between MJD 59280 and 59315 to be MJD$_{{\tt max}}=$ $59302.35 \pm 0.14$ ($r_{max}$). We use this as our reference time in the remainder of the paper.
Throughout this work, we assume a standard $\Lambda$CDM cosmology with $H_0=67.8$~km~s$^{-1}$ Mpc$^{-1}$ and $\Omega_{m} = 0.307$~\citep{Planck_13}.

The paper is structured as follows: Sect.~\ref{sec:data} describes the data acquisition and reduction.
In Sect.~\ref{sec:analysis} we present our photometric and spectroscopic analysis methodology, with the results presented in Sect.~\ref{sec:results}. We discuss our results from the analysis  in Sect.~\ref{sec:discuss}, and conclude by formulating a complete physical scenario for SN~2021foa in Sect.~\ref{sec:scenario}.

\vspace{1cm}
\section{Observations} \label{sec:data}

Our photometric and spectroscopic follow up observations of SN~2021foa, including archival and public data are described below. 

\subsection{Photometric data} \label{subsec:photdat}

We obtained optical photometry with the Sinistro imagers on the Las Cumbres Observatory (LCO) 1-m telescope network \citep{Brown_2013} in $UBVRI,up,gp,rp,ip,z$ bands, starting March 15, 2021 through August 17, 2021. 
Initial automatic processing of the imaging data, including instrument signature removal, pixel-level corrections and astrometric calibration, was performed by the LCO {\tt BANZAI} pipeline \citep{McCully_2018}. Thereafter, images were processed using {\tt photpipe} pipeline \citep{Rest_2005, Jones_2021}. We measured the flux of SN\,2021foa from the LCO images using an updated version of {\tt DoPhot}~\citep{Schechter_1993}, and this photometry was calibrated using $u$-band SDSS~\citep{Alam_2015} together with $griz$ Pan-STARRS1 photometric standards observed in the vicinity of SN~2021foa. 
We calibrated the photometry of R and I bands on the Cousins photometric system. 

\begin{figure*}[hpt!]
\epsscale{1.2}
\plotone{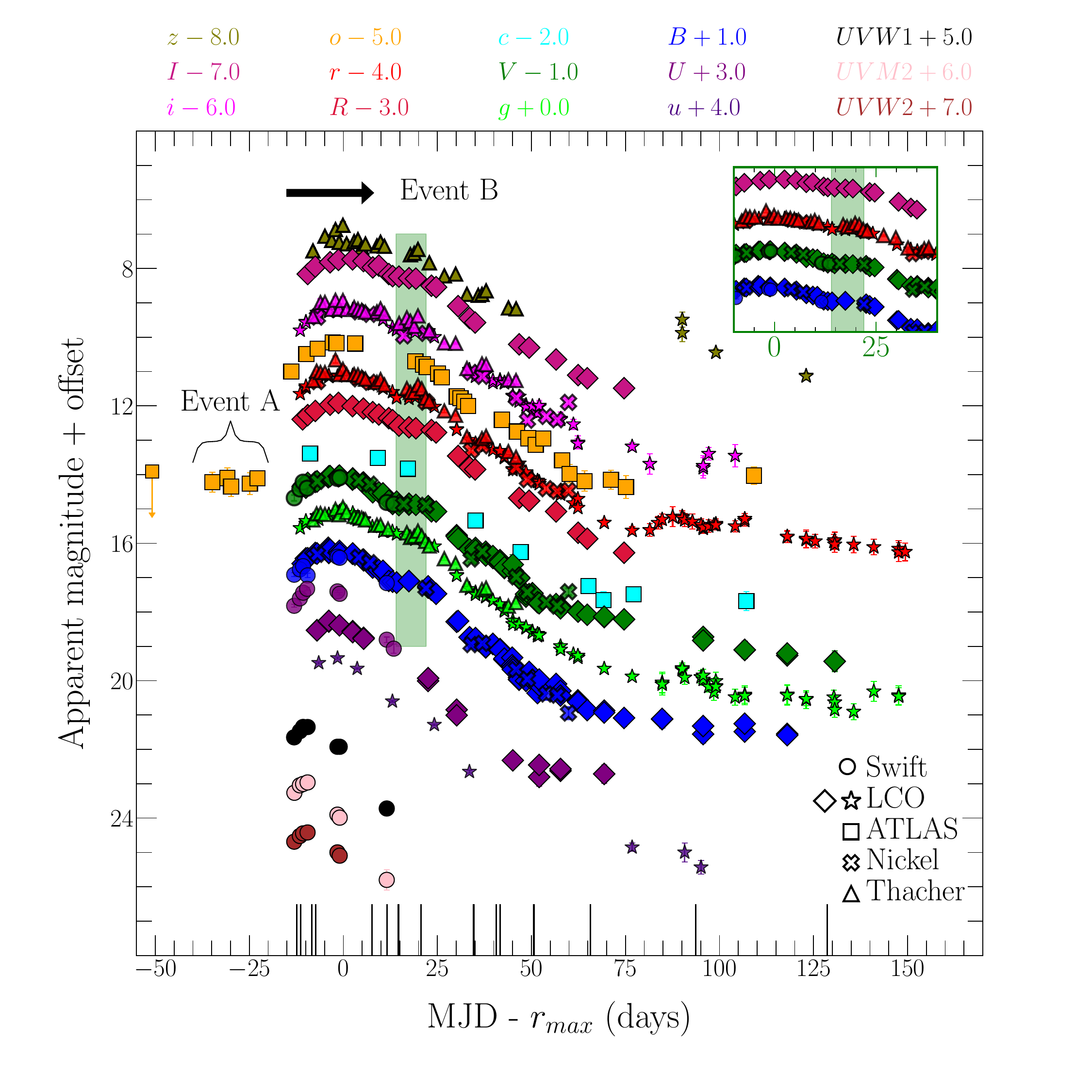}
\caption{Photometry of SN~2021foa.
{\it Swift UVW1, UVM2, UVW2}, Thacher $griz$, Nickel $BVri$, and LCO/Sinistro $UBVRI,up,gp,rp,ip,z$ data (AB system) are symbol and colour coded as described in the legend.  
All light curves are offset by a constant, for clarity. 
The gray vertical lines mark our spectroscopic data at their respective epochs relative to 
$r_{max}$ at MJD$_{{\tt max}}= 59302.35$ days. 
Event A encompasses the time where the light-curve of SN~2021foa shows precursor emission. 
Uncertainties are smaller than the size of the markers. 
The green band delimits the phases where a short plateau is observed (from $+14$ to $+22$ days). 
The inset highlights that same phase region. 
}
\label{fig:lc}
\end{figure*}

Observations with the Ultraviolet and Optical Telescope \citep[UVOT;][]{Roming_2005} on the {\it Neil Gehrels {\it Swift} Observatory} were reported in \citet{Reguitti_2022}. UVOT data of SN~2021foa  were taken between 
March 16, 2021 and May 10, 2021. Following the methodolgy described in \citet{Brown_2014}, 
we use {\tt uvotsource} from the {\tt HEASoft v6.26} package to perform aperture photometry within a 3\arcsec\ aperture centered on SN~2021foa. We measured the total background flux at the location of SN~2021foa from frames obtained on June 17 2022, when any residual light from the fading SN is well below the sky-background. The background emission was then subtracted from all previous observations. We detect emission of SN~2021foa at a $>3\sigma$ level in all UVOT bands in the 2021 observations.

SN~2021foa was also observed by the Asteroid Terrestrial-impact Last Alert (ATLAS) System~\citep{ATLAS_Tonry} between 
February 22 to July 16 and March 20 to July 14 in $o$- and $c$-bands, respectively. Following the procedure described by~\citet{Davis2022ann}, we obtained the binned light curve data calculated as a 3$\sigma$-cut weighted mean for each night. In contrast to~\citet{Reguitti_2022,Reguitti_precursor}, we did not find any significant detection in the $c$-band prior to February 22 (see Sect.~\ref{subsec:pre}).

Additional images of SN~2021foa were obtained in $BVri$ bands with the 1-m Nickel telescope at Lick Observatory and in $up, gp, rp, zp$ bands with the Thacher 0.7-m telescope in Ojai, CA~\citep{Thacher_Swift}. The images from Nickel telescope were calibrated using bias and sky flat-field frames following standard procedures. PSF photometry was performed, and photometry was calibrated relative to Pan-STARRS1 photometric standards \citep{Flewelling16}.
Similarly, photometry of the images from Thacher telescope was obtained using {\tt DoPhot} and calibrated with the $griz$ Pan-STARRS1 catalog.

Fig.~\ref{fig:lc} shows our photometric data of SN~2021foa, including LCO/Sinistro ($UBVRI,up,gp,rp,ip,z$), our re-reduced and host galaxy subtracted UVOT ($UVW1, UVW2, UVM2, U, B$ and $V$) data, ATLAS ($c$ and $o$), Nickel ($BVri$) and Thacher ($griz$) data. 
All photometric data are summarized in Tab.~\ref{tab:photo}.

\subsection{Spectroscopic data} \label{subsec:specdat}

\begin{figure*}[ht!]
\epsscale{1.2}
\plotone{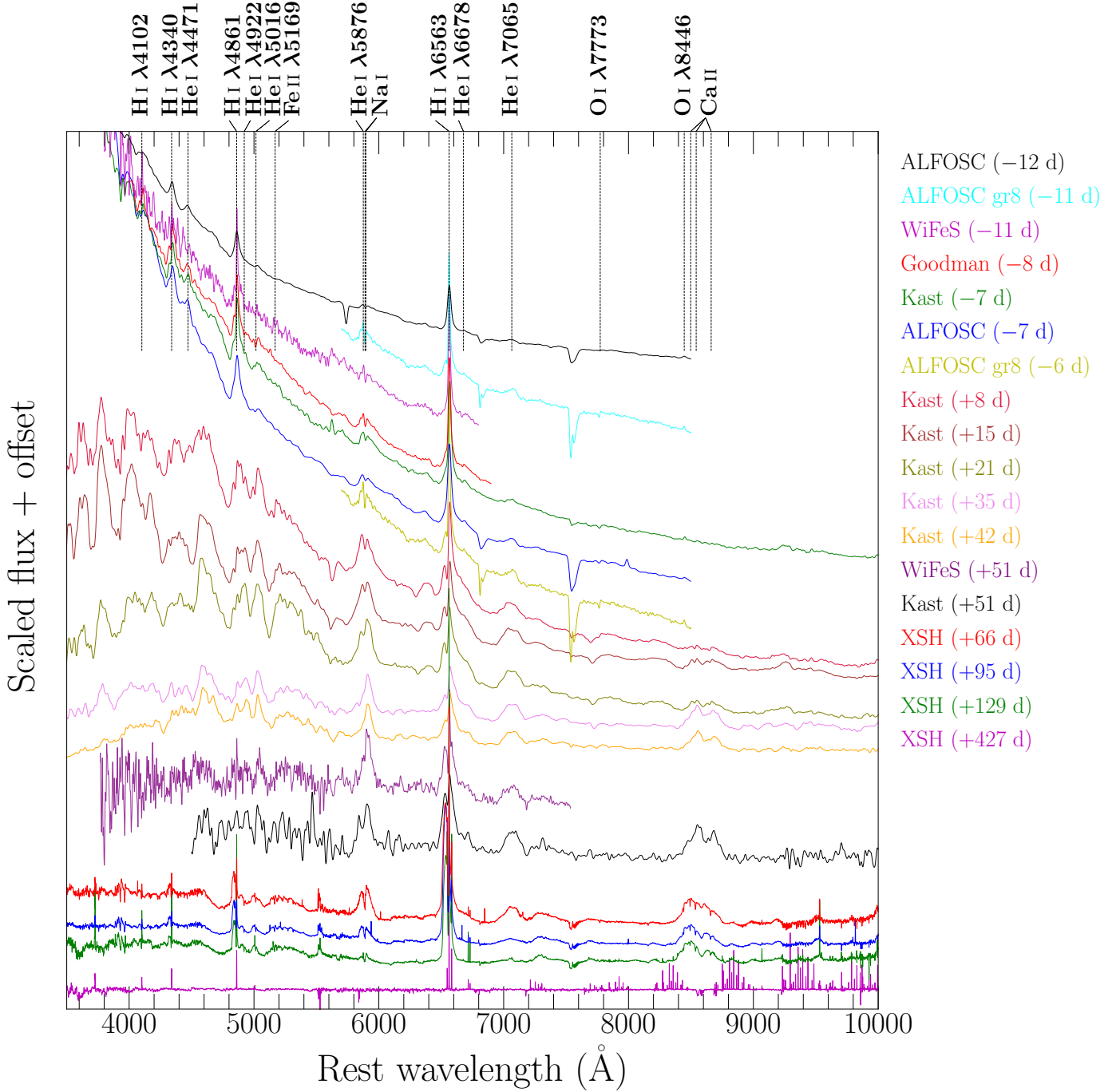}
\caption{
 Low and medium (XSH;VLT/X-shooter) resolution optical spectra of SN~2021foa (Sect.~\ref{subsec:specdat}, Tab.~\ref{tab:spec}).
Black, dashed vertical lines indicate the rest-frame wavelength of the strongest lines observed in SN~2021foa, mostly \ion{H}{i} and \ion{He}{i} lines. For visualization purposes, we smoothed all the spectra with a rolling Gaussian with kernel size of $\sim 15$ \AA. 
} 
\label{fig:spectra_optical}
\end{figure*}

Fig.~\ref{fig:spectra_optical} shows our extensive optical spectroscopic follow-up observations of SN~2021foa obtained within $-12$ days and $+427$ days. Data were obtained with the Kast dual-beam spectrograph~\citep{Miller_Kast} on the Lick Shane 3-m telescope at $-7$, $+8$, $+15$, $+21$, $+35$, $+42$ and $+51$ days, the Goodman spectrograph~\citep{Clemens_Goodman} on the NOIRLab 4.1-m Southern Astrophysical Research (SOAR) telescope at Cerro Pachón at $-8$ day, the Alhambra Faint Object Spectrograph and Camera (ALFOSC) on the Nordic Optical Telescope (NOT) at $-11$, $-7$ and $-6$  days and the Wide Field Spectrograph (WiFeS) at the Australian National University (ANU) 2.3-m telescope located at Siding Spring Observatory~\citep{Dopita_2007} at $-11$ and $+51$ days.

The Kast observations are performed with the blue side 452/grism, 300/7500 grating, d58 dichroic and 2\arcsec\ slit. Goodman observations were carried out using the 400 lines/mm grating with the M1 wavelength setting ($300-705$ nm). To reduce the Kast and Goodman spectra, we used the {\tt UCSC Spectral Pipeline}\footnote{\url{https://github.com/msiebert1/UCSC_spectral_pipeline}}  \citep{Siebert2019}. 
The ALFOSC spectra were taken with a  1.0\arcsec\ slit and
grisms 4, and 8. For all reduction, extraction and calibration steps, we used standard {\tt IRAF}\footnote{\url{https://iraf-community.github.io}} routines using {\tt PYRAF}\footnote{\url{https://github.com/iraf-community/pyraf}}.

The Wide Field Spectrograph (WiFeS) is an integral-field spectrograph with a field-of-view of $38 \times 25$ arcsec. 
SN~2021foa was observed using a RT-560 beam splitter and a B3000 and R3000 diffraction gratings that covers the $3200-5900$ \AA\ and $5300-9800$ \AA\  wavelength ranges. All observations had a Y=2 binning readout mode, corresponding to a $1\times 1$ sq. arcsec spaxel.
Each observation was reduced using {\tt PyWiFeS}~\citep{PyWiFeS}. 
We extract an isolated part of the sky for background determination and subtraction.

\begin{figure*}[ht!]
\epsscale{1.2}
\plotone{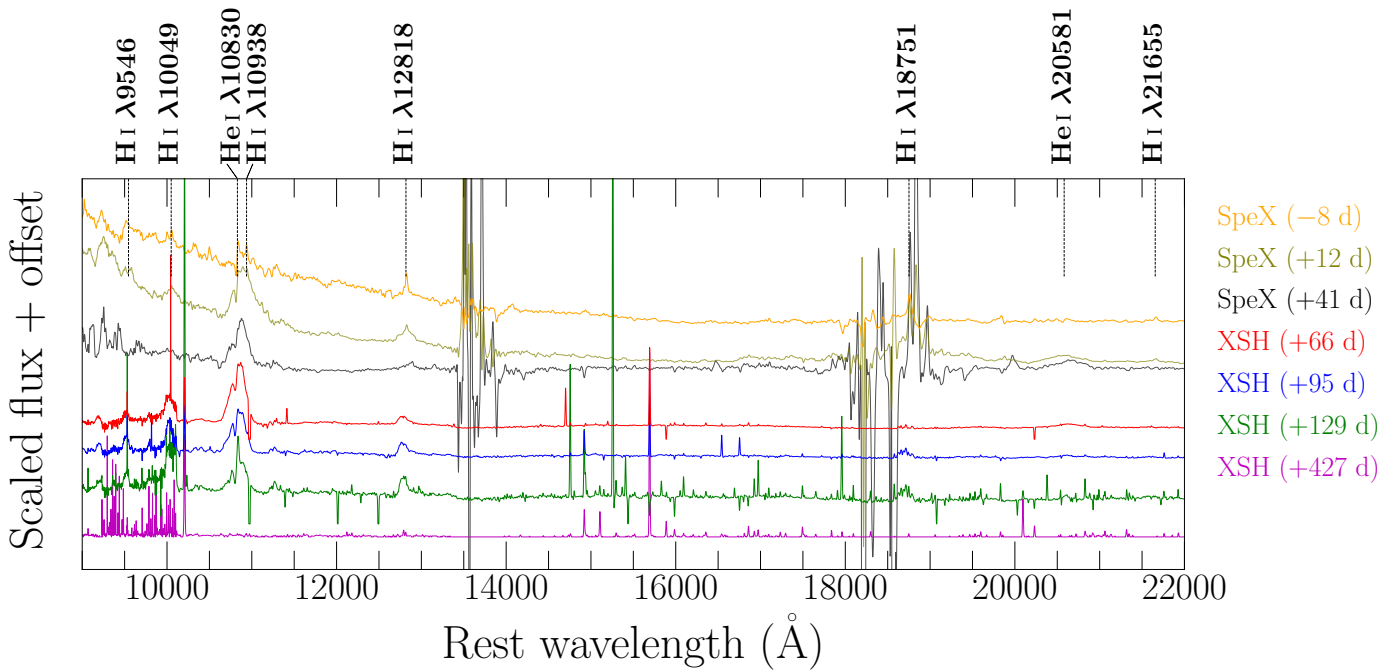}
\caption{Low and medium (XSH;VLT/X-shooter) resolution NIR spectra of SN~2021foa (see Sect.~\ref{subsec:specdat}, Tab.~\ref{tab:spec}). For visualization purposes, the Xshooter spectra was trimmed from 2.2 to~$2.5\mu$m due to low signal-to-noise in that region. All spectra were smothed with a rolling Gaussian with kernel size of $\sim 15$ \AA.}
\label{fig:spectra_nir}
\end{figure*}

Fig.~\ref{fig:spectra_nir} presents our NIR spectra of SN~2021foa which were obtained with the SpeX spectrograph~\citep{Rayner_2003_SpeX} mounted on the 3-m NASA Infrared Telescope Facility (IRTF) at $-8$, $+12$ and $+41$ days. In this mode with the 0.8\arcsec\ slit, the spectral resolving power is $R \approx 1000$. The SN was observed in an ABBA dithering pattern with an A0V star observed immediately before or after science observations for telluric correction. We also obtain observations of internal flat field and arcs calibration lamps at the science pointing. We reduced the data using \texttt{spextool} \citep{Cushing04}, which performed flat fielding, wavelength calibration, background subtraction, and spectral extraction. We then performed telluric correction using \texttt{xtellcor} \citep{Vacca03}.

Medium resolution spectra were obtained with the X-shooter echelle spectrograph~\citep{Vernet_Xshooter} mounted at the Very Large Telescope (VLT) at the European Southern Observatory (ESO) on Cerro Paranal, Chile. The data are presented in Figs.~\ref{fig:spectra_optical} and~\ref{fig:spectra_nir}. 
The X-shooter instrument covers the wavelength range of $0.3-2.5 \mu$m in three  arms, the ultraviolet and blue (UVB), visual (VIS) and near-infrared (NIR) wavelength ranges. The slit widths (and resolving power) for UBV, VIS and NIR for these observations were 0.9\arcsec\ ($R=5900$), 1.0\arcsec\ ($R=8900$) and 1.0\arcsec\ ($R=5600$), respectively.  
UVB, VIS and NIR arms were reduced with {\tt esoreflex 2.11.5}\footnote{\url{https://www.eso.org/sci/software/esoreflex/}}~\citep{Freudling_esoreflex} pipeline individually. Then, using a custom {\tt Python} code, UVB and VIS arms were combined in {\tt STARE} mode, while the NIR arm was combined in {\tt NOD} mode. Special efforts have been made to perform a detailed background subtraction around strong host galaxy emission lines such as H$\alpha$~(see Fig.~\ref{fig:hostlines}). 

To improve the flux-calibration from the mentioned pipelines, we mangle all low resolution spectra to the interpolated photometry from {\tt EXTRABOL}\footnote{\url{https://github.com/villrv/extrabol}}, as described in Sect.~\ref{subsec:lcbol}.
The UBV and VIS arms of the first three VLT/X-shooter spectra ($+$66, $+$95 and $+$129 days) were also flux-calibrated in this way. Due to the lack of NIR photometry at these epochs, we used the calibration provided by {\tt esoreflex}. The same approach was applied for the last spectrum at $+$427 days, and as this late-time spectrum includes minimal contamination from the SN, we use narrow H$\alpha$ to determine the redshift ($z=0.0086183$).

\subsection{Extinction}\label{subsec:ext}
The Milky Way extinction was obtained from~\citet{S&F_2011}; $E_{(B-V)_{\rm MW}} = 0.072$ for $R_V$ = 3.1 and the host galaxy extinction,  $E_{(B-V)_{\rm host}} = 0.129$, was adopted from \citet{Reguitti_2022}. 
We confirmed this value from measurements of the equivalent width (EW) of
the \ion{Na}{i}D ($\lambda 5890,5896$) absorption lines detected in our medium-resolution VLT/X-shooter spectra (see Sect.~\ref{sec:specana}) together with the extinction--EW relation from \citet{Turatto_2003}. We obtain an EW of $0.85\pm 0.36$ \AA, similar to measurements  
by~\citet{Reguitti_2022}. 

\vspace{1cm}

\section{Analysis Methodology}\label{sec:analysis}

The rich photometric and spectroscopic data set shows
that SN~2021foa has a prolonged precursor emission similar to Type IIn SN~2009ip, but spectroscopically closer to transitional IIn/Ibn objects, with strong \ion{He}{i} emission lines in the optical and near-infrared wavelengths. 
To understand the physical mechanism that drive the similarities and differences
and to put together a coherent physical scenario that can explain all observables we employ a range of analysis methods. 
First, we model the multiband photometry to obtain information about the progenitor and the CSM associated to SN~2021foa, such as the ejecta and CSM masses, and progenitor radius.
Secondly, we investigate the evolution of the most prominent lines (e.g., \ion{H}{i}, \ion{He}{i}, \ion{Ca}{ii}) of SN~2021foa by fitting their broad and narrow velocity components in emission and absorption. 
Such analysis provides insights of the complexity of the CSM while also highlights the transitional nature of SN~2021foa among IIn and Ibn classes.
Lastly, we fit two black body functions to the X-shooter spectra from optical to NIR wavelengths to estimate the temperature and amount of dust present at the respective epochs in SN~2021foa.

\subsection{Photometry} \label{sec:photana}

\begin{figure}[hbt!]
\epsscale{1.2}
\plotone{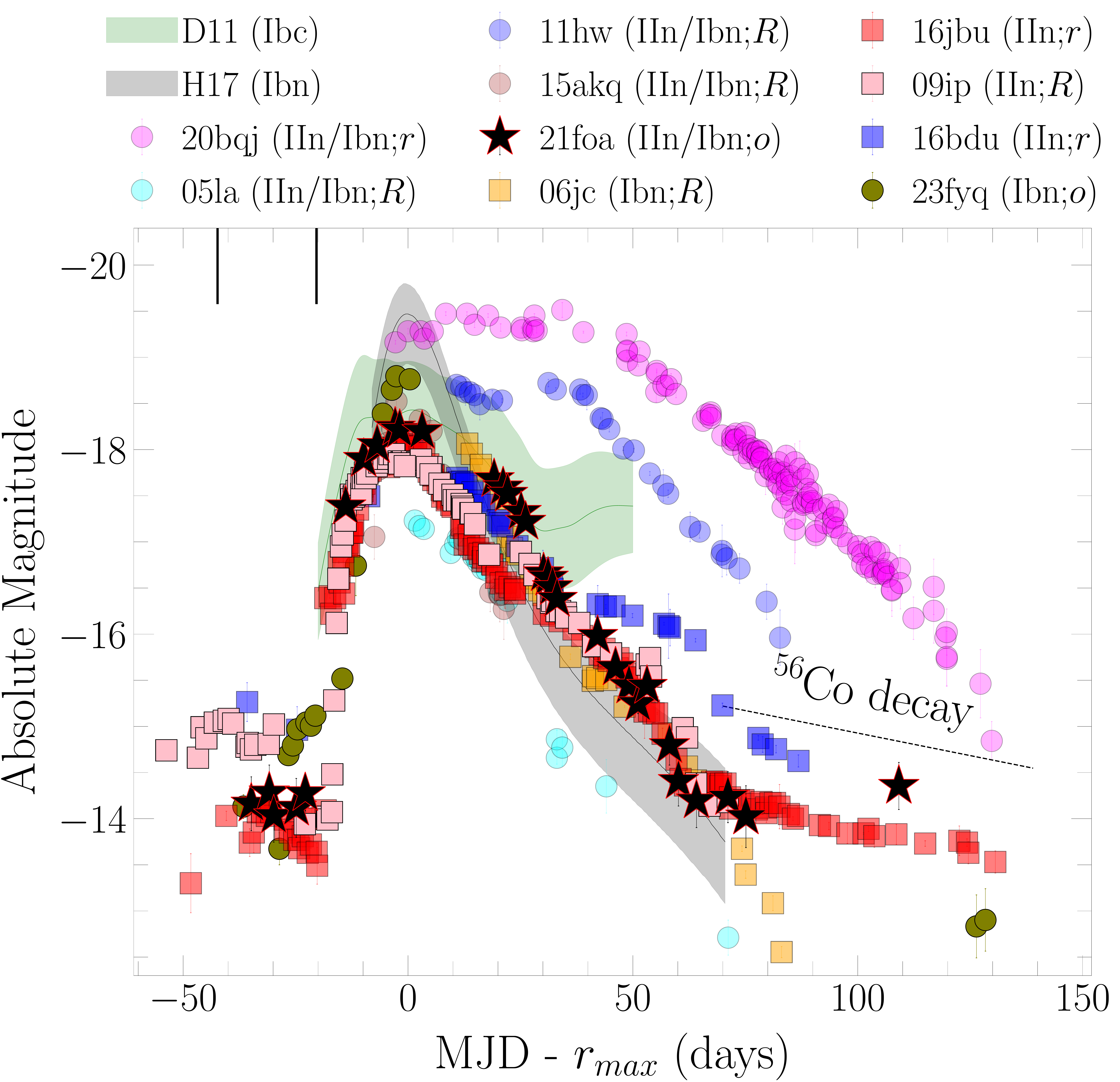}
\caption{$R/r$ band light curve comparison of interacting IIn and Ibn SNe.
Light curves are shown for SNe 2005la~\citep{Pastorello_2008_2005la}, 2011hw~\citep{Smith_2012_2011hw,Pastorello_2015_2011hw}, iPTF15akq~\citep{Hosseinzadeh_2017}, 2020bqj~\citep{Kool_2021_2020bqj}, 2021foa~\citep{Reguitti_2022}, 2009ip-like SNe 
2009ip~\citep{Pastorello_2013_09ip,Fraser_2013_09ip}, 2016jbu~\citep{Kilpatrick_2018,Brennan_2022a} and 2016bdu~\citep{GalYam_2005gl,Pastorello_2018_09iplike} including the Type Ibn SN~2006jc~\citep{Pastorello_2007_2006jc} and
SN~2023fyq~\citep{Brennan_23fyq,Dong_23fyq}. All data are symbol and color coded according to the legend. The gray shaded region corresponds to the Type Ibn template of \citep[][H17]{Hosseinzadeh_2017}. The green shaded area is the weighted mean $\pm\sigma$ of the photometrical sample of Type Ibc SNe in \citet[][D11]{Drout11}. The reference epoch of SN~2016jbu is with respect to $V$-band maximum. 
Vertical lines delimit the event A of SN~2021foa in Fig.~\ref{fig:lc}.
}
\label{fig:Rabs}
\end{figure}

We begin with an analysis of the photometry of  SN~2021foa, including a comparison to well-studied sources to inform the time-scales and energetics of the explosion, which in turn constrain the progenitor system. Fig.~\ref{fig:Rabs} shows the ATLAS $o$-band light curve of SN~2021foa compared to $R$-band-like ($o$, $r$ and $R$) light curves of other interacting SNe. 
Our comparison sample consists of transitional Type IIn/Ibn SNe~2005la, 2011hw, iPTF15akq
and 2020bqj.
We also include SNe AT2016jbu, 2005gl and 2016bdu, members of the 2009ip-like class with a photometric resemblance to SN~2021foa. The prototypical Type Ibn SN~2006jc and SN~2023fyq~\citep[ATLAS photometry using {\tt ATClean}, ][]{ATClean} serve as representation of a well studied, interacting He-rich SN. 
Additionally, we compare SN~2021foa against the Type Ibn template from~\citet{Hosseinzadeh_2017}, which represents the average, homogeneous photometric evolution of this SN class as well as the weighted mean ($\pm \sigma$) of the photometry of all stripped-envelope, hydrogen poor Type Ibc SNe in~\citet{Drout11}.
Interestingly, there is a plateau of $\approx$ one week duration starting from $\approx$ two weeks after maximum light in SN~2021foa~(see inset in~Fig.~\ref{fig:lc}). A similar plateau lasting for about two weeks can also be observed for SN~2016jbu~\citep{Brennan_2022a}.

SN~2009ip-like SNe are characterized by outbursts of $M_{R}\sim -11\pm 2$ mag several years before the explosion, 
faint emission mimicking SN impostors of $M_{R}\sim -13 \pm 2$ mag (event A) and a peak magnitude of $M_{R}\sim -18.5 \pm 0.5$ 
~\citep[event B,][]{Pastorello_2018_09iplike,Brennan_2022b}.
For SN~2021foa, we set an upper limit of $M_{o}\approx -13.4$ mag for any outburst within $\sim 5$ years before explosion (Sect.~\ref{subsec:precursor}),
while event A ($M_{r}\approx -14$ mag) and event B ($M_{r}\approx -18$ mag) agree with the values of 09ip-like class. 
The peak magnitude of Ibn SNe is, on average, much larger than that of SN~2021foa~\citep[$M_{R}=-19.47_{+0.54}^{-0.32}$,][]{Hosseinzadeh_2017}. 
Since rise/decay-times/slopes depend on the assumed explosion date and the interaction between the CSM and the SN ejecta, it is difficult to quantify the changing behaviour of the light curves of 09ip-like SNe. Nonetheless, the decay slopes from +0 to +30 days past $r$-band maximum of SN~2009ip, AT~2016jbu and SN~2021foa~
($\sim 0.04$, $\sim 0.07$ and~$\sim 0.04$ mag day$^{-1}$,  Farias et al., in prep) are smaller than the average decay rate of SN Ibn~\citep[$0.1$ mag day$^{-1}$,][]{Hosseinzadeh_2017}.  
We note that the light curve of SN~2021foa shows a clear re-brightening in $r$-bands-like after $\approx +80$ days. 

\subsubsection{Constructing a Bolometric light curve} \label{subsec:lcbol}

To obtain the bolometric light curve we fit the spectral energy distribution (SED) with a blackbody (BB) model using {\tt EXTRABOL}~\citep{ebol_cite}. 
{\tt EXTRABOL} models the light curves in flux space as a function of wavelength and phase with a 2D Gaussian process (typically with zero mean and stationary kernels in wavelength and phase, though these are configurable). A model SED (a BB model in the case of SN~2021foa) is fit to match the posterior mean of the Gaussian process, and the resulting SED surface is marginalized over  wavelengths to produce a quasi-bolometric light curve. 
We combined our optical observations in Sect.~\ref{subsec:photdat} with $J$, $H$ and $griz$ band photometry from \citet{Reguitti_2022} to construct the SED from $B$ band to near-infrared wavelengths. We computed the bolometric light curve only for epochs for which data were available in all selected passbands, as the 2D Gaussian process regresses to the mean when not constrained by observations.

\begin{figure}[hpt!]
\epsscale{1.2}
\plotone{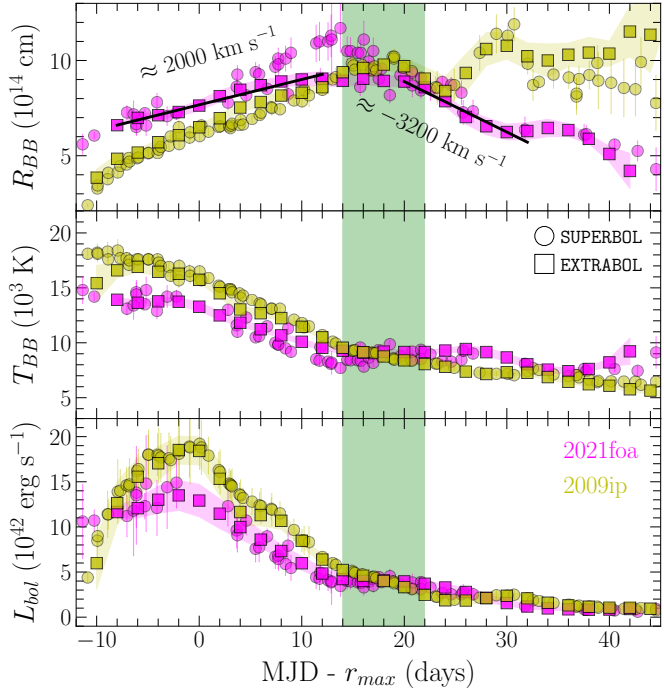}
\caption{{\it Upper}, {\it middle} and {\it bottom} panels show the evolution of the radius ($R_{BB}$), temperature ($T_{BB}$), and bolometric luminosity ($L_{bol}$) of SN~2021foa (magenta) and SN~2009ip (yellow) obtained through blackbody fits to the SED of SN~2021foa using {\tt EXTRABOL} (squares) and {\tt SUPERBOL}~\citep[circles, ][]{Nicholl_2018}. The green shaded area represent the epochs of the short plateau in the light curves (see~Fig.~\ref{fig:lc}).
} 
\label{fig:bb}
\end{figure}

Fig.~\ref{fig:bb} shows the evolution of the photospheric BB radius ($R_{BB}$), temperature ($T_{BB}$), and the resulting bolometric LC (L$_{bol}$) using {\tt EXTRABOL}. 
Additionally, we computed the bolometric light curves using {\tt SUPERBOL} to show the consistency of our method, assuming a constant color extrapolation at early and late times.  
For comparison, we also included the bolometric light curve of SN~2009ip, using {\it UVW2, UVM2, UVW2, U, B, R, V, I} bands to construct the SED.
We address that, even if our fits reproduce the overall evolution of SN~2009ip shown in~\citet{Margutti_2014_2009ip}, they fitted a two-blackbody function ({\it hot} and {\it cold} components) to the SED up to NIR bands.
As expected, the evolution of $R_{BB}$, $T_{BB}$ and $L_{bol}$ of both SNe share similarities such as a monotonic increase of $R_{BB}$ before $r$-band maximum and the phase at which $L_{bol}$ reaches maximum ($\approx -2$ days). Notoriously, the evolution of $R_{BB}$ drastically changes, possibly due to inclusion of UV bands to construct the SED of SN~2009ip. 

\subsubsection{Modeling the light curve with the \texttt{MOSFiT} Framework} \label{subsec:lcmodel}

In order to investigate the progenitor properties and its mass loss history, we adopt the radioactive decay (RD) model of $^{56}$Ni + circumstellar interaction (CSI) from the {\tt MOSFiT} \citep{mosfit} framework\footnote{\url{https://mosfit.readthedocs.io/en/latest/}} based on \citet{Chatzopoulos_2012,Chatzopoulos_2013}. 
The RD + CSI model has been extensively used to model LCs for different SNe Ibn~\citep{Karamehmetoglu_2017,Kool_2021_2020bqj,BenAmi_2022}. {\tt MOSFiT} estimates the posterior probability of a set of parameters from the RD+CSI model given the photometric data and physically-informed priors. 

\subsubsection{Parameterization of the RD+CSI Model}

The RD model accounts for the radioactive decay of $^{56}$Ni through three parameters: the nickel fraction, $f_{\rm Ni}$, the $\gamma$-rays opacity of the ejecta, $\kappa_{\gamma}$, and an optical opacity, $\kappa$.  Additionally, {\tt MOSFiT} includes a model for the SN ejecta-CSM interaction, with several physical quantities for the ejecta and the CSM listed as free parameters~\citep[see][for a detailed description]{Villar_2017}. 
This includes the total mass of the ejecta ($M_{\rm ej}$) and the inner and outer density profiles of the ejecta ($\rho_{\rm in,ej} \propto r^{-\delta}$, $\rho_{\rm out,ej} \propto r^{-n}$). The main parameters to describe the CSM are the inner radius of the CSM ($R_0$), the CSM density ($\rho_{\rm CSM}$) and a density profile of the CSM ($\rho_{\rm CSM} \propto r^{-s}$). For the latter, $s=0$ corresponds to a constant density (shell-like) CSM, while $s=2$ denotes a wind-like model. Additionally, {\tt MOSFiT} fits for the explosion time ($t_{\rm exp}$) relative to the first photometric observation~\citep{Kool_2021_2020bqj}. {\tt MOSFiT} also includes some nuisance parameters, such as the minimum allowed temperature of the photosphere before it recedes, $T_{\rm min}$. Finally, a white noise term $\sigma$, is included and added in quadrature to the photometric uncertainties. \texttt{MOSFiT} uses this white noise term to quantify the quality of the fitting. Acceptable values for this parameter are $\sigma < 0.2$~\citep{Nicholl_2017}.

\begin{figure*}[hbt!]
\epsscale{1.2}
\plotone{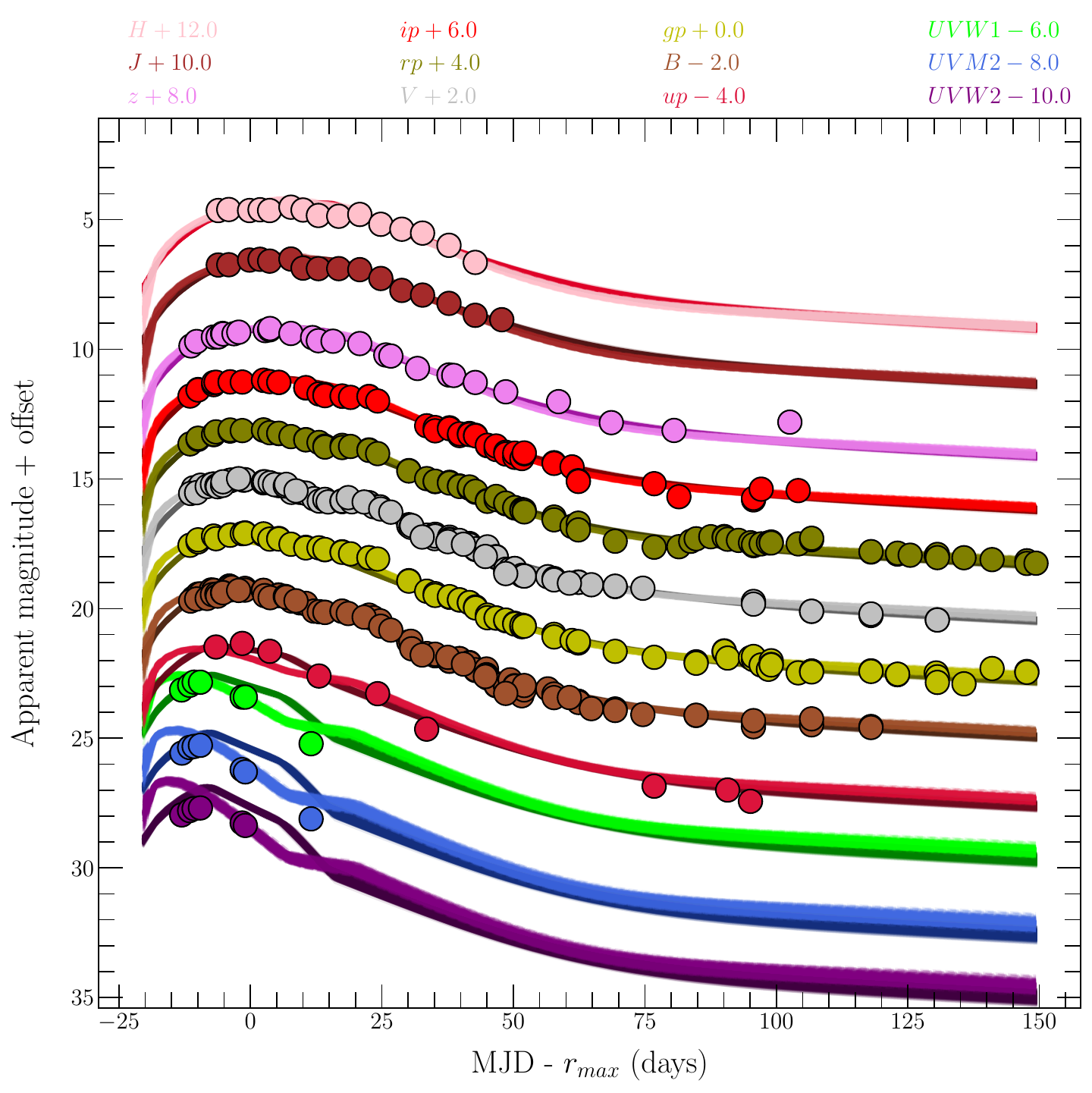}
\caption{Collection of light curves of SN~2021foa in several bands reported in this work and ~\citet{Reguitti_2022}. Colour symbols are the photometric data. Light and bold colour lines are 1000 sampled  {\tt MOSFiT} light curves from the posterior distribution for the shell ($s=0$) and wind-like CSM configuration ($s=2$), respectively.} 
\label{fig:mosfitLCS_2_21foa}
\end{figure*}\

\subsubsection{RD+CSI Modeling Choices and Priors}
\label{sss:mdchoice}
While this is a high-dimensional model, the fundamental physical parameters of interest to this study are $f_{\rm Ni}$, $M_{\rm ej}$, $M_{\rm CSM}$, $E_{\rm kin}$ and $R_{0}$. We fix three parameters of the model: $\delta=1$, $n=12$ and $s$. The assumed value of $n=12$ is typically used to characterize RSG-like progenitors rather than for BSG/WR stars, but $\delta$ and $n$ are not sensitive to derive the fundamental parameters~\citep{Villar_2017,Kool_2021_2020bqj}. In order to investigate a shell and wind-like CSM configurations, we fixed $s=0$ and $s=2$, respectively. With these choices, eleven free parameters remain: $f_{\rm Ni}$, $\kappa$, $M_{\rm ej}$, $M_{\rm CSM}$, $\rho_{\rm CSM}$, $R_{0}$, $E_{\rm kin}$, $t_{\rm exp}$~(i.e. fundamental parameters) $T_{\rm min}$, $\kappa_\gamma$ and $\sigma$ (the nuisance parameters). To constrain the explosion time of SN~ 2021foa, 
we only considered observations from $-20$ days onwards, i.e., we discard event A photometry as the RD+CSM model was not designed to model precursor emission. We used top-hat priors on the remaining parameters.  We note that other physical parameters including $\kappa_{\gamma}$, $\kappa$ and $t_{\rm exp}$ may be considered nuisance parameters as these are quantities that are not observable. These physically-informed modeling choices and priors significantly reduce the dimensionality of the problem, and help ensure that the inferred quantities are meaningful. 

Following \citet{Kool_2021_2020bqj}, we elected to use the dynamic nested sampling implemented in {\tt DYNESTY}~\citep{Speagle_2020}. 
Nested samplers estimate the evidence ($\mathcal{Z}$) i.e., the marginalized likelihood for the data given the model. Given that our prior distributions are constant for $s=0$ and $s=2$, we can directly compare the evidence reported by {\tt DYNESTY} to check which model is preferred given our data (i.e., the model with larger evidence).

\begin{deluxetable}{lcccc}
\tablecaption{Fundamental parameters of the {\tt MOSFiT} (CSM-Ni) models of a shell-like ($s=0$), wind-like ($s=2$).}
\tablehead{
\colhead{Par\tablenotemark{\footnotesize \rm a}} & \colhead{Units} & \colhead{Prior} &
\multicolumn{2}{c}{
Model}\\
\cline{4-5}
\colhead{} & \colhead{} & \colhead{} & \colhead{$s=0$} & \colhead{$s=2$}}
\startdata
$f_{\rm Ni}$ & $10^{-2}$ & $10^{-3} - 1.0$ & $7.85^{+0.41}_{-0.38}$ & $1.85^{+0.09}_{-0.09}$ \\
$\kappa$ & cm$^{2}$ g$^{-1}$ & $0.1-0.4$ & $0.396^{+0.003}_{-0.006}$ & $0.34^{+0.037}_{-0.043}$ \\
$E_{\rm kin}$ & $10^{51}$ erg & $0.01-20$ & $0.2^{+0.01}_{-0.01}$ & $1.79^{+0.07}_{-0.06}$ \\
$M_{\rm CSM}$ & $M_{\odot}$ & $0.1 - 30.0$ & $0.39^{+0.01}_{-0.01}$ & $0.07^{+0.01}_{-0.01}$ \\
$M_{\rm ej}$ & $M_{\odot}$ & $1.0-30.0$ & $1.91^{+0.08}_{-0.08}$ & $8.35^{+0.4}_{-0.4}$ \\
$R_{0}$ & $10^{14}$cm & $0.02 -20.0$ & $0.15^{+0.58}_{-0.11}$ & $0.7^{+0.14}_{-0.13}$ \\
$\rho_{\rm CSM}$ & $10^{-12}$g cm$^{-3}$ & $10^{-3} - 10^{4}$ & $0.65^{+0.03}_{-0.02}$ & $52.86^{+28.86}_{-15.76}$ \\
$t_{0}$ & days & $-30 - -20$ & $-25.3^{+0.07}_{-0.04}$ & $-21.45^{+0.15}_{-0.17}$ \\
$\sigma$ & mag & $10^{-5} - 10$ & $0.199^{+0.006}_{-0.005}$ & $0.193^{+0.005}_{-0.005}$ \\
\cline{1-5}
$M_{\rm Ni}$ & $10^{-2}{\rm M}_{\odot}$ & $-$ & $1.5^{+0.1}_{-0.1}$ & $1.54^{+0.11}_{-0.1}$ \\
$v_{\rm ej}$ & $10^{3}$ km~s$^{-1}$ & $-$ & $4.15^{+0.12}_{-0.12}$ & $5.98^{+0.19}_{-0.17}$ \\
\cline{1-5}
$\log \mathcal{Z}$ & $-$ & $-$ & $789.8\pm 0.2$ & $817.3 \pm 0.2$ \\
\enddata
\tablecomments{$^{a}$ Parameters of the radioative decay+CSI model are the nickel fraction~($f_{\rm Ni}$), 
optical opacity~$(\kappa)$, kinetic energy ($E_{\rm kin}$), ejecta mass ($M_{\rm ej}$), CSM inner radius ($R_0$), CSM density $(\rho_{\rm CSM})$, explosion epoch with respect to $r_{max}$ ($t_0$), nickel mass ($M_{\tt Ni}$), ejecta velocity ($v_{\tt ej}$) and the evidence, ($\log \mathcal{Z}$), detailed in Sect.~\ref{sss:mdchoice}.}
\label{tab:mosfit}
\end{deluxetable}

In order to reduce the computational cost without oversampling the observations per epoch, we provided {\tt MOSFiT} with the following bands: $UVW2,UVM2,UVW1,U,B,V$ (UVOT), $u,B,g,V,r,i,z$~\citep{Reguitti_2022}, $o,c$ (ATLAS) and $up,gp,rp,ip$ (LCO). 
Fig.~\ref{fig:mosfitLCS_2_21foa} 
shows the {\tt MOSFiT} model-light curves for $s=0$ and $s=2$ for a collection of bands considered, together with the observed data. The parameters of the model are summarized in Tab.~\ref{tab:mosfit} and discussed in Sect.~\ref{sec:mosfit_discussion}.


\subsection{Spectroscopy} \label{sec:specana}
Our multi-epoch spectroscopy 
permits a detailed analysis of the evolution of the prominent 
emission and absorption line profiles of SN~2021foa~(see~Figs.~\ref{fig:spectra_optical} and~\ref{fig:spectra_nir}) over a large range of epochs. %
We select the spectra with highest resolution, including those published by~\citet{Reguitti_2022} at $+35$, $+43$ days. For simplicity, we refer to any 
`velocity component' of a line profile as `component'. 

\begin{figure*}[hbt!]
\epsscale{1.2}
\plotone{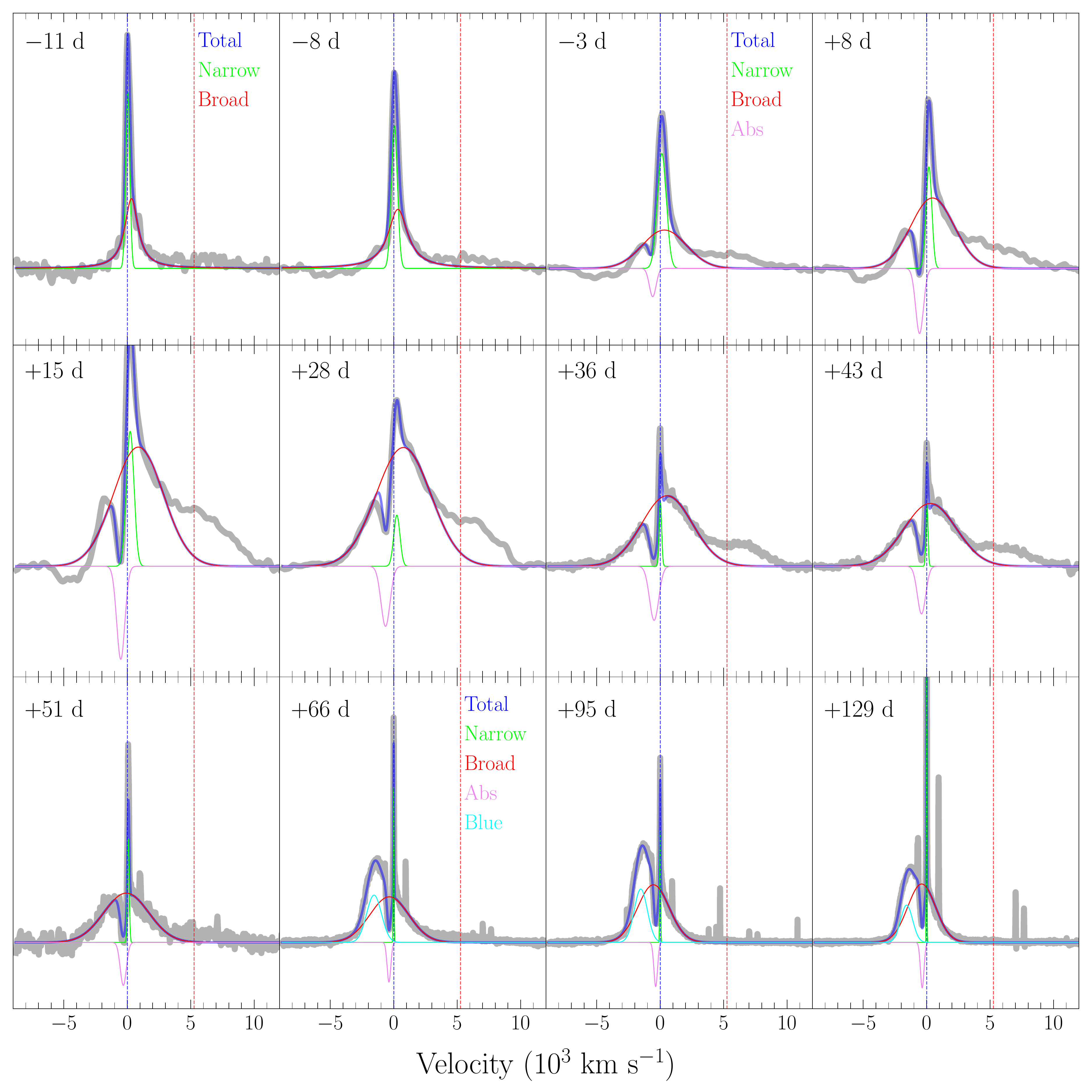}
\caption{Modeling of the H$\alpha$ profiles. Zero velocity corresponds to the rest wavelength of H$\alpha$ line. The first two epochs were fitted using a combination of a Gaussian and Lorentzian profile (red, green), while the spectra from $-3$ to $+51$ days are fit by three Gaussians, accounting for the narrow and broader emission (green and red), and the absorption (purple). For the medium resolution spectra ($+66$, $+95$ and $+129$ days), an additional Gaussian component has been added to fit the profile (turquoise). Blue and red dashed vertical lines mark the velocities of the rest frame wavelength of H$\alpha$ and \ion{He}{i}~$6678$\AA, respectively.
}
\label{fig:modelHa}
\end{figure*}

\subsubsection{Spectral Line fitting}\label{sec:linefit}

In order to quantify the evolution of the different  components of the Balmer lines 
(H$\alpha$, H$\beta$, H$\gamma$ and H$\delta$, panels A, B, C and D in Fig.~\ref{fig:vel_all_HHe}), 
we fit the lines with multiple Gaussian and Lorentzian functions (see Fig.~\ref{fig:modelHa}). Prior to this, we fit the continuum in the respective wavelength region of the emission lines (around $\pm 5000$ km s$^{-1}$) using a 1D polynomial and subtract it.  
Furthermore, to account for either low instrumental resolution or a low signal-to-noise ratio, we convolved the spectral region of the emission lines with a Gaussian kernel that has a standard deviation $\leq 2$ $\times$ the dispersion ($\sim$ resolution) of the instrument (see Tab.~\ref{tab:spec}).
We detail the line profile fitting method for  H$\alpha$ as an example. 

For the first two epochs ($-11$ and $-8$ days), we simultaneously fit the complex
H$\alpha$ emission line with two functions. Since the H$\alpha$ exhibits broad wings, we fit a Lorentzian profile to the broad base of the emission line and a Gaussian profile to the narrow component. The introduction of the Lorentzian profile was made to account for any electron scattering wings at early times~\citep{Fransson_2010jl}.
For all epochs between $-3$ and $+51$ days we fit the H$\alpha$ emission line with three Gaussian functions, one each for the narrow absorption, narrow emission as well as for the broad emission profile.   
At all late epochs ($+66$ to $+129$ days) we add another Gaussian profile to fit the strongly blue-shifted emission of H$\alpha$. 
For all other \ion{H}{i} emission lines we follow a similar method with two minor differences: (i) the central wavelength used to extract the line profile (ii) the number of Gaussian functions to fit the line profiles.  For example, for Pa$\beta$, a single Gaussian profile can well fit the emission line which neither exhibits narrow emission nor absorption. 

For the \ion{He}{i} lines, we only quantify the evolution of the strongest emission lines in the optical (\ion{He}{i}~$\lambda 5876$~{\AA} and $7065$~\AA). We fit these symmetric features with a single Gaussian at all epochs (see Sect.~\ref{app:hei_5876}). This is the simplest model that is consistent with the observations. 

We also identify narrow absorption lines of \ion{Fe}{ii} $\lambda\lambda 5169,5276,5317$, (likely) \ion{He}{i} $\lambda\lambda 4922,5016$, \ion{Ca}{ii} $\lambda\lambda 8498,8542,8662$ and \ion{O}{i}~$\lambda 8446$ (see Figs.~\ref{fig:vel_all_HHe} and~\ref{fig:all_other_lines}). To quantify the velocity evolution of the region harbouring these elements, we fit a single Gaussian profile to the absorption lines. 

Finally, we calculate the total line flux of all \ion{H}{i} emission lines from the continuum subtracted line profile fits. 
We exclude the emission lines from the host galaxy at late epochs ($> +40$ days).
The total emission line flux of the \ion{He}{i}~$\lambda 5876${\AA} line complex is estimated from the continuum subtracted data directly.  the emission line complex in the velocity range $\pm$6\,000 km/s.  

\subsection{Modeling the Late-time VLT/X-shooter Spectra}\label{sec:dustemis}

\begin{figure}[hbt!]
\epsscale{1.2}
\plotone{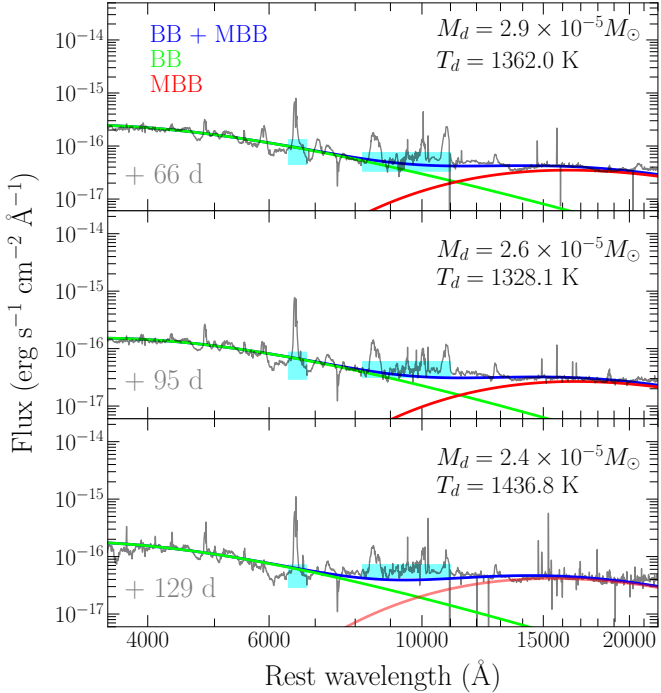}
\caption{SN and dust fit to VLT/X-shooter spectra at $+66$ days ({\it upper panel}), $+95$ days ({\it middle panel}) and $+129$ days ({\it lower panel}). The SN hot black-body function (green curves -- dominant below $\approx8,000$ \AA) is fit simultaneously with a modified black-body function to describe the NIR dust emission (red curves) to the observed spectra (black). The blue curves show the sum of the two individual components.
Cyan rectangles visualise the spectral regions excluded for the fit due to the large number of emission lines.
}
\label{fig:dust}
\end{figure}

A signature of either pre-existing or newly formed dust is the thermal (near) infrared excess emission of hot dust grains 
over the SN continuum emission. Such thermal emission has been observed in several types of core collapse SNe~\citep[see][for a review]{Gall_2011}. 
To test if dust is present in SN~2021foa, we fit the continuum emission from optical to the NIR wavelengths of our late-time SN~2021foa VLT/X-shooter spectra at $+66$, $+95$ and $+129$ days. We exclude the spectral regions containing strong emission lines such as H$\alpha$ and the \ion{He}{i}~$\lambda 10830 +$ and Pa$\gamma$ complexes prior to fitting as indicated in Fig.~\ref{fig:dust}. 

We then simultaneously fit a BB function to the hot SN photosphere and a modified BB function \citep{1983QJRAS..24..267H} to the near IR excess emission to account for any hot dust emission as:
\begin{equation}
F_{\mathrm{\lambda}} (\lambda) = \frac{R_{SN}^2}{D_{\mathrm{L}}^{2}}\, B_{\mathrm{\lambda}}(\lambda, T_{\mathrm{SN}}) + \frac{M_{\mathrm{d}}}{D_{\mathrm{L}}^{2}} \,    
\kappa_{\mathrm{abs}}(\lambda, a) \, B_{\mathrm{\lambda}}(\lambda, T_{\mathrm{d}}) 
\label{EQ:MBB}
\end{equation}
with $D_{\mathrm{L}}$ the luminosity distance to the SN, $R_{SN}$ the radius of the photosphere and $B_{\mathrm{\lambda}}(\lambda, T)$ the Planck function at temperatures of the photosphere, $T = T_{\mathrm{SN}}$ and dust, $T = T_{\mathrm{d}}$. 
We assume optically thin dust, with a dust mass $M_d$, which is distributed spherically symmetrically around the SN. For the dust mass absorption coefficient, $\kappa_{abs}(\lambda)$ we adopt the formalism of \citet{2017ApJ...849L..19G}, where $\kappa_{abs}(\lambda)$ follows a $\lambda^{-\beta}$ power law in the VLT/X-shooter NIR wavelength range. 
Thus,  $\kappa_{abs}(\lambda)$  can be parameterized as 
$A_d (\lambda/1 \mu$m$)^{-\beta}$, with $A_d = 1.0$ $\times$ 10$^{4}$~cm$^2$ g$^{-1}$ and $\beta=1.5$. This formalism mimics small sized ($\lesssim 0.1 \mu$m) carbonaceous grains \citep[e.g.,][]{1991ApJ...377..526R}. 

\section{Analysis Results}
\label{sec:results}

\subsection{Estimating Photosphere Expansion Velocity from the Bolometric Light Curve}\label{sec:bolometric}

The evolution of $T_{BB}$ follows the bolometric LC. 
The peak values of $L_{bol}$ and T$_{BB}$, reached between $-5$ and $0$ days, are $\approx 1.7\times 10^{43}$~erg~s$^{-1}$ and $\approx 15000$ K, respectively. The temperature evolution flattens between $14-23$ days as indicated by the green shaded region in Fig.~\ref{fig:bb}. This coincides with the short LC-plateau that is observable in several optical passbands (see Fig.~\ref{fig:lc} and Sect.~\ref{sec:photana}).
Subsequently, $T_{BB}$ decreases to $8000$ K at day $+35$.    
On the other hand, $R_{BB}$ is largest, with about 10$^{15}$~cm at the beginning of the short LC-plateau. 

From the estimated slope of the $R_{BB}$ evolution until this maximum at $\sim +10$ days, we infer an expansion velocity of the photosphere of $\approx 2000$~km~s$^{-1}$. Thereafter, $R_{BB}$ recedes at a velocity of $\sim -3200$~km~s$^{-1}$ to a radius of $\approx 6\times 10^{14}$~cm at day $\sim +30$, the same level as at $-10$ days.

\subsection{Estimating Progenitor Properties from the RD+CSI Model}\label{sec:mosfit_discussion}

We use the parameters of the RD+CSI model inferred with the {\tt MOSFiT} framework~(Sect.~\ref{subsec:lcmodel}, Tab.~\ref{tab:mosfit}) to constrain progenitor and SN properties. 
From the evidence values ($\log \mathcal{Z}$), $s=2$ model is preferred over $s=0$.
We obtain an ejecta mass of $\approx 2.0$~M$_{\odot}$ for a $s=0$ (CSM-shell-like) model. 
This value is close to the maximum mass of  $\approx 1.2 \, {\rm M}_{\odot}$, as estimate for any progenitor (single or in a binary system) of Type Ibn SNe ~\citep[][]{Dessart_2022}.
However, the ejecta mass inferred for $s=2$ is $\approx 8$ M$_{\odot}$. This is lower than the average value of $\approx 16$  M$_{\odot}$, as obtained for Type Ibn SN~2019uo~\citep[][]{Gangopadhyay_2020},  2020bqj~\citep[][]{Kool_2021_2020bqj} and PS15dpn~\citep{wang2020}.  

The CSM mass is $\sim 0.4$ and $\sim 0.1 ~{\rm M}_{\odot}$ for the $s=0$ and $s=2$ models, respectively. This is consistent with the CSM mass estimates of 0.016 M$_{\odot}$ and $< 0.1 ~{\rm M}_{\odot}$ for SN~2016jbu~\citep[0.016 M$_{\odot}$]{Brennan_2022b} and typical Ibn SNe~\citep[][]{Maeda_Moriya2022}, respectively.

The resulting CSM inner radius, $R_0$, and the CSM density $\rho_{\rm CSM}$ for the $s=0$ model are $\sim$ 2 $\times$ 10$^{13}$ cm and $\sim 6\times 10^{-13}$ g~cm$^{-3}$. 
For the $s=2$ model, $R_0$ $\sim$ 7 $\times$ 10$^{13}$ cm and $\rho_{\rm CSM}$ $\sim 5\times 10^{-11}$ g~cm$^{-3}$. 

Following \citet{Chatzopoulos_2012} approach, we can derive the value of $R_{f}$, the outer radius of the CSM. For the preferred model ($s=2$), $R_{f}\approx 1.5\times 10^{14}$ cm, while for $s=0$,  $R_{f}\approx 7\times 10^{14}$ cm. 

While the \texttt{MOSFiT} framework is flexible, the fundamental parameters of the RD+CSI model are not always directly expressed as standard literature quantities, such as nickel mass and mass loss rate. For the $s=2$ model,
we can estimate the mass loss rate, $\dot{M}$ using the formalism of~\citet{BenAmi_2022} and \citet{Kool_2021_2020bqj} as:
\begin{equation}~\label{eq:massloss}
   \dot{M}(r) = 4\pi \rho_{\rm CSM} r^{2} v_{w}, 
\end{equation}
where $v_{w}$ is a steady CSM wind velocity. Additionally, for $s=0$ and $s=2$ models, we calculate the nickel mass as $M_{\rm Ni}$ = $M_{\rm ej} \times f_{\rm Ni}$ and ejecta velocity $v_{\rm ej} = \sqrt{\frac{10}{3} \frac{E_{\rm kin}}{M_{\rm ej}}}$.

Using Eq.~\ref{eq:massloss} to estimate a mass loss rate, and assuming a wind velocity, $v_{w} \sim 400$~km~s$^{-1}$ as inferred from the velocity of the absorption minima of the \ion{H}{i} lines at late times, together with values of $R_0$ and $\rho_{\rm CSM}$ as resulting from the $s=2$ model we obtain a mass loss rate $\sim 2 ~{\rm M}_{\odot} {\rm yr}^{-1}$. This mass loss rate is not expected for typical LBV and WR winds~\citep[$\sim 10^{-4} - 10^{-5} ~\rm{M}_{\odot} {\rm yr}^{-1}$]{smith_review}, but has also been estimated for the Type IIn SNe iPTF13z~\citep{Nyholm_2017}, 2017hcc~\citep{Smith_2017hcc} and the Ibn SN~2019kbj~\citep{BenAmi_2022}.
From X-ray observations, mass-loss rates of SN~2009ip and 09ip-like SN~2010mc are estimated below $10^{-1}$ M$_{\odot}$yr$^{-1}$~\citep{Ofek_2009ip,Ofek_2010mc,Boian_2015bh}, while Type Ibn SN~2022ablq has an upper limit of $0.5$ M$_{\odot}$yr$^{-1}$~\citep{Pellegrino_2022ablq}.
%

\subsection{Determining CSM Structure from the Line Evolution}\label{subsec:linevol}

\begin{figure}[hbt!]
\epsscale{1.2}
\plotone{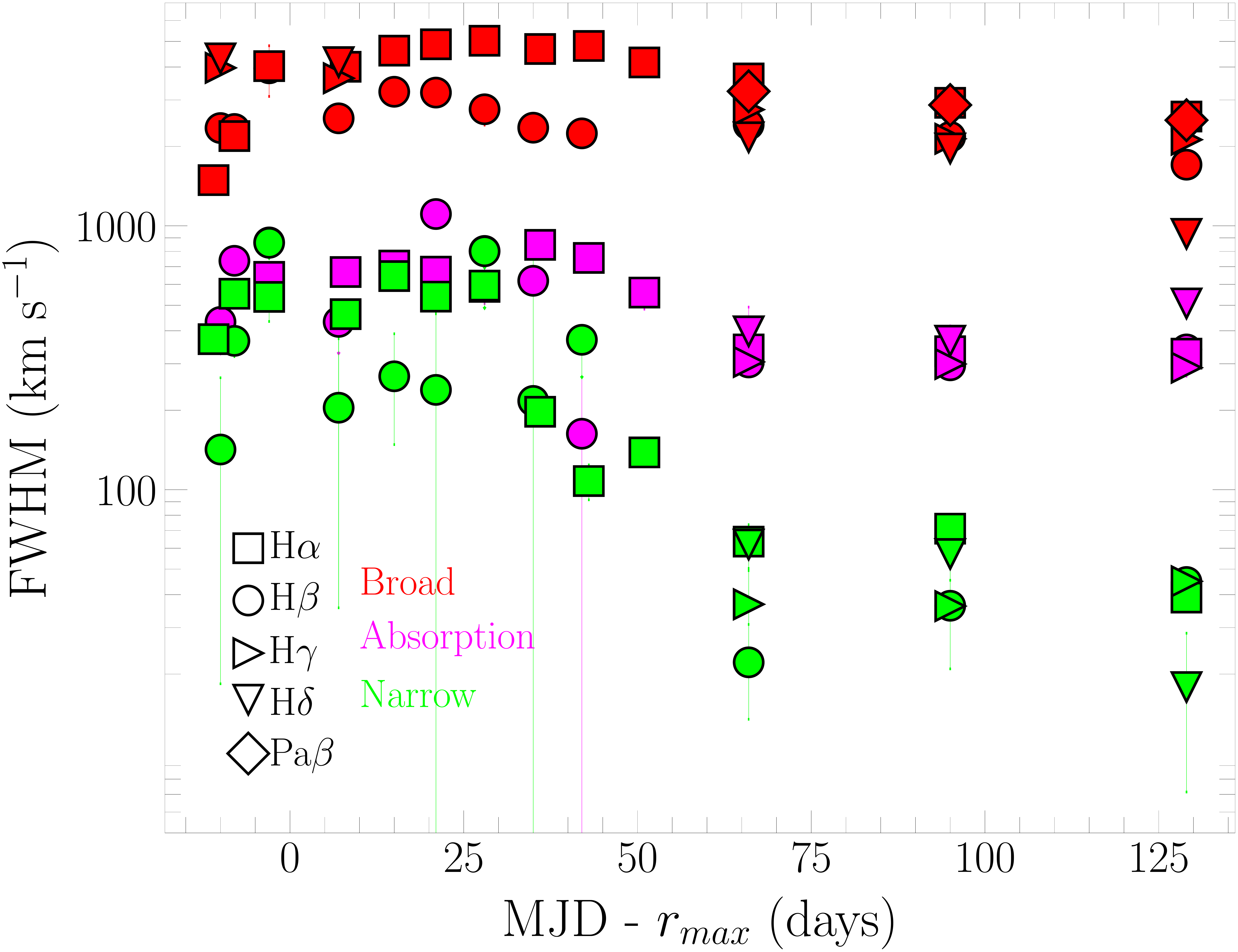}
\caption{Evolution of the FWHM of \ion{H}{i} lines. 
The broad and narrow emission and the absorption components are color-coded (red, green and magenta). The values of the FWHM are corrected for the resolution of the instrument. If the line was unresolved, an upper limit of the FWHM is estimated as the resolution of the instrument.
}
\label{fig:HFWHM}
\end{figure}

Fig.~\ref{fig:HFWHM} presents the evolution of the FWHM of the strongest \ion{H}{i} lines, inferred from the spectral modeling described in Sect.~\ref{sec:linefit} and summarized in Tab.~\ref{tab:FWHM}. For the first $+40$ days, the FWHM of the broad and the absorption components remain constant at about $4\,000$~km~s$^{-1}$ and $800$~km~s$^{-1}$, respectively.

Between $+40$ and $+60$ days the FWHM of all absorption components decrease by $\sim 50$ \%, remaining constant at this level until about 130 days. This suggests a second structural component in the CSM. 
The FWHM of the narrow emission component of H$\alpha$ and H$\beta$ is about $600$~km~s$^{-1}$ during the first 20 days. Thereafter, it rapidly declines to about 60~km~s$^{-1}$ at which it remains constant from $+66$ days onwards. 

\begin{figure}[hbt!]
\epsscale{1.2}
\plotone{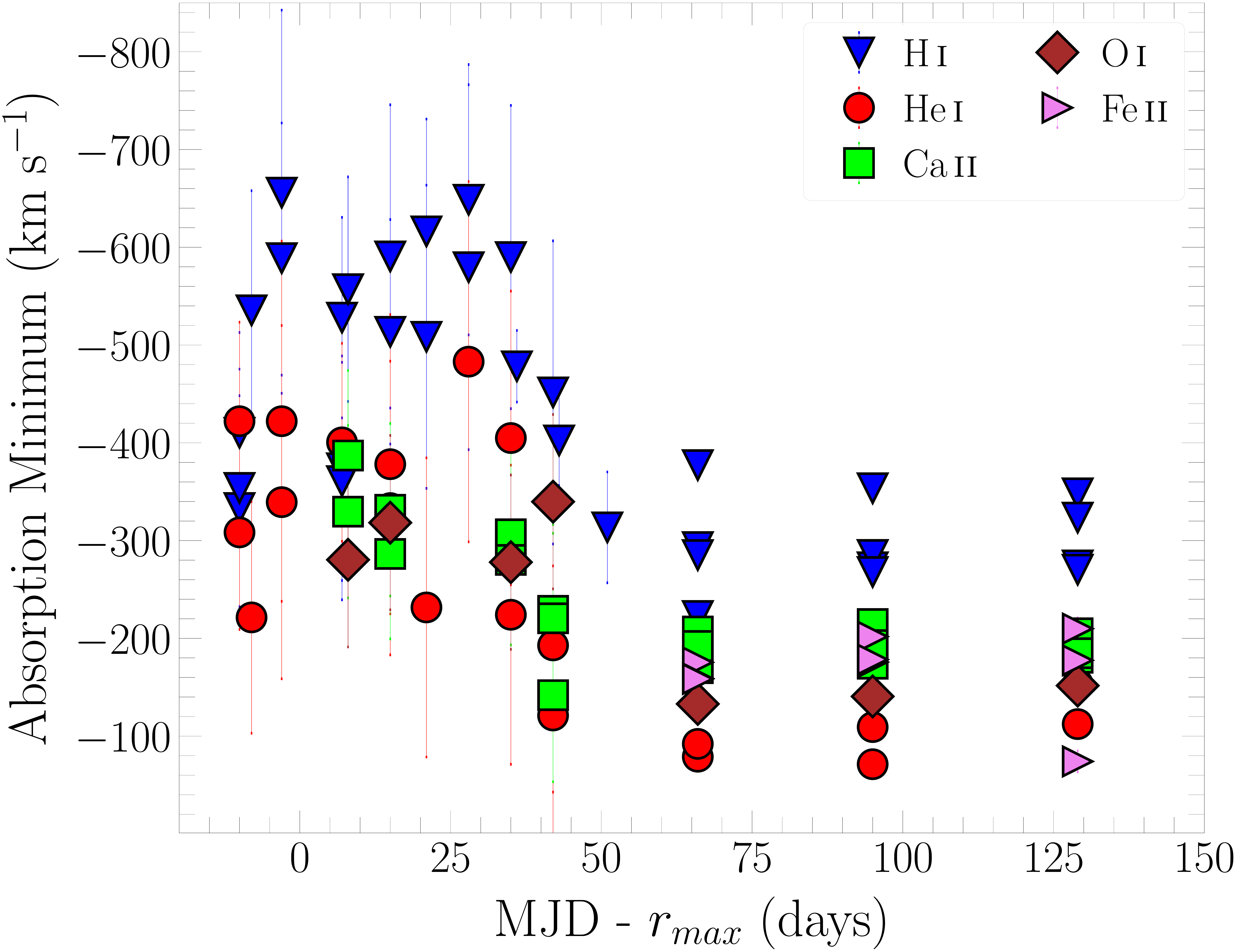}
\caption{Evolution of the velocity of the absorption minima of \ion{H}{i}~(H$\alpha$, 
H$\beta$, H$\gamma$ and H$\delta$), \ion{He}{i}~($\lambda\lambda 4922,5016,$), \ion{Ca}{ii}~($\lambda\lambda 8498,8542,8662$), \ion{Fe}{ii}~($\lambda\lambda 5169,5276,5317$) and \ion{O}{i}~$\lambda 8446$ lines. 
}
\label{fig:Othermin}
\end{figure}

Fig.~\ref{fig:Othermin} shows the evolution of the velocity of the absorption minima of \ion{He}{i}, \ion{Ca}{ii}, \ion{O}{i} and \ion{Fe}{ii} and H$\beta$, summarized in~Tab.~\ref{tab:absmin}.
It is evident that the velocities of the \ion{He}{i}, \ion{Ca}{ii}, \ion{O}{i} and \ion{Fe}{ii} absorption lines are at velocities which are $\gtrsim 200~{\rm km~s}^{-1}$ lower than H$\beta$. 
The spectra exhibit a sudden decline of the narrow absorption velocities to about 100~km~s$^{-1}$ past $+60$ days for all transitions (see Fig.~\ref{fig:HFWHM} and Fig.~\ref{fig:Othermin}).

%


\subsubsection{Line Fluxes and Ratios and the Curious Case of SN~2021foa's ``Flip-Flop''}

\begin{figure}[hbt!]
\epsscale{1.2}
\plotone{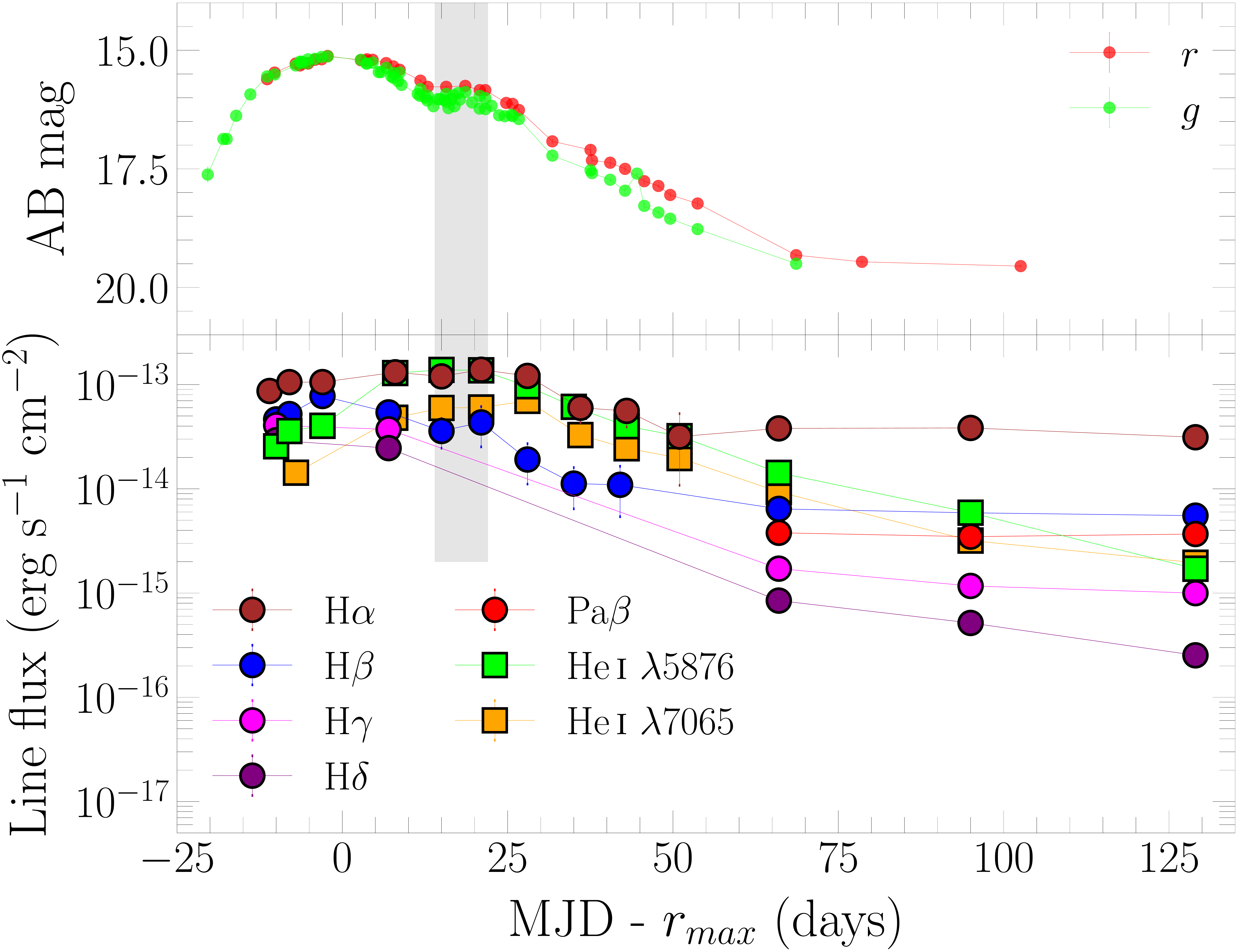}
\caption{ 
{\it Upper panel:} Light curves of $g$ and $r-$bands.
{\it Lower panel:} Evolution of the total emission flux of \ion{H}{i} and \ion{He}{i} lines. The 
shaded region encompasses the epochs of the short photometric plateau at $\approx +15$ days. 
}
\label{fig:HHeflux}
\end{figure}
Fig.~\ref{fig:HHeflux} shows the evolution of the calculated emission line fluxes of the strongest \ion{H}{i} and \ion{He}{i} lines (summarized in Tab.~\ref{tab:flux}). The line fluxes of \ion{H}{i} including Pa$\beta$ and \ion{He}{i} evolve in a similar fashion up to $+50$ days. They increase up to $+12$ days after which they stay constant until about $+22$ days. This agrees with the plateau phase of the $r$- and $g$-band light curves, as shown for comparison in the upper panel of Fig.~\ref{fig:HHeflux}. At later epochs ($> +60$ days), all \ion{He}{i} and H$\delta$ line fluxes decrease while most H line fluxes remain constant (or increases, as evident with H$\alpha$).
Interestingly, at these same epochs, the decline in $r$-band halts (see Fig.~\ref{fig:lc}).
We find that the \ion{He}{i}~$\lambda 5876$ and H$\alpha$ lines reach approximately the same line flux of 10$^{-13}$ erg~s$^{-1}$~cm$^{-2}$ around $+20$ days, consistent with the findings of~\citet{Gango_2024}. This is in contrast to \citet{Reguitti_2022} who estimate that the \ion{He}{i} $\lambda 5876$ flux is about half of that of H$\alpha$ at this epoch (see Fig.~\ref{fig:proof_fluxratio}).  

\begin{figure}[hbt!]
\epsscale{1.2}
\plotone{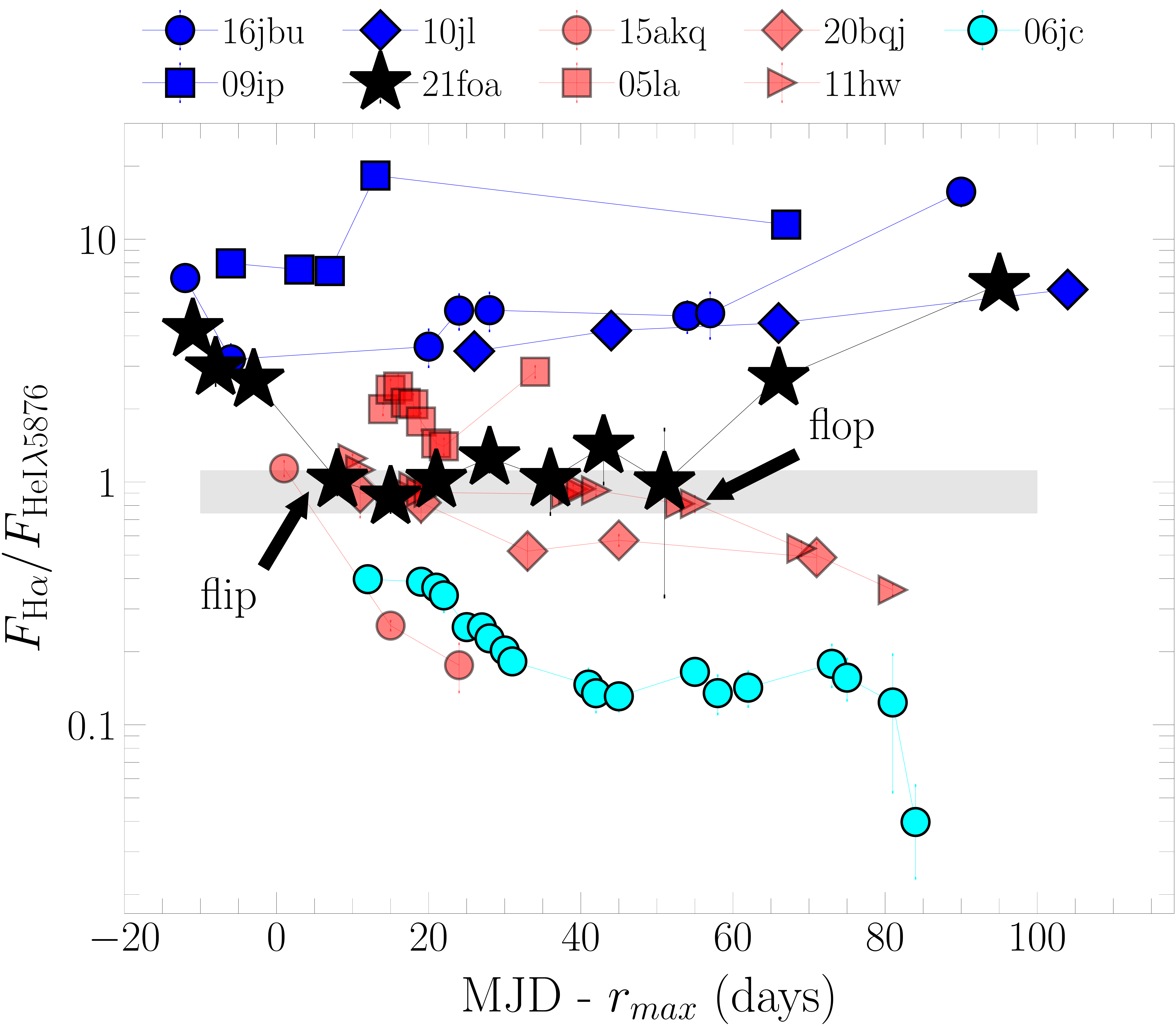}
\caption{Flux ratio of H$\alpha$/\ion{He}{i}~$\lambda 5876$ for the transitional IIn/Ibn objects, SN Type Ibn 2006jc and SNe Type IIn SN~2016jbu, SN~2009ip and SN~2010jl. Each class is color-coded as red (IIn/Ibn), cyan (Ibn) and blue (IIn). The gray band encompasses the mean and one standard deviation of the line flux ratios of SNe 2011hw, 2020bqj and 2021foa within $+10$ to $+55$ days. Fluxes are averaged in a 3-day bin. 
}
\label{fig:ratio}
\end{figure}

The flux of \ion{He}{i}~$\lambda 7065$ is as strong as that of H$\beta$ for most of the epochs. Fig.~\ref{fig:ratio} shows the evolution of the H$\alpha$/\ion{He}{i}~$\lambda 5876$ line ratio for SN~2021foa in comparison to other transitional IIn/Ibn SNe. These include SNe 2005la, 2011hw, 2020bqj and iPTF15akq. We also include the H$\alpha$/\ion{He}{i} ratio for the Ibn prototype SN~2006jc and the Type IIn SN~2009ip, SN~2010jl and SN~2016jbu. For consistency, we re-computed the line ratios for these objects similar to those of SN~2021foa (see Sect.~\ref{sec:linefit}). We find discrepancies in the line ratios of SN~2006jc and SN~2011hw a factor of two between~\citet{Smith_2012_2011hw} and our measurements (see Sect.~\ref{app:06jc_11hw}). We describe our method to determine the line ratios in Sect.~\ref{app:line_trans}, and attribute this discrepancy to the mis-estimation of the local continuum surrounding H$\alpha$ and \ion{He}{i}~$\lambda 5876$ lines in previous work. 

As shown in Fig.~\ref{fig:ratio},  all transitional IIn/Ibn SNe have H/He ratios larger than $\approx 1$ at early epochs ($< +10$ days). Thereafter, these SNe transit into a more He prominent regime (the ``flip''), with a mean H$\alpha$/\ion{He}{i} ratio of 0.93 $\pm$ 0.19 up to about 50 days past peak brightness. After day 50, the line-flux ratio of all other transitional IIn/Ibn SNe drops below one. The evolution of the H$\alpha$/\ion{He}{i} $\lambda 5876$ ratio for the transitional IIn/Ibn SNe is clearly different from the classical IIn and 09ip-like events (blue markers), as well as  Ibn events such as SN~2006jc (cyan markers). Type IIn SN~2016jbu, SN~2009ip and SN~2010jl are hydrogen dominated (H$\alpha$/\ion{He}{i} $>$ 1) at all epochs. For the Ibn SN~2006jc, the line emission of \ion{He}{i}~$\lambda 5876$ dominates at all epochs. It is the presence of a ``flip'' from hydrogen-dominated to a line ratio of $\approx1$ that truly determines if an object is a member of the class of transitional IIn/Ibn. 

However, uniquely for SN~2021foa even amongst transitional IIn/Ibn\footnote{SN~2005la also shows the {\it flop}, although the H$\alpha$ line at $\approx +30$ days is overestimated~(see Sect.~\ref{app:line_trans}).}, 
the line-ratio \emph{flips back} (the ``flop'') to hydrogen-dominated after the end of the plateau of the line-flux ratio (day $+50$), and following the re-brightening of H line emission. This double transition ``flip-flop'' behavior highlights the diversity of these transitional supernovae, which ultimately arises from the varied mass loss histories of their massive star progenitors. We discuss the implications of this observation in Sect.~\ref{sec:discuss}.

\subsection{Constraining the Photosphere and Dust Emission Properties from the Late-time VLT/X-shooter Spectra }\label{sub:dustX}

The fits of the SN~2021foa VLT/X-shooter spectra at $+66$, $+95$ and $+129$ days (see Sect.~\ref{sec:dustemis}), constrain the radius and temperature of the photosphere, as well as the mass and temperature of the dust. These results are presented in Tab.~\ref{tab:dustbb}. 

\begin{deluxetable}{ccccc}
\tablecaption{Fitted parameters for the BB + MBB model to the VLT/X-shooter spectra.}
\tablehead{
\colhead{Epoch}
& \colhead{$R_{\rm SN}$} & \colhead{$T_{\rm SN}$} & \colhead{$M_d$} & \colhead{$T_d$} \ \\
\colhead{}
& \colhead{$10^{14}$ cm} & \colhead{$10^{3}$ K} & \colhead{$10^{-5} {\rm M}_{\odot}$} & \colhead{$10^{3}$ K}
}
\startdata
$+66$ & $1.95(01)$ & $9.14(03)$ & $2.88(03)$ & $1.36(002)$ \\
$+95$ & $1.83(01)$ & $8.47(02)$ & $2.59 (02)$ & $1.33(002)$ \\
$+129$ & $1.49(01)$ & $9.54(06)$ & $2.40(04)$ & $1.44(004)$ 
\enddata
\tablecomments{Fitting uncertainties are given in $10^{-1}$ units of each column. }
\label{tab:dustbb}
\end{deluxetable}

We infer the radius of the SN photosphere at $+66$ and $+95$ days of $\approx 2.0\times 10^{14}$~cm and the temperature $\approx 9000$ K. We can compare this to the inferred $R_{BB}$ and $T_{BB}$ (Sect.~\ref{subsec:lcbol}) at the last epoch, $\approx +35$ days, of our {\tt EXTRABOL} light curve modeling.  The {\tt EXTRABOL} results predict that $R_{BB}$ and $T_{BB}$ decline with time (see Fig.~\ref{fig:bb}), and our inferred radius is within the bounds of a linear extrapolation of the {\tt EXTRABOL} prediction at $+35$ days. We stress that we do not use any {\tt EXTRABOL} extrapolation to constrain the late-time photosphere properties. As such a linear extrapolation from $+35$ days to $+66$ and $+95$, while the simplest possible model, is likely unphysical. Indeed, the large inferred photospheric temperature from the VLT/X-shooter spectra indicates a shallower evolution.  

At all epochs, the dust mass and temperature inferred for SN~2021foa are $\approx 3 \times 10^{-5} {\rm M}_{\odot}$ for a carbonaceous dust composition and $\approx 1\,400$ K, respectively. However, the inferred low dust temperature indicates that a silicate dust composition can be possible. Adopting a silicate dust composition ($A_d = 0.2$ $\times$ 10$^{4}$~cm$^2$ g$^{-1}$), the inferred dust mass increases by about a factor of 5 at similar dust temperature.  

\vspace{1cm}
\section{Interpreting the Observations and Analysis Results}\label{sec:discuss}

As with all members of the transitional IIn/Ibn, SN~2021foa exhibits characteristics of both Type IIn and Type Ibn supernovae, albeit with key differences to both classes. SN~2021foa exhibits a short ($\sim$ 10 days) plateau in the optical light curves about two weeks past peak. However the plateau length is shorter than that of other transitional SNe such as SN~2011hw and SN~2020bqj  \citep[$\sim 50$ days,][]{Kool_2021_2020bqj}. Similarly, while there are clear similarities between the spectra of SN~2021foa and Type Ibn supernovae after $+22$ days, neither SN~2021foa nor any other transitional IIn/Ibn SNe follow the Ibn template $R$-band light curve. 

However, two aspects make SN~2021foa unique:
\begin{itemize}
    \item SN~2021foa is the first clear example of transitional IIn/Ibn that transitions back -- a ``flip-flop'' 
    \item SN~2021foa exhibited prominent precursor emission about 50--20 days before peak brightness, as is common for SN~2009ip-like objects~ (Fig.~\ref{fig:Rabs})
\end{itemize}

The photometric resemblance of SN~2021foa with 2009ip-like transients may point to a common progenitor system as already suggested by~\citet{Reguitti_2022}. However, the mechanism of producing the precursor emission and the light curve plateau for SN~2021foa is likely different. 

Our spectroscopic analysis of SN~2021foa shows that
prominent \ion{He}{i}~$\lambda\lambda 5876,7065$ and \ion{Ca}{ii} IR emission lines have a broad ($\sim 6000$ km s$^{-1}$) component.
Contrary, the velocities of the broad components in all \ion{H}{i} lines do not surpass $5000$~km~s$^{-1}$.

While such a velocity agree with the average bulk ejecta velocities of most core collapse SNe~\citep{Gutierrez_IIvel}, the velocities inferred from the Balmer lines of SN~2009ip~\citep[$\sim 10\,000$~km~s$^{-1}$]{Pastorello_2013_09ip} and SN~2016jbu~\citep[$\sim 7000$~km~s$^{-1}$]{Kilpatrick_2018,Brennan_2022a} are higher.
Furthermore, all \ion{H}{i} and the \ion{He}{i}~$\lambda 5016$  lines have narrow emission lines ($\sim$ $600 - 200$ km s$^{-1}$). Such narrow lines are characteristic of classical Type IIn and Ibn SNe such as SN~2010jl \citep{Gall_2010jl} and SN~2006jc, and originate from an extended CSM. However, most \ion{He}{i} lines in the spectra of SN~2021foa lack a narrow emission component. 

\begin{figure}[hbt!]
\epsscale{1.2}
\plotone{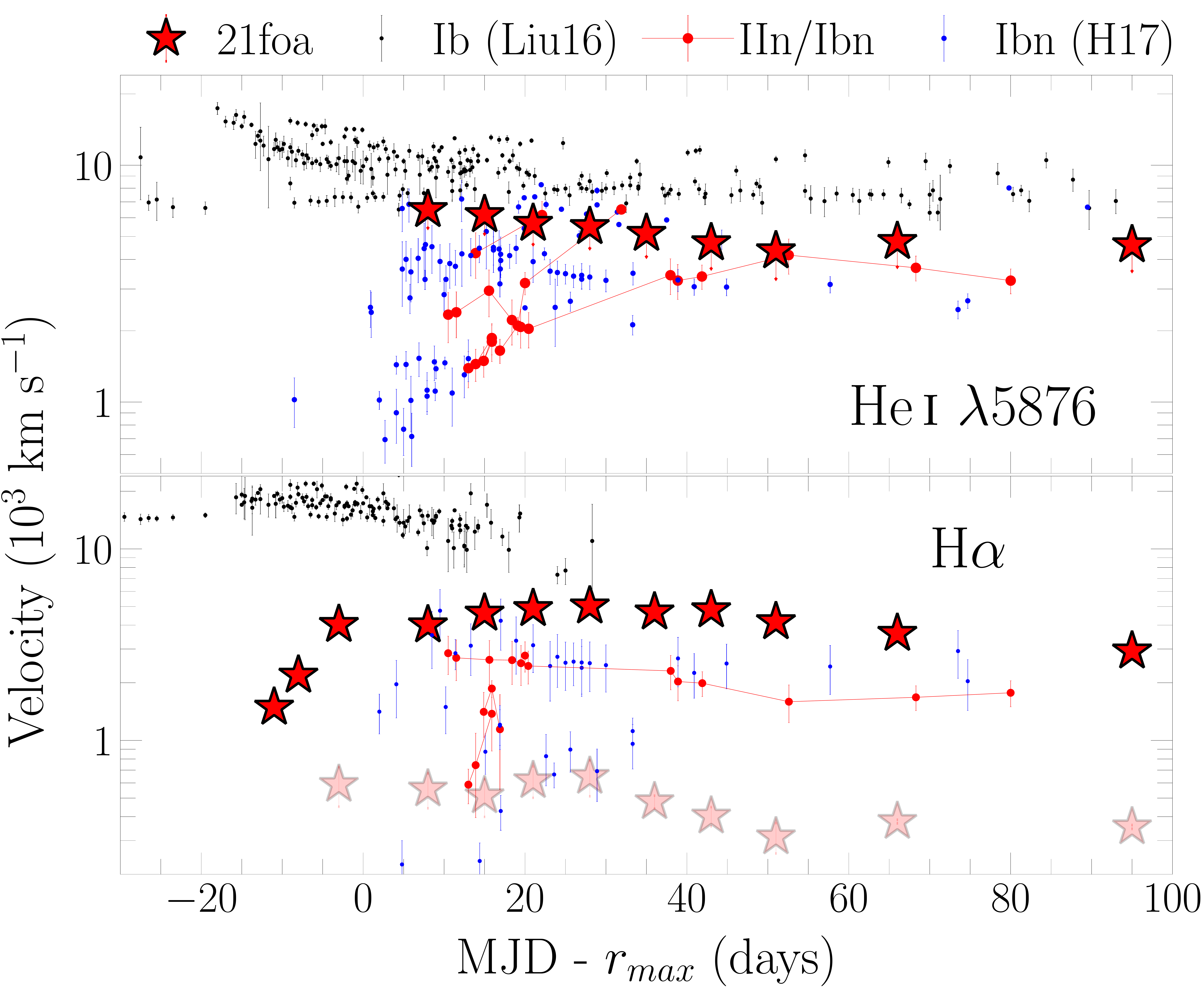}
\caption{{\it Upper panel:} Evolution of the velocity of the \ion{He}{i}~$\lambda 5876$ for Type Ib~\citep[][Liu16]{Liu_2016} and Ibn and IIn/Ibn SNe~\citep[][H17]{Hosseinzadeh_2017}. 
{\it Lower panel:} Same as {\it upper panel}, showing the evolution of the velocity of H$\alpha$ broad (bold) and narrow-absorption (light) components.
}
\label{fig:vel_ib}
\end{figure}
Fig.~\ref{fig:vel_ib} visualises the velocity evolution of H$\alpha$ and \ion{He}{i} $\lambda 5876$ of SN~2021foa in comparison to a sample of He-rich stripped-envelope Type Ib and IIn/Ibn and Ibn SNe from \citet{Liu_2016} and \citet{Hosseinzadeh_2017}. Evidently, the velocities measured for SN~2021foa are inconsistent with those measured for these individual SN types. 

SN~2021foa should be considered a hybrid helium-hydrogen-rich CSM interacting SN, and points to the diversity of the class of transitional IIn/Ibn supernovae. This diversity ultimately arises from the myriad of mass loss histories of massive stars. SN~2021foa's precursor emission together with our detailed photometric and spectroscopic record for SN~2021foa allows us to further constrain the properties of the progenitor environment.      

\subsection{The Origin of Strongly Blue-shifted Emission Lines}

\begin{figure}[hbt!]
\epsscale{1.2}
\plotone{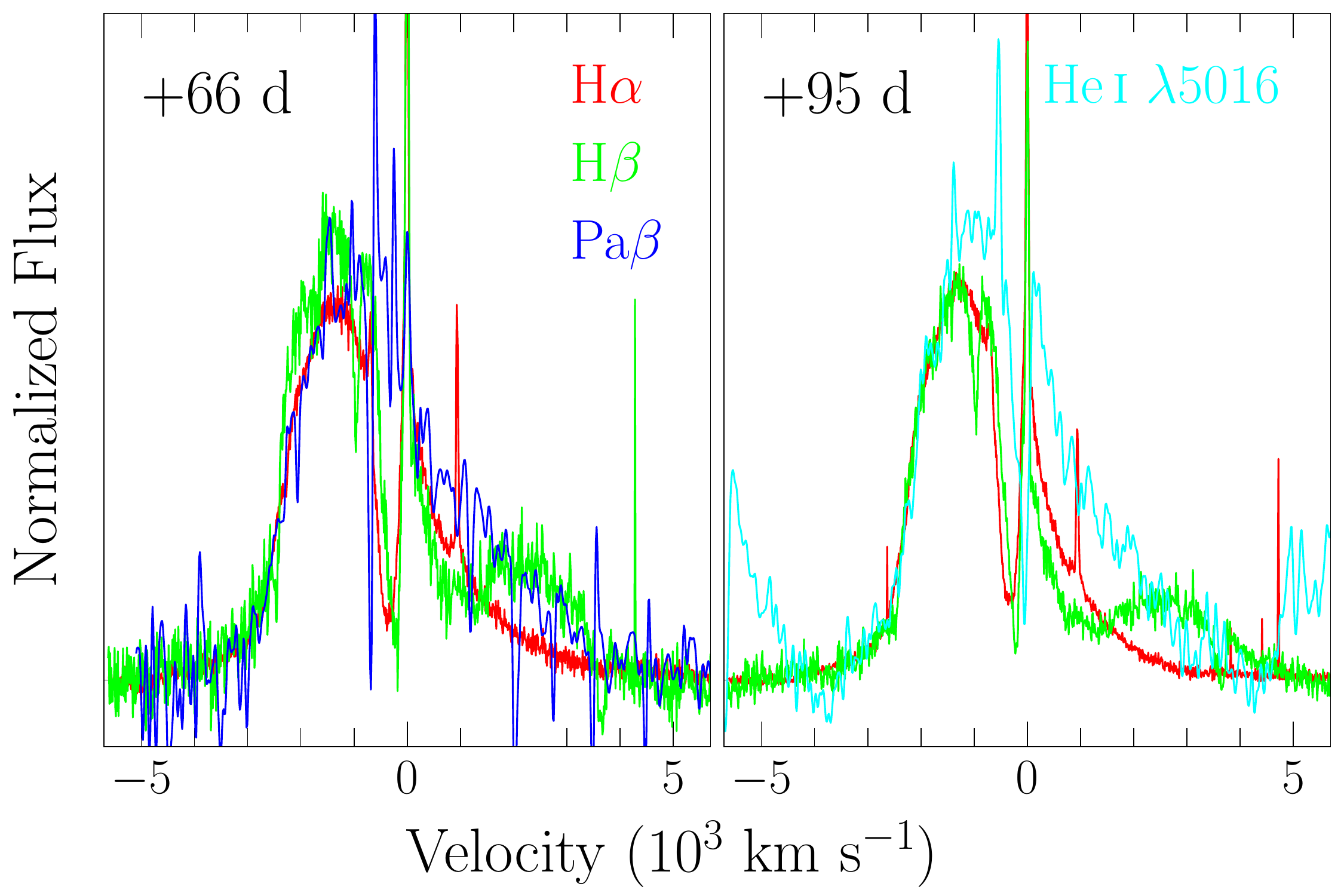}
\caption{{\it Left panel:} Normalized VLT/X-shooter spectra of H$\alpha$, H$\beta$ and Pa$\beta$ at $v\approx -1800$~km~s$^{-1}$ (day +66). {\it Right panel:} Normalized VLT/X-shooter spectra (+95) at $v\approx -1800$~km~s$^{-1}$, comparing H$\alpha$, H$\beta$ and \ion{He}{i}~$\lambda 5016$. 
} 
\label{fig:HHecomparison}
\end{figure}

The most intriguing spectroscopic signature of SN~2021foa is the persistent blue-shift of the peak of all \ion{H}{i} and \ion{He}{i}~$\lambda 5016$ emission lines past $+66$ days~(Fig.~\ref{fig:HHecomparison}). There are three possible origins of these blue-shifted profiles that can either be i) an asymmetric CSM or SN ejecta, ii)  an effect of dust extinction, or iii) occultation by the optically-thick photosphere of photons coming from a close line-forming region. We discuss each of these scenarios below.
We note that a radiatively accelerated CSM has been proposed to explain blue-shifted asymmetries observed in SN~2010jl~\citep{Fransson_2010jl}. Nevertheless, we do not consider that scenario in this work since the  blue-shifted profiles in SN~2021foa are persistent at late times, while the acceleration is stronger at peak luminosity~\citep{Smith_2017hcc}.

\subsubsection{Asymmetric CSM}
\label{subsubsec:asym}

An asymmetric geometry of either the CSM or the ejecta can lead to strongly blue/red-shifted emission line profiles, as has been suggested for e.g., Type IIn SNe~2009ip~\citep{Margutti_2014_2009ip},~2016jbu~\citep{Brennan_2022b} and~2013L ~\citep{Andrews_2013L}, Type Ibn SNe~2006jc~\citep{Foley_2006jc}, ~2019wep~\citep{Gango_2022} and ~2015G~\citep{Shivvers_2015G}. Indeed, the presence of either an asymmetric CSM or ejecta is confirmed by spectropolarimetric observations of supernovae, such as the  Type IIn SN~1998S~\citep{Leonard_1998S},  SN~2009ip~\citep{Mauerhan_2009ip_pol}, SN~2010jl~\citep{Patat_2010jl_pol} and SN~2017hcc~\citep{Kumar_2017hcc_pol}, among others~\citep[see ][for the complete Type IIn sample]{Bilinski_IIn_pol}.
Unfortunately, we do not have spectropolarimetric observations for SN~2021foa, and cannot conclusively confirm or rule out an asymmetric CSM / ejecta. Nevertheless, a disk-like CSM configuration is possible, with a high density H-rich 
material moving towards the observer. 
The CSM from the receding side in this configuration has lower density since the red emission does not re-brights as much as blue\'s.
In this scenario, the blue-shifted emission would be mostly dominated by the interaction of the SN ejecta with the dense CSM material. This scenario has been proposed for Type IIn~PTF11iqb~\citep{Smith_PTF11iqb} to explain the highly asymmetric red-shifted H$\alpha$ profiles observed at $>+500$ days.

Alternatively, \citet{Thone_2017} posit that the blue-shifted profiles observed in Type IIn (SN~2009ip-like) SN~2015bh at $>+126$ days are explained as shocked emission from a single CSM shell expelled at $-2000$ km s$^{-1}$ about $-50$ days prior to the main explosion event. Given our observations of SN~2021foa, we can compare the the scenario of \citet{Smith_PTF11iqb} and the single shell suggestion for SN~2015bh~\citep{Thone_2017}. Our observations indicate that the shell would need to be asymmetric in a specific direction towards the observer to produce the blue-shifted profile (see Fig.~\ref{fig:HHecomparison}). 

Furthermore, our observations, particularly the decrease in line-velocities at late-times, suggest that there are \emph{multiple} H-rich CSM shells out to larger radii, rather than a single shell. Each of those shells can deviate from simple spherical symmetry. Thus, the composite of all these shells would likely lead to line profiles inconsistent with what is observed here for SN 2021foa.  

While the disc-like scenario 
is more likely than the single shell scenario, a pre-existing H-rich, high-density CSM towards the observer must also show emission at early-times. 
However, as shown in Fig.~\ref{fig:modelHa}, the H$\alpha$ line is well modeled by symmetric profiles throughout its evolution. 
Therefore, the high-density CSM must be placed further out of a spherical symmetric CSM.

Thus, while we cannot unambiguously rule out an asymmetric CSM as an explanation for the blue-shifted line profiles without spectropolarimetry, the scenario is unlikely as it requires special fine-tuning of the CSM properties (e.g., density and location) to be consistent with our observations. Next, we consider newly formed dust and occultation by the photosphere as a potential origin of the blue-shifted emission.

\subsubsection{Newly formed dust}\label{subsubsec:dust}

Newly formed dust located either close to the emission line formation region or within it causes a blue-shift of the peaks of emission lines. Simultaneously, a red-blue asymmetry of the emission line profiles due to absorption of photons from the receding side of the SN is produced~\citep{Lucy_1989, Gall_2010jl, Bevan_1987A} . Thus, the blue side of the line profile remains unaffected, while the red side of the line gets extinguished (red-blue asymmetry). 

Early dust formation in a cool dense shell (CDS) causing a red-blue asymmetry of emission lines has been observed in, for e.g., SN 2006jc,  SN 2010jl or SN~2017hcc~ \citep[e.g. ][]{,Smith_dust_2006jc,Chugai_dust_2006jc,Gall_2010jl, Chugai_dust_2010jl_2018, Smith_2017hcc,Bevan_2010jl}. However, the newly formed dust is composed of both large and small grains \citep{Gall_2010jl, Smith_2017hcc}. For the latter case, the blue-shift of the emission line profiles and red-blue asymmetry exhibits a measurable wavelength dependence, with bluer emission lines exhibiting larger blue-shifts than redder emission lines. 
As shown in Fig.~\ref{fig:HHecomparison}, the emission line peaks of the normalised profiles of H$\alpha$, H$\beta$ and Pa$\beta$ for SN~2021foa are nearly identical. This rules out newly formed dust in a CDS as the origin of the blue-shifted \ion{H}{i} and \ion{He}{i} emission lines.  

Additionally, ejecta dust formation typically starts around one year after explosion when the temperature of the ejecta has cooled to less than about 1600 -- 2000 K, which are the sublimation temperatures of  silicate and carbonaceous dust, respectively ~\citep[][and references therein]{Gall_2011}. For SN~2021foa we have a strong temperature constraint from our X-shooter modeling at late times. As shown in Tab.~\ref{tab:dustbb}, the temperature of the photosphere remains at $\approx 9000$~K -- a factor of 4 too hot to form dust grains. Thus, ejecta dust formation can be ruled out as the origin of blue-shifted emission line profiles as well as the observed NIR excess emission in SN~2021foa. 

Consequently, the observed thermal dust emission (see Sect.~\ref{sub:dustX}) must originate from surviving pre-existing dust at large distances. The amount of dust inferred from our modified BB fits is consistent with dust masses derived at early epochs in other core collapse SNe \citep{Gall_2011, Gall_2018,dust_ibn}. Further, the surviving pre-existing dust must be at radii $> 10^{17}$ cm, while the emission line forming region is at lower radii ($< 10^{15}$ cm). Hence, also this dust does not cause a blue-red asymmetry of the emission lines. Our observations therefore conclusively rule out newly formed dust as a source of the blue-shifted line profiles.

\subsubsection{Occultation by the photosphere}\label{subsubsec:occ}

Occultation can be an alternative explanation for the origin of the non-wavelength dependent blue-shifts and red-blue asymmetry of the emission line profiles of SN~2021foa. In such a scenario, the emission from the line-forming region at the receding end of the CSM/SN ejecta is occulted by the optically thick continuum photosphere \citep{Chevalier_1976,Smith_2012_blue, Dessart_2015}. This has been observed in non-interacting Type II SNe~\citep{Anderson_II} and suggested for some 
Type IIn such as SN~2010jl~\citep{Fransson_2010jl}, SN~2021adxl~\citep{Brennan_2023} or SN~2013L~\citep{Taddia_2013L}. 

In the case of occultation, the line-forming region producing the intrinsically symmetric emission lines needs to be very close to the photosphere, else the effect of occultation is minimal as discussed for e.g., SN~2010jl, where the wavelength dependent blue-shifts are likely due to newly formed dust in the CDS (see Sect.~\ref{subsubsec:dust}).

For SN~2021foa, occultation is likely, because we neither observe a change of the red-blue asymmetry and blue-shifts with either wavelength or time ($\sim$ 66 -- 129 days).
Furthermore, from our BB fits to the VLT/X-shooter data (Sect.~\ref{sub:dustX}) we find that the photospheric radius remains at around $2\times 10^{14}$ cm, which is similar to the location of the outer radius of the CSM using both {\tt MOSFiT} models~(\ref{sec:mosfit_discussion}).

However, if occultation occurs, the blue-shift and red-blue asymmetry of the emission lines should decrease over time since the photosphere continues receding, blocking less photons with time. Unfortunately, we do not have any data coverage of SN~2021foa beyond $+129$ days. Thus, we cannot unambiguously determine, if occultation by the photosphere is indeed, a viable explanation for the observed blue-shifts, but it is the most natural scenario that is consistent with all of our observations.

\subsection{Precursor emission of 2009ip-like objects}\label{subsec:precursor}

SN~2021foa has shed off most of its hydrogen envelope, as evident from the spectra (Sect.~\ref{subsec:specdat}). Over its lifetime, the progenitor of SN~2021foa created the multiple CSM layers with different velocities as evident from our emission and absorption line analysis (Sect.~\ref{subsec:linevol}, Figs.~\ref{fig:HFWHM} and~\ref{fig:Othermin}). 

The formation of multiple discrete CSM layers requires episodes of strong mass loss as eruptions or steady winds.  Indeed, precursor luminous outbursts have been observed months-to-years before terminal explosion for SN~2006jc, SN~2015bh and 2016jbu~\citep{Foley_2006jc,Thone_2017,Brennan_2022a}. 

Additionally, our high mass loss rate estimates ($2.0$ M$_{\odot}$ yr$^{-1}$) suggest that SN~2021foa suffered from intense mass loss prior to explosion. 
Our analysis of the ATLAS $o$-band light curve data from $\sim 5$ years prior to the SN~2021foa terminal explosion (Sect.~\ref{subsec:pre}), shows that SN~2021foa had no eruption brighter than $20$ mag over 5 years prior to explosion. 
Under the $s=2$ scheme, {\tt MOSFiT} gives an outer radius of the CSM of $\approx 1.5\times 10^{14}$ cm. 
Assuming a wind velocity of $400$ km s$^{-1}$, all the CSM was expelled $\sim 12$ years ago, during a period of $\sim$ half a year.  
Given the lack of observations of SN~2021foa prior year 2019, we cannot confirm such an event. 
However, SN~2015bh showed numerous outbursts throughout 20 years before the event B in 2015~\citep{Thone_2017}.

Several possibilities to explain the precursor emission of 2009ip-like objects have been suggested.
For SN~2009ip,~\citet{Mauerhan_2009ip_finalexp} argued that the explosion occurred at the onset of event A as a weak Type II SN, while the brighter outburst is mainly powered by SN ejecta-CSM interaction. A similar scenario was suggested for SN~2015bh~\citep{Elias_2016}.
In contrast,~\citet{Pastorello_2013_09ip} and~\citet{Margutti_2014_2009ip} have proposed that event A of SN~2009ip is an eruption similar to those observed in the years before. Then, the event B is either due to interaction of the material expelled at event A with previous eruptions (colliding shells) or the expanding SN ejecta itself (terminal explosion).
Shell-shell interaction is one of the suggested scenario to expain the precursor emission of SN~2015bh~\citep{Thone_2017}.

For SN~2021foa, our {\tt MOSFiT} calculations were only performed for event B photometry i.e., assuming that the true CC-SN occurred at the end of event A, and that interaction with one CSM (RD+CSM) is sufficient to reproduce the light curve.
Based on these assumptions, the low $^{56}$Ni mass obtained with {\tt MOSFiT}, typical for SN~2009ip-like SNe, suggest that event B is most likely powered by shock breakout of the CSM + SN ejecta-CSM interaction rather than only radioactive decay.

\begin{figure*}[hbt!]
\epsscale{1.}
\plotone{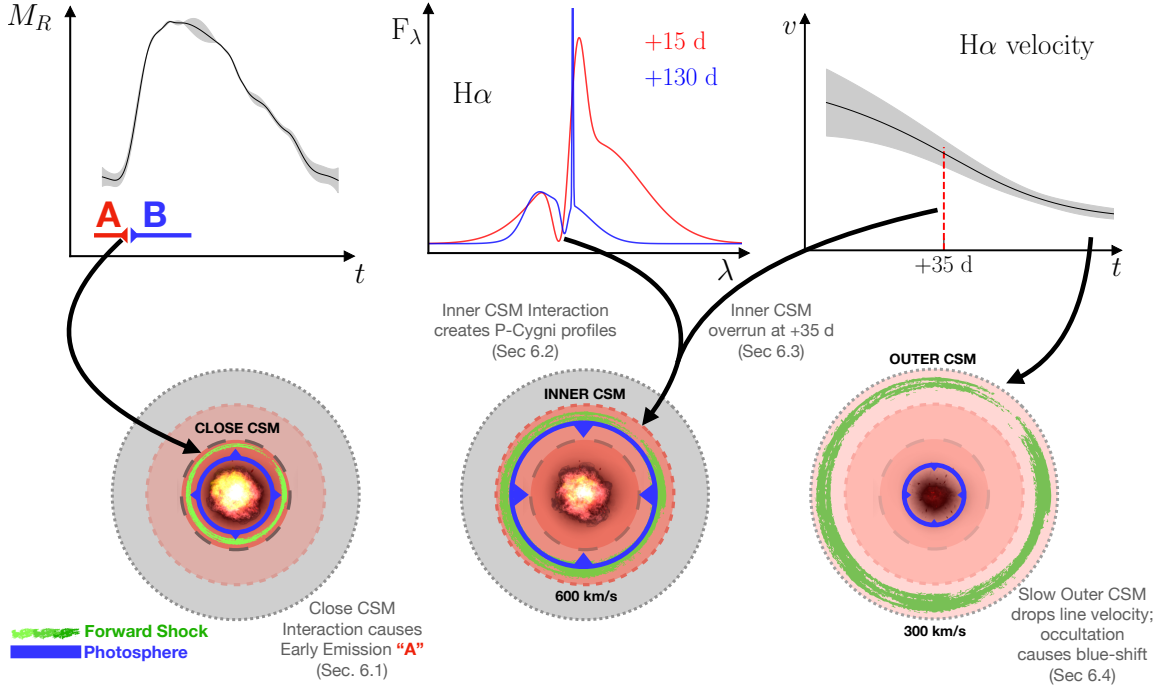}
\caption{Proposed scenario for SN~2021foa, sketching the temporal evolution of the forward shock and the photospheric radius.
In this scenario, SN~2021foa is embedded in a H-rich CSM composed by multiple shells. 
The precursor emission (event A) is originated by the interaction of the SN forward shock with an
unobserved, close CSM. 
From day $-10$ to $+35$, a second, inner CSM with a characteristic velocity of $\approx 600$ km s$^{-1}$ is observed through the P-Cygni profile of the Balmer lines. 
The He-rich ejecta, hidden at early epochs, is visible at day $+15$ after the photospheric radius reaches its maximum. The strong \ion{He}{i} lines at this epoch sets up the {\it flip} in Fig.~\ref{fig:ratio}.
A rapid decline of the H$\alpha$ velocity at day $+35$ marks the phase where the forward shock overran most of the inner CSM.
After day $+50$, the velocity of the H$\alpha$ line remains constant at $\approx 350$ km s$^{-1}$. 
We attribute this velocity to an outer, CSM shell. 
The re-brightening of Balmer lines at late times ($+66$--$+130$ days) due to hydrogen recombination of a shocked SN ejecta-CSM (either the close, inner or the outer CSM), gives rise to the {\it flop} in Fig.~\ref{fig:ratio}. 
The apparent blue-shift of several lines at late times is either due to an asymmetric CSM or an effect of occultation by the optically thick photosphere close to the line forming region of these lines, located at $\sim 2\times 10^{14}$ cm.
}
\label{fig:final_scenario}
\end{figure*}

\vspace{1cm}
\section{Building a Complete Picture of SN~2021foa}\label{sec:scenario}

In this section, we summarize the key features from Sect.~\ref{sec:discuss} and build a complete, cohesive model for SN~2021foa. 

\begin{itemize}
\item SN~2021foa resembles 2009ip-like SNe~\citep{Reguitti_2022}. In particular, the precursor emission (event A) of SN~2021foa starting about $\sim -50$ days prior to the peak of the light curve (event B) and the presence of a short plateau of a few days after the peak are very similar to SNe~2016jbu~\citep{Kilpatrick_2018,Brennan_2022a} and~2009ip~\citep{Pastorello_2013_09ip,Margutti_2014_2009ip}.
\item For SN~2021foa, the velocity of the minimum of the narrow absorption component of \ion{H}{i}, \ion{He}{i}, \ion{Fe}{ii} , \ion{Ca}{ii} and \ion{O}{i} lines decreases from $\sim 600$ at day $+15$ to $\lesssim 300$ km s$^{-1}$ at day $+60$.
\item Intriguing and strongly blue-shifted emission lines of \ion{H}{i} and \ion{He}{i} lines emerge at late times ($>+66$ days) in  SN~2021foa.
\item SN~2021foa exhibit a distinctive phase where the \ion{He}{i}~$\lambda 5876$ emission line is as strong as H$\alpha$. This line-ratio-plateau is observed in the transitional IIn/Ibn SNe.
\item Unambiguously, SN~2021foa has a ``flip-flop'' nature, transitioning from a IIn before peak brightness (flip) to a He-dominated (Ibn-like) SN for about $+30$ days, and returning to a IIn past $+66$ days (flop). 
\end{itemize}

\subsection{The Luminosity of Event A and Explosion Date}

As for most of the 2009ip-like events, the explosion time is uncertain and ultimately dependent on the physical mechanism employed to explain the luminosity of event A. Our spectroscopic observations strongly support a scenario where multiple CSM shells are expelled at different times prior to explosion. We sketch this scenario in Fig.~\ref{fig:final_scenario}. In this scenario, the closest CSM is created by material ejected from prior outbursts or the terminal explosion. Consequently, this material is close to the progenitor system and is rapidly overrun by the forward shock from the SN, powering the emission at event A. 

This scenario requires that the explosion date be at event A. For convenience, we place the assumed explosion at $\approx -25$~days, consistent with our results from {\tt MOSFiT} fits~(Tab.~\ref{tab:mosfit}), and the lack of  prior outbursts (see Sect.~\ref{subsec:precursor}). 
We stress that the final scenario is not dependent on the precise explosion date. Even in the scenario that the outburst occurred earlier than event A, the key feature of our model for SN~2021foa is that event B itself is powered by SN ejecta-CSM interaction rather than radioactive decay. 

\subsection{The Dynamics of the Forward Shock from Event B to the Plateau}
The broad components of \ion{H}{i}, \ion{He}{i} and \ion{Ca}{ii} IR emission lines suggest an ejecta velocity of $\sim 6000$~km~s$^{-1}$. The maximum velocity inferred from the blue wings of the broad component, indicative of the shock velocity ($v_{\rm FS}$), does not surpass 10\,000~km~s$^{-1}$ at all epochs. Therefore, we assume the velocity of the forward shock as $v_{\rm FS} =$ 10\,000~km~s$^{-1}$. 
The first estimate of the photospheric radius is at $\approx 6\times 10^{14}$~cm at day $-10$. At this epoch, the forward shock is located at $\approx 10^{15}$~cm.  Hence, the photosphere lies within a shocked CSM region.
The interaction between the SN forward shock and this close CSM might be the underlying powering mechanism of event A. 
Before peak, we only observe emission from the H-rich CSM, thus explaining the weak \ion{He}{i} lines in the spectra.  
From $-10$ up to $+30$ days, the P-Cygni profile of the Balmer lines suggest a velocity of the inner CSM layer  (Fig.~\ref{fig:final_scenario}) to be $600$~km~s$^{-1}$.
Around day $+15$, the photospheric radius reaches its maximum ($\approx 10^{15}$~cm) while the forward shock is located at $\approx 3.5 \times 10^{15}$~cm. 
Around this epoch, 
we observe the {\it flip}; the He-rich ejecta is becoming visible and \ion{He}{i} lines get stronger.

\subsection{Evolution Post-plateau to Late-time}
After the plateau phase ($+25$ days), the photospheric radius starts to recede to a radius of about $6\times 10^{14}$~cm. In the meanwhile, the forward shock continues to propagate outwards, reaching a radius of about $5\times 10^{15}$~cm, assuming a constant forward shock velocity. At day $+35$, we first observe the decline of the velocity of the absorption minimum of the Balmer lines~(\ion{H}{i} in Fig.\ref{fig:Othermin}). This decline implies that the forward shock has overrun the inner CSM entirely.
Nevertheless, though weaker, the continuous interaction between the SN ejecta and the CSM still powers the emission of \ion{H}{i} and \ion{He}{i} lines~(Fig.~\ref{fig:HHeflux}). 
This is evident in the plateau phase of the H$\alpha$/\ion{He}{i}~$\lambda 5876$ ratio~(Fig.~\ref{fig:ratio}). 

Past $+50$ days, the line fluxes of \ion{H}{i} lines, from H$\delta$ to Pa$\beta$, increase (or stay constant), while \ion{He}{i} lines keeps decreasing~(Fig.~\ref{fig:HHeflux}).
At this epoch, we observe the {\it flop} in Fig.~\ref{fig:ratio}.

\subsection{Late-time Evolution}
At $+130$ days, the velocities of the absorption minima of \ion{H}{i}, \ion{He}{i}, \ion{Ca}{ii}, \ion{O}{i} and \ion{Fe}{ii} are below $\sim350$~km~s$^{-1}$. This is consistent with a slow moving outer CSM layer at at a distance of $>1.3\times 10^{16}$~cm and thus, has not been overrun yet by the forward shock ($v_{\rm FS}=10^{4}$~km~s$^{-1}$). 
The likely origin of the low velocity, outer CSM is from slow winds at early stages in the evolution of the progenitor star, while the inner CSM shells are from eruptions or faster winds closer to the explosion. In the multiple-shells scenario, the slowly receding photosphere ($\sim 2\times 10^{14}$~cm from $+60$ to $+130$ days) lies within the inner CSM, close to the line forming region of the intermediate/broad components ($\approx 4000$ km s$^{-1}$). In this scenario, the blue-shifted emission lines emerging after the plateau phase are the result of occultation~(Sect.~\ref{subsubsec:occ}) of the emission line region
by a dense and optically thick CSM shell. Finally, at late times, the H-rich inner CSM recombines, explaining the re-brightening of the Balmer lines (Fig.~\ref{fig:HHeflux}). Alternatively, as suggested for PTFiqb~\citep{Smith_PTF11iqb}, the interaction between the SN ejecta and a dense, outer CSM approaching to the observer at $-2000$ km s$^{-1}$, could also be viable option to explain the late re-brightening and asymmetry observed in the \ion{H}{i} lines.

\subsection{Final remarks}

The early discovery and follow-up of SN~2021foa exhibits the imprints of CSM interaction on the SN’s light curve and the evolution of its spectral features. 
From our light curve analysis (Sect.~\ref{sec:mosfit_discussion}), we obtain a CSM mass $\leq 1 ~{\rm M}_{\odot}$, ejecta mass $\approx$ 8 M$_{\odot}$ and mass loss rate of $2$ M$_{\odot}$yr$^{-1}$ for a wind-like ($s=2$) scenario. 
This mass loss rate is higher than values typical found for Wolf-Rayet or LBV stars, favoured progenitors of Type Ibn and IIn SNe, respectively. 
From our detailed analysis of the evolution of the line profiles (Sect.~\ref{subsec:linevol}), we conclude that SN~2021foa had a rich mass-loss history, forming multiple CSM shells before the terminal explosion. This CSM configuration, while rare, does share key similarities with other scenarios proposed for interacting SNe such as SN~2015bh~\citep{Elias_2016}.
We stressed out that the assumption of a one-shell CSM in {\tt MOSFiT} is insufficient to explain the precursor emission of SN~2021foa, and incompatible with a CSM composed by multiple-shells.
Nevertheless, the overall behaviour of the LC during event B might be approximated well by this assumption. 
Therefore, the estimated {\tt MOSFiT} physical parameters might still be valid if the interaction with the close/inner CSM in our scenario is the main contributor to the luminosity of event B. 

SN~2021foa adds to the number of SNe with truly complex CSM structures which challenges our understanding of extreme mass-loss mechanisms in massive stars, opening up the possibilities of different progenitor scenarios for strongly interacting CC-SNe.

\vspace{1cm}
\section{acknowledgments}\label{sec:ack}

We thank the S.\ Raimundo, O.\ Rodriguez, and the anonymous referee for constructive comments that helped to improve the manuscript.

This work is supported by a VILLUM FONDEN Young Investigator Grant (project number 25501) and by research grants (VIL16599,VIL54489) from VILLUM FONDEN. 
C.R.A.\ is supported by the European 1172 Research Council (ERC) under the European Union’s 1173 Horizon 2020 research and innovation programme (grant 1174 agreement No.\ 948381).
V.A.V.\ acknowledges support from NSF through grant AST--2108676.
This work is supported by the National Science Foundation under Cooperative Agreement PHY--2019786 (The NSF AI Institute for Artificial Intelligence and Fundamental Interactions, \url{http://iaifi.org/}).
Parts of this research were supported by the Australian Research Council Discovery Early Career Researcher Award (DECRA) through project number DE230101069.
The UCSC team is supported in part by NASA grant 80NSSC20K0953, NSF grant AST--1815935, the Gordon \& Betty Moore Foundation, the Heising-Simons Foundation, and by a fellowship from the David and Lucile Packard Foundation to R.J.F.
R.Y.\ received support from a Doctoral Fellowship from the University of California Institute for Mexico and the United States (UCMEXUS) and a NASA FINESST award (21--ASTRO21--0068)
postdoctoral fellowship.
GN gratefully acknowledges NSF support from AST-2206195 for this work. GN is also supported by NSF CAREER grant AST-2239364, supported in-part by funding from Charles Simonyi, and NSF OAC-2311355, DOE support through the Department of Physics at the University of Illinois, Urbana-Champaign (13771275), and support from the HST Guest Observer Program through HST-GO-16764. and HST-GO-17128 (PI: R. Foley). This work was performed in part at Aspen Center for Physics, which is supported by National Science Foundation grant PHY-2210452.

This investigation is based on observations made with ESO Telescopes at the La Silla Paranal Observatory under programme ID 107.22RH (PI C.\ Gall) and 109.23K3 (PI D.\ Farias).
Research at Lick Observatory is partially supported by a generous gift from Google.
Based in part on observations obtained at the Southern Astrophysical Research (SOAR) telescope, which is a joint project of the Minist\'{e}rio da Ci\^{e}ncia, Tecnologia e Inova\c{c}\~{o}es (MCTI/LNA) do Brasil, the US National Science Foundation’s NOIRLab, the University of North Carolina at Chapel Hill (UNC), and Michigan State University (MSU).
The data presented here were obtained in part with ALFOSC, which is provided by the Instituto de Astrofisica de Andalucia (IAA) under a joint agreement with the University of Copenhagen and NOT.
We acknowledge the use of public data from the Swift data archive.
This work makes use of data taken with the Las Cumbres Observatory global telescope network.  The LCO group is funded by National Science Foundation (NSF) grants AST--1911151 and AST--1911225.
This work has made use of data from the Asteroid Terrestrial-impact Last Alert System (ATLAS) project. ATLAS is primarily funded to search for near earth asteroids through NASA grants NN12AR55G, 80NSSC18K0284, and 80NSSC18K1575; byproducts of the NEO search include images and catalogs from the survey area.  The ATLAS science products have been made possible through the contributions of the University of Hawaii Institute for Astronomy, the Queen's University Belfast, and the Space Telescope Science Institute.
This work is based in part from data obtained at the Infrared Telescope Facility, which is operated by the University of Hawaii under contract 80HQTR24DA010 with the National Aeronautics and Space Administration.

YSE-PZ was developed by the UC Santa Cruz Transients Team. The UCSC team is supported in part by NASA grants NNG17PX03C, 80NSSC19K1386, and 80NSSC20K0953; NSF grants AST-1518052, AST-1815935, and AST-1911206; the Gordon \& Betty Moore Foundation; the Heising-Simons Foundation; a fellowship from the David and Lucile Packard Foundation to R.J.\ Foley; Gordon and Betty Moore Foundation postdoctoral fellowships and a NASA Einstein Fellowship, as administered through the NASA Hubble Fellowship program and grant HST-HF2-51462.001, to D.O.\ Jones; and an NSF Graduate Research Fellowship, administered through grant DGE-1339067, to D.A.\ Coulter.

\vspace{5mm}
\facilities{
{\it Swift} (UVOT), LCOGT (Sinistro), Shane (Kast, Nickel), VLT (X-shooter), NOT (ALFOSC), Siding Spring (WiFeS), SOAR (Goodman), IRTF (SpeX), ATLAS, Thacher. Computational facility: High Performance Cluster (HPC) of the University of Copenhagen.}

\software{astropy~\citep{2013A&A...558A..33A,2018AJ....156..123A},{\tt esoreflex}~\citep{Freudling_esoreflex},
{\tt EXTRABOL}~\citep{ebol_cite},
{\tt SUPERBOL}~\citep{Nicholl_2017},
matplotlib~\citep{matplotlib},
{\tt MOSFiT}~\citep{mosfit},
numpy~\citep{numpy},
specutils~\citep{specutils}, YSE-PZ~\citep{YSE_PZ}.
}

\appendix

\renewcommand{\thefigure}{A\arabic{figure}}
\renewcommand{\theHfigure}{A\arabic{figure}}
\renewcommand{\thetable}{A\arabic{table}}
\renewcommand{\theHtable}{A\arabic{table}}

\setcounter{figure}{0}
\setcounter{table}{0}

\section{Host emission at H$\alpha$}\label{app:hostha}

\begin{figure}[hbt!]
\epsscale{1.1}
\plotone{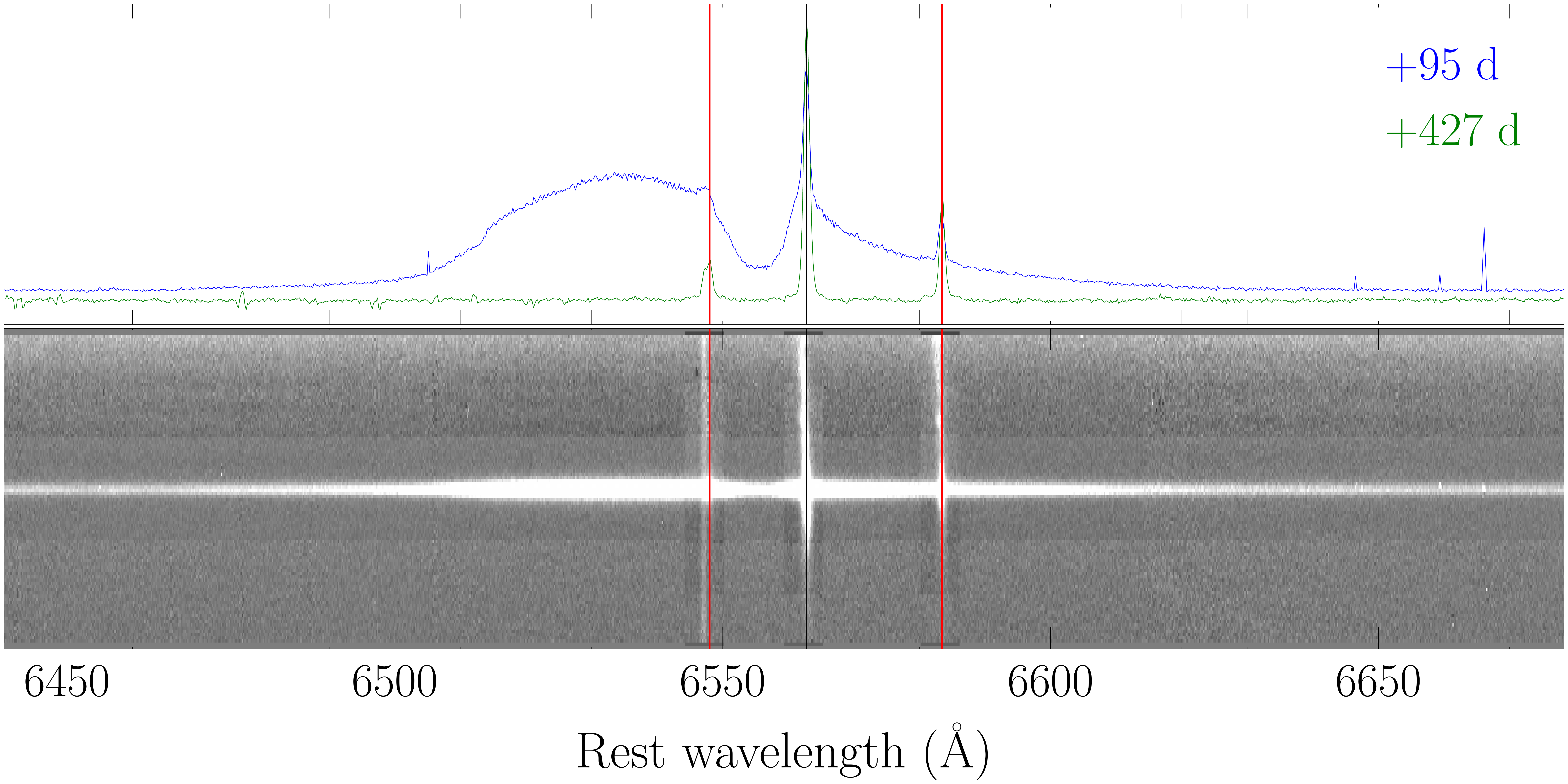}
\caption{VLT/X-shooter spectral region surrounding H$\alpha$ of SN~2021foa. {\it Upper panel:} Spectra at $+$95 and $+$427 days past peak. The black and red lines are located at the rest wavelength of H$\alpha$ and [\ion{N}{ii}]~$\lambda\lambda 6548,6583$, respectively. {\it Bottom panel:} the 2D spectrum at day $+$95 within the same spectral region.
}
\label{fig:hostlines}
\end{figure}

The narrow emission component of Balmer lines in the VLT/X-shooter spectra is dominated by the host galaxy emission rather than photo-ionized unshocked CSM. 
Fig.~\ref{fig:hostlines} demonstrates the challenge to perform an accurate reduction at the spectral range covering H$\alpha$ profile due to the contamination of the host emission.
Despite special efforts to correctly address this issue, the line flux of H$\alpha$~(Fig.~\ref{fig:HHeflux}) at $+95$ days might be overestimated.

\section{Pre-SN emission}\label{subsec:pre}

To investigate the pre-explosion activity of SN~2021foa, we followed the analysis described in~\citet{Wang_2023} using {\tt ATClean}~\citep{ATClean,Sofia_ATCLEAN}. We obtain the ATLAS forced photometry of $o$- and $c$-bands covering $\sim 5$ years up to event A at the position of SN~2021foa. 
Additionally, we performed forced photometry of eight control light curves within a distance of $17''$ to the SN. On average, the flux of these control light curves is expected to be zero. 
To emphasize the emission of a potential eruption, we defined a figure of merit (FOM) as the signal-to-noise ratio (SNR) convolved with a rolling Gaussian with a fixed kernel size determined by typical timescale of an eruption, $5 <\tau_G < 100$ days~\citep{Ofek_2010mc,Strotjohann_2021}.
The same rolling Gaussian was also applied to the control light curves to determine the FOM of the control light curves. 
By setting up a detection threshold
FOM$_{\rm limit}$, we expect that most of the FOM of the control light curves lies below this limit. If it is not the case, then there are unaccounted sources of contamination within the field of SN~2021foa.
Any real detection in the pre-SN LC must have a FOM larger than FOM$_{\rm limit}$. 
In order to establish a magnitude limit to detect eruptions of a given peak magnitude, we added three simulated Gaussian bumps ($p_1$, $p_2$ and $p_3$) with increasing amplitudes to one control light curve. 
A non-detection of any of these peaks translates into an upper limit of a real pre-SN eruption throughout the ATLAS coverage. 

For the particular case of SN~2021foa, we first convolved both the SNR of the control and SN LCs with a rolling Gaussian with a kernel size of  $\tau_G = 30$ days, close to the duration of the precursor emission observed in SN~2021foa.
Furthermore, we added three simulated Gaussian bumps, with peak magnitudes of $p_1 = 21.4$, $p_2 = 20.21$, $p_3 = 19.45$ mag, and fixed standard deviation of 25 days to the control LC \# 4.
Finally, we set up the detection limit as FOM$_{\rm limit} = 15$. With this same value,~\citet{Wang_2023} recovered $80\%$ of the eruptions larger than 20 mag for Type Ibn SN~2020nxt. 
For SN~2021foa, we recover $70\%$ of the eruptions larger than $o$-band peak magnitude $\approx 20$ mag.

\begin{figure}[hbt!]
\epsscale{.8}
\plotone{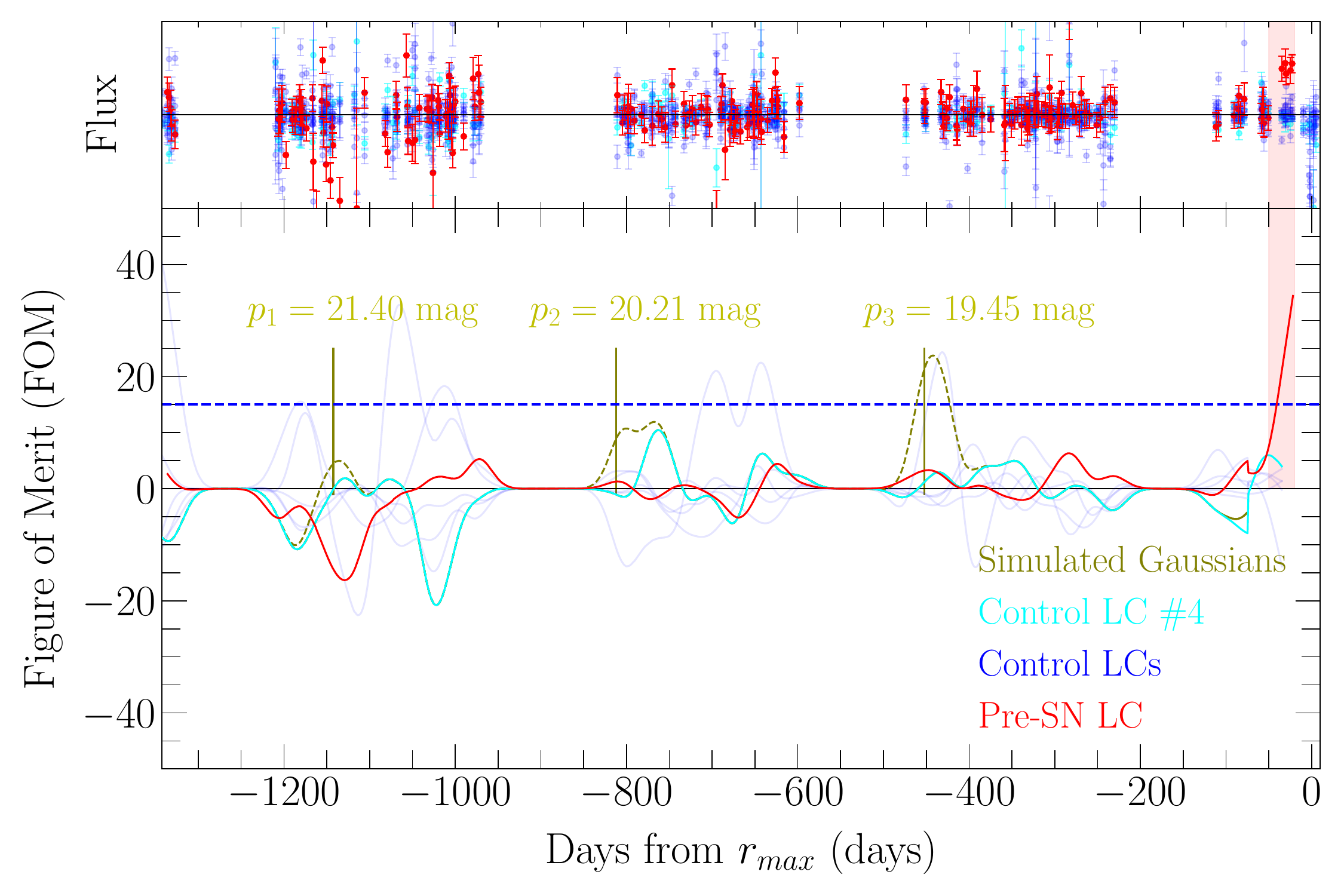}
\caption{Upper panel: light curve of SN~2021foa (red) and the eight control LCS (blue) before event A (red). Bottom panel: Figure of merit (FOM) of the pre-SN LC (red), eight control LCs (blue), control LC \#4 (cyan) and the simulated Gaussians + control LC \#4 (green) over ATLAS coverage ( $\sim 3$ years) up to event A of SN~2021foa. All these FOM curves were obtained after the convolution of the SNR of each LCs with a rolling Gaussian with kernel size of $\tau_G = 30$ days. The simulated Gaussians have peak magnitudes of 21.4 mag, 20.21 mag and 19.35 respectively and a fixed standard deviation of 15 days. The detection limit for our analysis was set up as FOM$_{\rm limit}=15$ (dashed line). Red shaded area encompasses the precursor emission of SN~2021foa.}
\label{fig:fom}
\end{figure}

The upper panel of Fig.~\ref{fig:fom} displays the forced photometry ATLAS light curves for SN~2021foa (red) and the control LCs (cyan and blue). We find no signature of any precursor emission associated with SN~2021foa. The bottom panel shows the results of our detection analysis. On average, the control light curves are below FOM$_{\rm limit}$. For the simulated Gaussians, only $p_3$ was successfully detected, while $p_2$ lies slightly below the threshold. Therefore, we can safely conclude that, similar to~\citet{Wang_2023}, no eruption is observed for SN~2021foa with a magnitude greater than or equal to $\sim 20$ mag. This translates into a detection limit of absolute magnitude of $M_{o}\approx -13.4$ mag. 
However, we cannot discard any pre-SN activity below this magnitude limit.  

\newpage
\section{Line profiles}\label{append_profiles}

\begin{figure*}[hbt!]
\epsscale{1.2}
\plotone{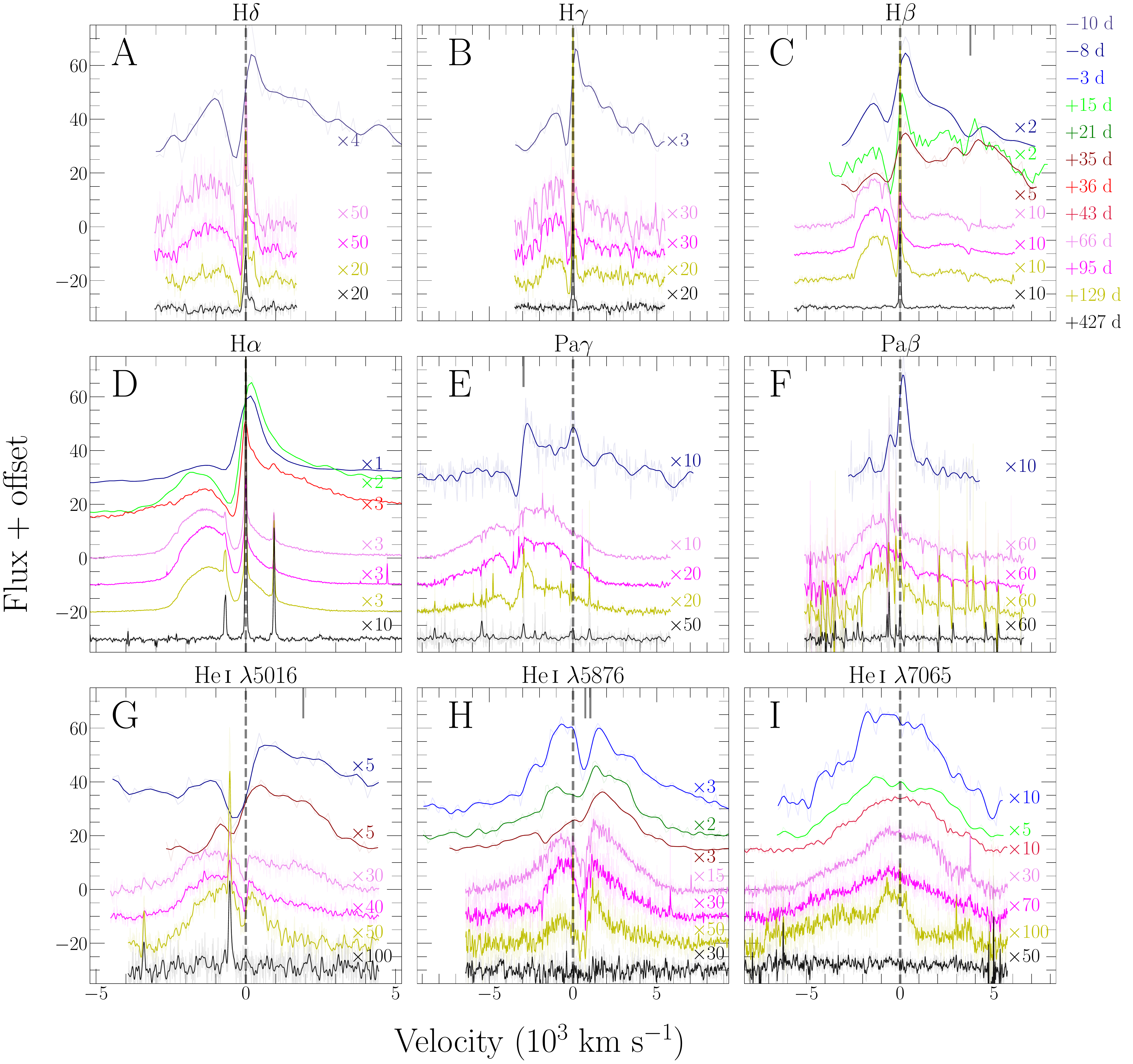}
\caption{Evolution of the the most prominent continuum-subtracted \ion{H}{i} and \ion{He}{i} lines of SN~2021foa. Dashed, vertical lines correspond to zero velocity. Solid, vertical lines in panels C, E, G and H correspond to the velocities at the wavelengths of \ion{He}{i}~$\lambda\lambda 4922,5048,10830$ and \ion{Na}{i}D, respectively. 
}
\label{fig:vel_all_HHe}
\end{figure*}

\subsubsection{{\rm H}$\alpha$}\label{app:ha}

Panel D in Fig.~\ref{fig:vel_all_HHe} shows the  evolution of H$\alpha$ between about one week prior $r$-band peak and to $+427$ days. The line is characterized by a narrow (FWHM $\sim 600$~km s$^{-1}$) and an intermediate  component with a FWHM that increases from $\sim 1500 - 4500$~km s$^{-1}$.
After maximum light, no major changes are observed aside from a decreasing flux at the red wing of the asymmetric profile.
After $+$60 days, the red-blue asymmetry has flipped, i.e. the blue wing has increased in strength over the red wing. 
The narrow emission component has faded after about $+51$ days, revealing the H$\alpha$ host galaxy emission (FWHM  $\lesssim 100$~km~s$^{-1}$) instead.
The apparent re-brightening of the blue peak of H$\alpha$ at $+$95 days is an artifact stemming from the extraction of the 2D spectrum at the location of H$\alpha$~(see Fig.~\ref{fig:hostlines}). 

\subsubsection{{\rm H}$\beta$}\label{app:hb}

The evolution of H$\beta$ line is shown in Panel C in Fig.~\ref{fig:vel_all_HHe}. 
The H$\beta$ line profile is similar to H$\alpha$ prior to peak magnitude. At later epochs, 
 H$\beta$ is blended with the strong \ion{He}{i}$~\lambda 4922$ emission.
Both H$\beta$ and \ion{He}{i}~$\lambda 4922$ exhibit narrow absorption lines with absorption velocities from $-10$ days onwards ($\sim - 600$ and $\sim -400$~km~s$^{-1}$, respectively). 
However, the narrow absorption line of \ion{He}{i} could be partially associated with \ion{Fe}{ii}$~\lambda 4924$ line of multiplet 42.
Similar to the H$\alpha$ line, the line flux of the H$\beta +$\ion{He}{i}$~\lambda 4922$ complex reaches a maximum around $+15$ days and decreases between $+28$ and $+66$ days. 
After this epoch, H$\beta$ shows a strong blue-shifted emission component while the velocity of the minimum of the absorption component reaches $\sim 300$~km~s$^{-1}$. The emission of \ion{He}{i}~$\lambda 4922$ is weak in comparison to H$\beta$, allowing us to disentangle both lines at these late epochs. Similar to H$\alpha$ and H$\beta$, \ion{He}{i}~$\lambda 4922$ also exhibits a blue-shifted emission component.

\subsubsection{\ion{He}{i}~$\lambda 5016$}\label{app:hei_5016}

In panel G (Fig.~\ref{fig:vel_all_HHe}), we show the evolution of \ion{He}{i} 
$\lambda 5016$ line. 
From $-10$ days onwards, this line shows a P-Cygni profile, with the absorption minimum at a velocity of $\sim -400$~km~s$^{-1}$.
Similarly to \ion{He}{i} $\lambda 4922$, the narrow absorption component can be associated with \ion{Fe}{ii}~$\lambda5018$ of multiplet 42.
The extension of the wings of the broad emission component may indicate a maximum velocity of $4000$~km~s$^{-1}$ (bulk velocity of $\sim 3000$~km~s$^{-1}$). 
After maximum light, the shape of the profile is broad and boxy-like. Assuming one component, the FWHM of this complex is about $3600$~km~s$^{-1}$. This is consistent with what is seen for H$\alpha$.
From $+95$ days onward, the flux at the red portion of the \ion{He}{i}~$\lambda 5016$ emission line profile rapidly decreases.
The region might be affected at early times by the emission of \ion{He}{i}$~\lambda 5048$ \AA.

\subsubsection{\ion{He}{i}~$\lambda 5876$}\label{app:hei_5876}

Panel H in Fig.~\ref{fig:vel_all_HHe} displays the evolution of \ion{He}{i} $\lambda 5876$, which is the strongest of all \ion{He}{i} lines in the entire VLT/X-shooter spectral wavelength range.  
There is no indication of a narrow P-Cygni profile as in other \ion{He}{i}~$\lambda\lambda 4922,5016$ lines. However, there is conspicuous, red-shifted absorption at all epochs up to $+129$ days, that likely is the \ion{Na}{i}D doublet from interstellar material along the line of sight. %
The emission line profile appears symmetric at early epochs (two weeks past maximum) in comparison to \ion{H}{i} lines. Thereafter, the line develops a red-shifted peak. From about $+66$ days onward it appears symmetric again, unlike other \ion{He}{i} and \ion{H}{i} lines. 

\subsubsection{\ion{He}{i}~$\lambda 7065$}\label{app:hei_7065}

In panel I (Fig.~\ref{fig:vel_all_HHe}), we show the evolution of \ion{He}{i}~$\lambda 7065$. 
 In analogy to \ion{He}{i}$~\lambda 5876$, the emission line profile of \ion{He}{i}$~\lambda 7065$ is of boxy-like shape with no evident narrow P-Cygni profile. 
 Additionally, at all epochs the line profile exhibits a blue shoulder at around $-$3000~km~s$^{-1}$. The origin of both emission features is unclear but may be due to another element.

\subsubsection{{\rm NIR} lines} \label{app:nir}

The NIR spectra exhibit emission lines of \ion{He}{i}~$\lambda 10830$ + Pa$\gamma$, Pa$\beta$ and \ion{He}{i}$~\lambda 20581$. \ion{H}{i}~$\lambda  18751$ (Pa$\alpha$) is detected, but it coincides with a telluric region. 
Panel E in Fig.~\ref{fig:vel_all_HHe} shows the \ion{He}{i}~$\lambda 10830$ + Pa$\gamma$ line complex, which is dominated by \ion{He}{i}~$\lambda 10830$.
We find that the absorption line at about $-4000$~km~s$^{-1}$ (with reference to Pa$\gamma$) must be attributed to \ion{He}{i}~$\lambda 10830$. 
This, because it is unlikely that a narrow Pa$\gamma$ absorption at a velocity of about $-4000$~km~s$^{-1}$ with a FWHM of only about $800$~km~s$^{-1}$ exists. 
Furthermore, the velocity of the absorption minimum, if associated with \ion{He}{i}~$\lambda 10830$, remains nearly constant at about $-600$~km~s$^{-1}$ at all epochs up to $+$129 days.

Panel F in Fig.~\ref{fig:vel_all_HHe} displays the evolution of Pa$\beta$. 
At $-8$ days, the emission line has a FWHM of only about $2000$~km~s$^{-1}$ and thus, is narrower than other \ion{H}{i} lines at that epoch. On top of that is a narrow P-Cygni profile with a FWHM of about $\sim 1000$~km~s$^{-1}$ and an absorption component with a minimum at $\sim -500$~km~s$^{-1}$.
From $+66$ to $+129$ days, the Pa$\beta$ line profile is nearly identical to the optical \ion{H}{i} lines.
However, Pa$\beta$ does not show a narrow absorption component, likely because it coincides with telluric lines at that position.  

\subsubsection{Other lines profiles}\label{app:other}

\begin{figure}[hbt!]
\epsscale{1.2}
\plotone{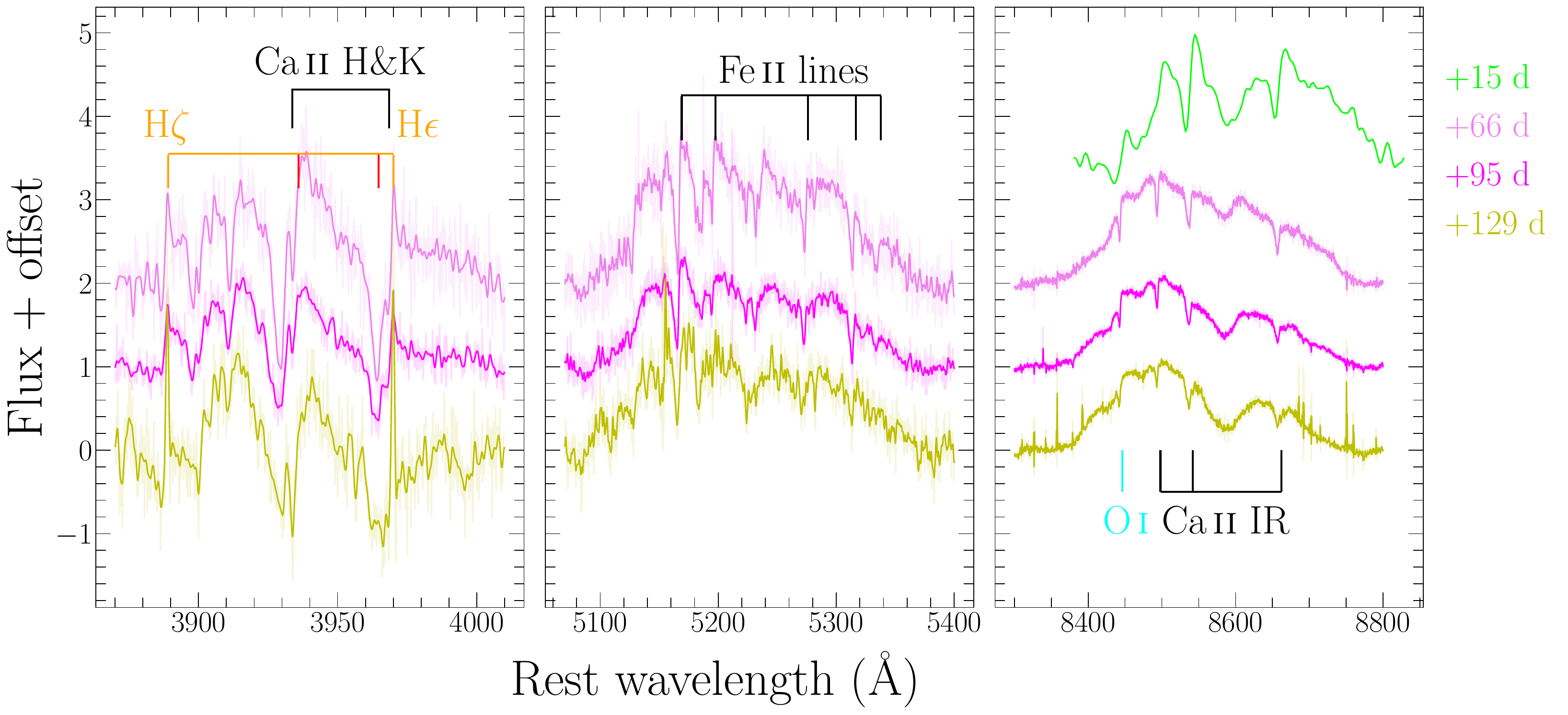}
\caption{{\it Left panel}: Evolution of the spectral region surrounding \ion{Ca}{ii} H$\&$K. The P-Cygni profile of the Balmer lines H$\zeta$ and H$\epsilon$ are prominent. {\it Middle panel}: 
Evolution of the spectral region surrounding the pseudo-continuum produced by the blended \ion{Fe}{ii} lines.
{\it Right panel}: 
Evolution of the \ion{Ca}{ii} NIR + \ion{O}{i}~$\lambda 8446$. All the regions in each panel are continuum-subtracted.
}
\label{fig:all_other_lines}
\end{figure}

Another complex spectral structure is prominent in the wavelength range of \ion{Ca}{ii} {\rm H\&K} ($\lambda\lambda 3934,3968$). Left panel of Fig.~\ref{fig:all_other_lines} shows the evolution of this structure at $+66$, $+95$ and $+129$ days.
The absorption complex at the position of \ion{Ca}{ii} {\rm H\&K} 
exhibit multiple components, some appear blue-shifted with respect to the SN redshift. 
Strong absorption features, potentially associated with these \ion{Ca}{ii} lines are at about $-350$~km~s$^{-1}$. Since there is no indication of any emission or absorption lines at these wavelengths in the $+427$ day spectrum, it is likely that the absorption complex originates from different CSM layers around the supernova. 
However, the mismatch between the low velocities from the absorption trough of  the \ion{Ca}{ii} IR and the \ion{Ca}{ii}~{\rm H\&K} counterparts is intriguing.  
This apparent discrepancy could be solved considering that \ion{Ca}{ii}~{\rm H\&K} could be misidentified due to the strong blending with \ion{He}{i}~$\lambda\lambda 3936,3965$~(vertical red lines in {\it left} panel).

Several \ion{Fe}{ii} (multiplet 42, 48 and 49) lines are shown in middle panel in Fig.~\ref{fig:all_other_lines} at $+$66 to $+$427 days. Strong blending of \ion{Fe}{ii} forest lines can create a pseudo-continuum as observed bluewards of 5700 {\AA}. This is not unusual for Type Ibn and other interacting SNe~\citep{Pastorello_LaSilla_2015}.  %
Strong P-Cygni absorption components at about $<-300$~km~s$^{-1}$ are detected for all \ion{Fe}{ii} lines.

Right panel in Fig.~\ref{fig:all_other_lines} displays the evolution of the \ion{Ca}{ii} $\lambda\lambda 8498,8542,8662$, possibly blended with \ion{O}{i} $\lambda 8446$. 
While it is difficult to disentangle these four lines 
we find that the bulk velocity of the \ion{Ca}{ii} + \ion{O}{i} complex does not surpass $6000$~km~s$^{-1}$. 
This is the maximum velocity of the red wing of \ion{Ca}{ii} $\lambda 8662$. 
Furthermore, at all epochs past $+15$ days, narrow P-Cygni absorption is observed
at a velocity of about $-400$~km~s$^{-1}$, which continuously decreases to $-200$~km~s$^{-1}$ between $+$40 and $+$129 days.

\subsubsection{On SN~2006jc and SN~2011hw}\label{app:06jc_11hw}

\begin{figure}[htp!]
\epsscale{.9}
\plotone{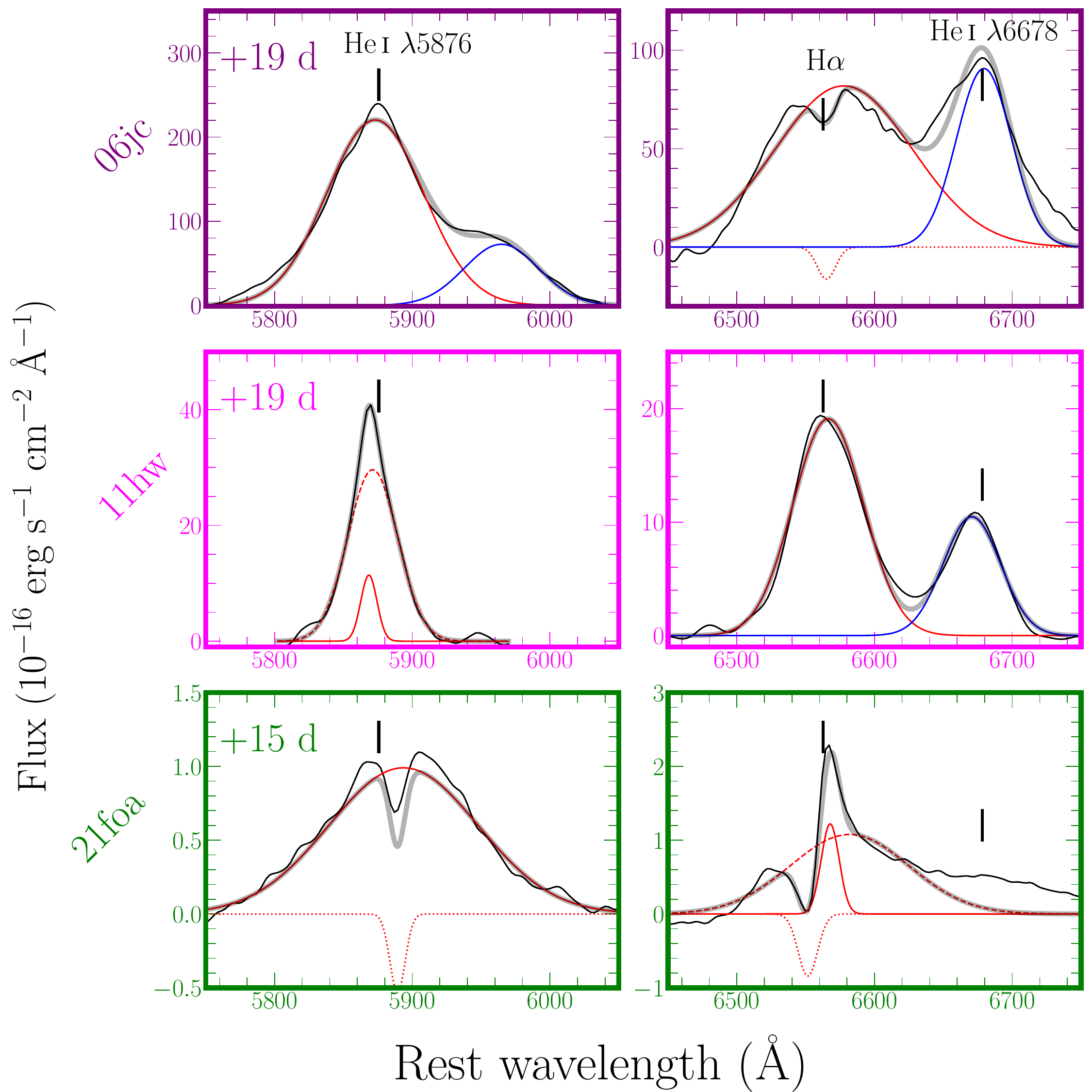}
\caption{Decomposition of continuum-subtracted line profiles of \ion{He}{i}$~\lambda 5876$ ({\it left column}) and H$\alpha$ ({\it right column}) of Type Ibn SN~2006jc ({\it upper row}), transitional Type IIn/Ibn SN~2011hw ({\it middle row}) and SN~2021foa ({\it lower row}) at two weeks after $r$-band maximum. Solid and dashed lines represent the fits to the emission components of the line profiles, while dotted lines correspond to the fit to any absorption trough. 
}
\label{fig:decomp_Ibn}
\end{figure}

In Fig.~\ref{fig:decomp_Ibn} we show the decomposition of the line profiles of \ion{He}{i}$~\lambda 5876$ ({\it left} column) and H$\alpha$ ({\it right} column) of three different Type Ibn SNe: the prototype SN~2006jc, the transitional SN~2011hw and SN~2021foa two weeks after $r$-band maximum. It is clear that our decomposition is in very good agreement with the total line-flux of each profile. Furthermore, in contrast to~\citet{Smith_dust_2006jc, Smith_2012_2011hw}, we show that our decomposition correctly deblends the H$\alpha$ and \ion{He}{i}~$\lambda 6678$.

\begin{figure}[hbt!]
\epsscale{.9}
\plotone{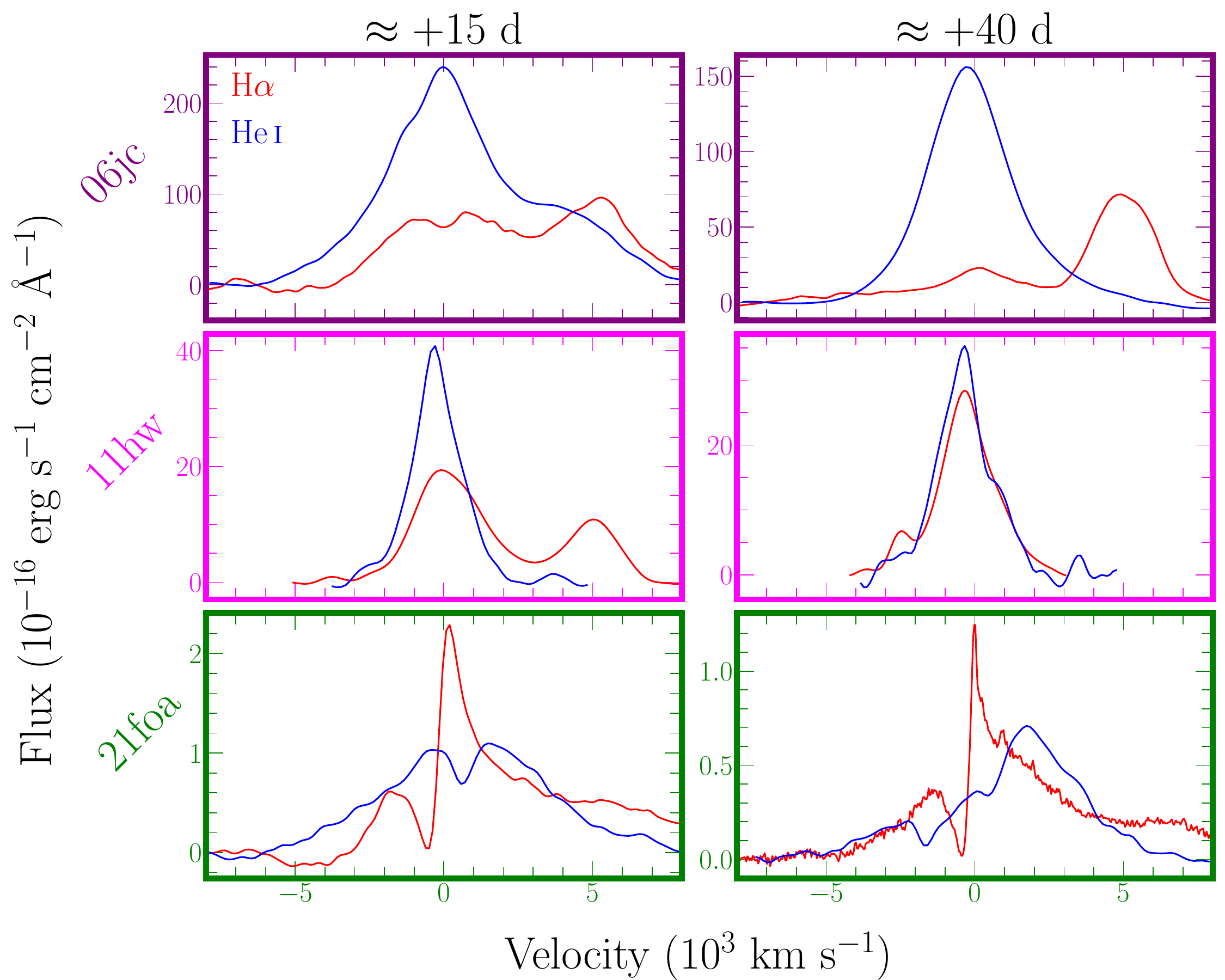}
\caption{Comparison of continuum-subtracted line profiles of H$\alpha$ (red) and \ion{He}{i}~$\lambda 5876$ (blue) of Type Ibn SN~2006jc ({\it upper row}), transitional IIn/Ibn SN~2011hw ({\it middle row}) and SN~2021foa ({\it lower row}) at $\approx 15$ ({\it left column}) and $\approx 40$ ({\it right column}) days past $r$-band maximum.  
}
\label{fig:proof_fluxratio}
\end{figure}

In Fig.~\ref{fig:proof_fluxratio}, we show the flux-calibrated H$\alpha$ (red) and \ion{He}{i}~$\lambda 5876$ (blue) profiles of SN~2006jc, SN~2011hw and SN~2021foa at$~\approx 15$ ({\it left panel}) and $\approx 40$ ({\it right panel}) days $r$-band maximum. These two epochs encompass the flux-line ratio plateau observed in Fig.~\ref{fig:ratio} for transitional IIn/Ibn SNe. For SN~2006jc, the flux-line of H$\alpha$ is smaller than that of \ion{He}{i} in both epochs. This difference is not observed in either SN~2011hw and SN~2021foa, where the flux-line ratio is $\approx 1$ for both epochs. Fig.~\ref{fig:proof_fluxratio} proves that the values obtained in~\citet{Reguitti_2022} of the line-flux ratio of H$\alpha$/\ion{He}{i}~$\lambda 5876$, $\approx 0.5$, are not consistent with our observations.

\subsubsection{Line decomposition in transitional objects}\label{app:line_trans}

For SN~2005la and iPTF15akq, a careful analysis was done to take into account the fact that these sources exhibit strong, broad P-Cygni absorption profiles for both H and or/He at early epochs. However, we note that this absorption was not accounted for in the line fluxes calculations.

For SN~2011hw, SN~2005la, SN~2020bqj and SN~2006jc, we deblended the H$\alpha$ from \ion{He}{i} $\lambda 6678$ line by fitting two different Gaussian profiles to each line. 
This results in line fluxes of H$\alpha$ and \ion{He}{i}$~\lambda 5876$ for SN~2006jc and SN~2011hw that are discrepant by a factor 2 from measurements in the literature \citep{Smith_dust_2006jc,Smith_2012_2011hw} using different methods.
In Fig.~\ref{fig:decomp_Ibn} we show that our decomposition of the line profiles of H$\alpha$ and \ion{He}{i}~$\lambda 5876$ light recovers the total line-flux at~$\approx$ two weeks after maximum.

\startlongtable
\begin{deluxetable}{lcccccc}
\tablecaption{Photometry of SN2021foa}
\tablehead{\colhead{MJD} & \colhead{Filter} &\colhead{Magnitude} & \colhead{Error} & \colhead{System} & \colhead{Instrument} & \colhead{Telescope}} 
\startdata
58147.50 & $o$ & 19.046 & 0.264 & AB & ACAM1 & ATLAS \\
58295.50 & $o$ & 19.403 & 0.357 & AB & ACAM1 & ATLAS \\
59019.50 & $o$ & 18.336 & 0.303 & AB & ACAM1 & ATLAS \\
59267.50 & $o$ & 19.224 & 0.291 & AB & ACAM1 & ATLAS \\
59271.50 & $o$ & 19.102 & 0.297 & AB & ACAM1 & ATLAS \\
59272.50 & $o$ & 19.351 & 0.288 & AB & ACAM1 & ATLAS \\
59277.50 & $o$ & 19.268 & 0.318 & AB & ACAM1 & ATLAS \\
59279.50 & $o$ & 19.111 & 0.190 & AB & ACAM1 & ATLAS \\
59288.50 & $o$ & 16.002 & 0.018 & AB & ACAM1 & ATLAS \\
59289.26 & $UVW1$ & 15.139 & 0.050 & Vega & UVOT & Swift \\
59289.27 & $U$ & 14.820 & 0.043 & Vega & UVOT & Swift \\
59289.27 & $B$ & 15.916 & 0.060 & Vega & UVOT & Swift \\
59289.29 & $UVW2$ & 15.957 & 0.061 & Vega & UVOT & Swift \\
59289.29 & $V$ & 15.678 & 0.086 & Vega & UVOT & Swift \\
59289.33 & $U$ & 15.573 & 0.078 & Vega & UVOT & Swift \\
59290.79 & $UVW2$ & 15.790 & 0.049 & Vega & UVOT & Swift \\
59290.79 & $V$ & 15.419 & 0.078 & Vega & UVOT & Swift \\
59290.79 & $U$ & 15.356 & 0.049 & Vega & UVOT & Swift \\
59290.79 & $U$ & 14.605 & 0.043 & Vega & UVOT & Swift \\
59290.80 & $UVW1$ & 14.958 & 0.048 & Vega & UVOT & Swift \\
59290.80 & $B$ & 15.759 & 0.067 & Vega & UVOT & Swift \\
59290.87 & $gp$ & 15.558 & 0.016 & AB & Sinistro & LCO \\
59290.87 & $rp$ & 15.649 & 0.017 & AB & Sinistro & LCO \\
59290.87 & $ip$ & 15.803 & 0.021 & AB & Sinistro & LCO \\
59291.52 & $B$ & 15.591 & 0.019 & AB & Sinistro & LCO \\
59291.52 & $V$ & 15.343 & 0.015 & AB & Sinistro & LCO \\
59291.52 & $R$ & 15.394 & 0.014 & AB & Sinistro & LCO \\
59291.64 & $UVW2$ & 15.715 & 0.054 & Vega & UVOT & Swift \\
59291.65 & $U$ & 14.439 & 0.045 & Vega & UVOT & Swift \\
59291.65 & $U$ & 15.316 & 0.054 & Vega & UVOT & Swift \\
59291.65 & $B$ & 15.667 & 0.079 & Vega & UVOT & Swift \\
59291.65 & $V$ & 15.220 & 0.085 & Vega & UVOT & Swift \\
59291.65 & $UVW1$ & 14.837 & 0.051 & Vega & UVOT & Swift \\
\enddata
\tablecomments{Full table available in electronic form.}
\end{deluxetable}
\label{tab:photo}

\startlongtable
\begin{deluxetable}{lllccc}
\tablecaption{Spectroscopy of SN2021foa}
\tablehead{\colhead{MJD} & \colhead{Phase (d)} & \colhead{Coverage (\AA)} & \colhead{Dispersion (\AA)} & \colhead{Instrument} & \colhead{Telescope}} 
\startdata
2021-03-18 & $-11$ & $3768-6939$ & $0.76-1.24$ & WiFeS & ANU \\
2021-03-18 & $-11$ & $5680-8580$ & $1.4$ & ALFOSC & NOT \\ 
2021-03-21 & $-8$ & $3877-7037$ & $1.98$ & Goodman & SOAR \\
2021-03-21 & $-8$ & $6845-25485$ & $1.19-3.55$ & SpeX & IRTF \\
2021-03-22 & $-7$ & $3345-10504$ & $2.51$ & Kast& Shane \\
2021-03-22 & $-7$ & $3768-8923$ & $3.35$ & ALFOSC & NOT \\
2021-03-23 & $-6$ & $5680-8580$ & $1.4$ & ALFOSC & NOT \\
2021-04-06 & $+8$ & $3345-10504$ & $2.51$ & Kast & Shane \\
2021-04-10 & $+12$ & $6845-25485$ & $1.19-3.55$ & SpeX & IRTF \\
2021-04-13 & $+15$ & $3345-10504$ & $2.51$ & Kast & Shane \\
2021-04-19 & $+21$ & $3345-10504$ & $2.51$ & Kast & Shane \\
2021-05-03 & $+35$ & $3345-10504$ & $2.51$ & Kast & Shane \\
2021-05-09 & $+41$ & $7091-25485$ & $1.19-3.55$ & SpeX & IRTF \\
2021-05-10 & $+42$ & $3345-10504$ & $2.51$ & Kast & Shane \\
2021-05-19 & $+51$ & $3768-6939$ & $0.76-1.24$ & WiFeS & ANU \\
2021-05-10 & $+51$ & $3345-10504$ & $2.51$ & Kast & Shane \\
2021-06-03 & $+66$ & $2964-24583$ & $0.19-0.59$ & X-shooter & VLT \\
2021-07-02 & $+95$ & $2964-24583$ & $0.19-0.59$ & X-shooter & VLT \\
2021-08-04 & $+129$ & $2964-24583$ & $0.19-0.59$ & X-shooter & VLT \\
2022-05-30 & $+427$ & $2964-24583$ & $0.19-0.59$ & X-shooter & VLT \\
\enddata
\end{deluxetable}
\label{tab:spec}

\begin{longrotatetable}
\centerwidetable
\begin{deluxetable*}{lcccccccccccccccccc}
\tablecaption{Full Width at Half-maximum of \ion{H}{i} and \ion{He}{i} lines of SN~2021foa.}
\tablehead{
\colhead{Epoch (days)} &
\multicolumn{4}{c}{
Narrow }& \colhead{} &
\multicolumn{4}{c}{
Absorption } & \colhead{} &
\multicolumn{6}{c}{
Broad} & \colhead{} &
\\
\cline{2-5}
\cline{7-10}
\cline{12-17}
\colhead{} &
\colhead{H$\delta$} & \colhead{H$\gamma$} & \colhead{H$\beta$} & \colhead{H$\alpha$} &
\colhead{} & 
\colhead{H$\delta$} & \colhead{H$\gamma$} & \colhead{H$\beta$} & \colhead{H$\alpha$} &
\colhead{} & 
\colhead{H$\delta$} & \colhead{H$\gamma$} & \colhead{H$\beta$} & \colhead{H$\alpha$} & \colhead{Pa$\beta$} & \colhead{\ion{He}{i}~$\lambda 5876$}
}
\startdata
$-11/10$ & $-$ & $-$               & $1.4(1.2)$ & $3.7(0.1)$ && $-$ & $-$               & $4.3(1.0)$ & $-$        && $-$ & $-$                 & $23.5(3.1)$ & $14.9(1.0)$ & $-$         & $-$ \\
$-3$     & $-$ & $-$               & $8.6(1.0)$ & $5.5(1.0)$ && $-$ & $-$               & $8.7(1.0)$ & $6.4(0.2)$ && $-$ & $-$                 & $39.5(8.6)$ & $40.3(3.0)$ & $-$         & $-$ \\
$+7/8$   & $-$ & $-$               & $2.0(1.7)$ & $4.6(0.4)$ && $-$ & $-$               & $4.3(1.0)$ & $6.7(0.3)$ && $-$ & $-$                 & $25.6(2.2)$ & $40.3(1.3)$ & $-$         & $64.8(-)$ \\
$+15$    & $-$ & $-$               & $2.7(1.2)$ & $6.5(0.6)$ && $-$ & $-$               & $6.6(0.4)$ & $7.1(0.4)$ && $-$ & $-$                 & $32.2(2.0)$ & $46.1(1.6)$ & $-$         & $61.8(-)$ \\
$+21$    & $-$ & $-$               & $2.4(4.7)$ & $5.4(0.8)$ && $-$ & $-$               & $11.1(1.3)$& $6.7(0.4)$ && $-$ & $-$                 & $32.0(2.4)$ & $48.7(1.3)$ & $-$         & $56.7(-)$ \\
$+28$    & $-$ & $-$               & $8.0(1.0)$ & $5.9(1.0)$ && $-$ & $-$               & $8.0(1.0)$ & $5.9(0.8)$ && $-$ & $-$                 & $27.7(3.6)$ & $50.3(0.6)$ & $-$         & $55.0(-)$ \\
$+35/36$ & $-$ & $-$               & $2.2(5.7)$ & $2.0(0.1)$ && $-$ & $-$               & $6.2(0.2)$ & $8.5(0.4)$ && $-$ & $-$                 & $23.5(1.9)$ & $46.8(0.4)$ & $-$         & $51.5(-)$ \\
$+42/43$ & $-$ & $-$               & $3.7(1.0)$ & $1.1(0.2)$ && $-$ & $-$               & $1.6(2.1)$ & $7.6(0.4)$ && $-$ & $-$                 & $22.4(1.4)$ & $48.0(0.6)$ & $-$         & $47.0(-)$ \\
$+51$    & $-$ & $-$               & $-$        & $1.4(0.1)$ && $-$ & $-$               & $-$        & $5.4(0.7)$ && $-$ & $-$                 & $-$         & $41.6(1.0)$ & $-$         & $43.5(-)$ \\
$+66$    & $0.6(0.1)$ & $0.4(0.1)$ & $0.2(0.1)$ & $0.6(0.0)$ && $4.0(0.9)$ & $3.0(0.2)$ & $3.0(0.2)$ & $3.4(0.1)$ && $21.7(1.1)$ & $27.6(1.1)$ & $24.2(0.4)$ & $36.2(0.4)$ & $32.3(0.7)$ & $47.6(-)$ \\
$+95$    & $0.6(0.1)$ & $0.4(0.2)$ & $0.4(0.0)$ & $0.7(0.0)$ && $3.7(0.5)$ & $3.0(0.1)$ & $3.0(0.2)$ & $3.3(0.1)$ && $19.7(0.9)$ & $21.4(0.6)$ & $21.7(0.3)$ & $29.3(0.2)$ & $28.7(0.6)$ & $46.1(-)$ \\
$+129$   & $0.2(0.1)$ & $0.4(0.0)$ & $0.4(0.0)$ & $0.4(0.0)$ && $5.1(0.4)$ & $2.9(0.2)$ & $3.4(0.1)$ & $3.3(0.2)$ && $9.3(0.8)$  & $21.2(1.0)$ & $17.0(0.8)$ & $26.0(0.4)$ & $25.1(1.0)$ & $-$ \\
\enddata
\tablecomments{Values of the FWHM are in units of $100$ km s$^{-1}$. All values are corrected by the resolution of the instrument ($\approx$ dispersion in Tab.~\ref{tab:spec}). Given the complexity of \ion{He}{i}~$\lambda 5876$ line profile, only upper limits are reported.}
\end{deluxetable*}
\label{tab:FWHM}
\end{longrotatetable}

\begin{longrotatetable}
\centerwidetable
\begin{deluxetable*}{lcccccccccccccccccc}
\tablecaption{Fitted absorption minimum of several spectral lines of SN~2021foa.}
\tablehead{
\colhead{Epoch (days)} &
\multicolumn{4}{c}{
\ion{H}{i}} & \colhead{} &
\multicolumn{2}{c}{
\ion{He}{i}} & \colhead{} &
\multicolumn{3}{c}{
\ion{Fe}{ii}} & \colhead{} &
\multicolumn{3}{c}{
\ion{Ca}{ii}} & \colhead{} &
\multicolumn{1}{c}{
\ion{O}{i}} & \colhead{}
\\
\cline{2-5}
\cline{7-8}
\cline{10-12}
\cline{14-16}
\cline{18-18}
\colhead{} & 
\colhead{H$\delta$} & \colhead{H$\gamma$} & \colhead{H$\beta$} & \colhead{H$\alpha$} &
\colhead{} & 
\colhead{$\lambda 4923$} & \colhead{$\lambda 5016$} & 
\colhead{} &
\colhead{$\lambda 5169$} & \colhead{$\lambda 5276$} &  \colhead{$\lambda 5317$} &
\colhead{} &
\colhead{$\lambda 8498$} & \colhead{$\lambda 8542$} &  \colhead{$\lambda 8662$} &
\colhead{} &
\colhead{$\lambda 8446$}
}
\startdata
$-10$    & $3.5(1.2)$ & $3.3(1.1)$ & $4.1(1.0)$ &     $-$    && $4.2(1.0)$ & $3.1(1.0)$ &&    $-$     &     $-$    &     $-$    && $-$        & $-$        & $-$        && $-$ \\
$-3$     &    $-$     &    $-$     & $6.6(1.9)$ & $5.9(1.4)$ && $4.2(1.8)$ & $3.4(1.8)$ &&    $-$     &     $-$    &     $-$    && $-$        & $-$        & $-$        && $-$\\
$+7/8$   & $3.6(1.2)$ & $3.7(1.1)$ & $5.3(1.0)$ & $5.6(1.1)$ && $4.0(1.0)$ &     $-$    &&    $-$     &     $-$    &     $-$    && $-$        & $3.3(0.9)$ & $3.9(0.9)$ && $2.8(0.9)$\\
$+15$    &    $-$     &    $-$     & $5.9(1.6)$ & $5.1(1.1)$ && $3.8(1.5)$ & $3.3(1.5)$ &&    $-$     &     $-$    &     $-$    && $-$        & $3.3(0.9)$ & $2.9(0.9)$ && $3.2(0.9)$\\
$+21$    &    $-$     &    $-$     & $5.1(1.6)$ & $6.2(1.1)$ && $2.3(1.5)$ &     $-$    &&    $-$     &     $-$    &     $-$    && $-$        & $-$        & $-$        && $-$\\
$+28$    &    $-$     &    $-$     & $5.8(1.9)$ & $6.5(1.4)$ && $4.8(1.8)$ &     $-$    &&    $-$     &     $-$    &     $-$    && $-$        & $-$        & $-$        && $-$\\
$+35/36$ &    $-$     &    $-$     & $5.9(1.6)$ & $4.8(0.4)$ && $2.2(1.5)$ & $4.0(1.5)$ &&    $-$     &     $-$    &     $-$    && $-$        & $3.1(0.9)$ & $2.8(0.9)$ && $2.8(0.9)$\\
$+42/43$ &    $-$     &    $-$     & $4.2(1.6)$ & $4.0(0.5)$ && $1.2(1.5)$ & $1.9(1.5)$ &&    $-$     &     $-$    &     $-$    && $2.3(0.9)$ & $2.2(0.9)$ & $2.6(0.9)$ && $3.4(0.9)$\\
$+66$    & $2.2(0.1)$ & $2.9(0.1)$ & $2.9(0.1)$ & $3.8(0.1)$ && $0.9(0.1)$ & $0.8(0.1)$ && $1.8(0.1)$ & $1.6(0.1)$ & $1.6(0.1)$ && $1.7(0.1)$ & $2.1(0.1)$ & $1.9(0.1)$ && $1.3(0.1)$\\
$+95$    & $2.7(0.1)$ & $2.9(0.1)$ & $2.7(0.1)$ & $3.5(0.1)$ && $1.1(0.1)$ & $0.7(0.1)$ && $2.0(0.1)$ & $1.8(0.1)$ & $1.8(0.1)$ && $1.7(0.1)$ & $2.1(0.1)$ & $1.9(0.1)$ && $1.4(0.1)$\\
$+129$   & $3.2(0.1)$ & $2.7(0.1)$ & $2.7(0.1)$ & $3.5(0.1)$ && $1.1(0.1)$ & $1.5(0.1)$ && $2.1(0.1)$ & $0.7(0.1)$ & $1.8(0.1)$ && $1.8(0.1)$ & $2.1(0.1)$ & $1.8(0.1)$ && $1.5(0.1)$\\
\enddata
\tablecomments{Values of the absorption minimum are reported in units of $-100$ km s$^{-1}$. Uncertainties are estimated as 
$c\cdot \Delta\lambda_D / \lambda_0$, with $c$ the speed of light, 
$\Delta\lambda_D$ the dispersion of the instrument in Tab.~\ref{tab:spec}, and $\lambda_0$ the center of the Gaussian profile.}
\end{deluxetable*}
\label{tab:absmin}
\end{longrotatetable}

\startlongtable
\begin{deluxetable*}{lccccccc}
\tablecaption{Line fluxes of \ion{H}{i} and \ion{He}{i} lines of SN~2021foa.}
\tablehead{
\colhead{Epoch (days)} &
\colhead{H$\delta$} & \colhead{H$\gamma$} & \colhead{H$\beta$} & \colhead{H$\alpha$} &
 \colhead{Pa$\beta$} & \colhead{\ion{He}{i}~$\lambda  5876$  } & \colhead{\ion{He}{i}~$\lambda  7065$  } }
\startdata
$-11/10$ & $29.3(2.4)$  & $40.7(2.6)$  & $46.4(4.7)$    & $86.9(7.3)$    &     $-$    & $25.6(-)$  &     $-$      \\
$-8/7$   &     $-$     &     $-$       & $52.2(3.6)$    & $105.6(15.6)$  &     $-$    & $35.9(-)$  & $14.3(1.3)$  \\
$-3$     &     $-$     &     $-$       & $77.7(4.4)$    & $105.7(11.6)$  &     $-$    & $40.3(-)$  &     $-$      \\
$+7/8$   &$24.6(2.2)$  & $37.2(3.0)$   & $54.2(3.3)$    & $131.3(16.3)$  &     $-$    & $129.2(-)$ & $48.1(1.7)$  \\
$+15$    &     $-$     &     $-$       & $35.9(11.4)$   & $120.0(15.7)$  &     $-$    & $138.5(-)$ & $59.7(1.8)$  \\
$+21$    &     $-$     &     $-$       & $43.4(18.1)$   & $139.0(15.8)$  &     $-$    & $137.1(-)$ & $60.3(2.0)$  \\
$+28$    &     $-$     &     $-$       & $19.1(8.0)$    & $121.2(14.7)$  &     $-$    & $96.4(-)$  & $69.8(1.8)$  \\
$+35/36$ &     $-$     &     $-$       & $11.2(4.8)$    & $59.9(16.7)$   &     $-$    & $61.3(-)$  & $32.8(0.7)$  \\
$+42/43$ &     $-$     &     $-$       & $10.9(5.5)$    & $56.0(16.7)$   &     $-$    & $39.9(-)$  & $24.5(0.5)$  \\
$+51$    &     $-$     &     $-$       &      $-$       & $31.7(20.9)$   &     $-$    & $32.0(-)$  & $19.8(1.1)$  \\
$+66$    & $0.8(0.0)$  &  $1.7(0.1)$   & $6.4(0.1)$     & $38.0(0.1)$    & $3.8(0.0)$ & $14.3(-)$  & $9.3(0.0)$   \\
$+95$    & $0.5(0.0)$  &  $1.2(0.0)$   & $5.9(0.0)$     & $38.4(0.1)$    & $3.5(0.0)$ & $5.9(-)$   & $3.2(0.0)$   \\
$+129$   & $0.3(0.0)$  &  $1.0(0.0)$   & $5.5(0.1)$     & $31.3(0.1)$    & $3.7(0.0)$ & $1.7(-)$   & $2.0(0.0)$   \\
\enddata
\tablecomments{Values are reported in units of $10^{-17}$ erg s$^{-1}$ cm$^{-2}$. Line fluxes of \ion{He}{i}~$\lambda 5876$ correspond to the integration of the data over $\approx \pm 5000$ km s$^{-1}$ with respect to 5876 \AA.}
\end{deluxetable*}
\label{tab:flux}

\bibliography{corecollapse}{}
\bibliographystyle{aasjournal}

\end{document}